\documentclass{aa}
\usepackage{graphicx}
\usepackage{amsmath}
\usepackage{txfonts}
\usepackage{hyperref}
\hypersetup{colorlinks, allcolors=blue}
\usepackage{orcidlink}
\usepackage{tabularx}
\usepackage{overpic}
\usepackage{multicol}
\usepackage{xcolor}
\renewcommand{\arraystretch}{1.3}

\makeatletter\newcommand{\setcaptype}[1]{\def\@captype{#1}}\makeatother

\begin{document}

   \title{The ESO SupJup Survey}
   \subtitle{XI. Atmospheric properties of six isolated M- and L-type dwarfs with CRIRES+}
   \titlerunning{SupJup XI: Atmospheric properties of isolated M \& L dwarfs}

   \author{
C. R. Malcolm\inst{\ref{instLeiden}}\orcidlink{0009-0004-1091-898X} \and
N. Grasser\inst{\ref{instLeiden}}\orcidlink{0009-0009-6634-1741} \and
I. A. G. Snellen\inst{\ref{instLeiden}}\orcidlink{0000-0003-1624-3667} \and
S. de Regt\inst{\ref{instLeiden}}\orcidlink{0000-0003-4760-6168} \and
D. Gonz\'alez Picos\inst{\ref{instLeiden}}\orcidlink{0000-0001-9282-9462} \and
Y. Zhang\inst{\ref{instCalTech}}\orcidlink{0000-0003-0097-4414} \and
T. Stolker\inst{\ref{instLeiden}}\orcidlink{0000-0002-5823-3072} \and
S. Gandhi\inst{\ref{instWarwick},\ref{instCEH}}\orcidlink{0000-0001-9552-3709} \and
P. Mollière\inst{\ref{instMPIA}}\orcidlink{0000-0003-4096-7067}\and
E. Nasedkin\inst{\ref{instTCD}}\orcidlink{0000-0002-9792-3121} \and
R. Landman\inst{\ref{instLeiden}}\orcidlink{0000-0002-7261-8083} \and
A. Y. Kesseli\inst{\ref{instIPAC}}\orcidlink{0000-0002-3239-5989}}

\institute{
Leiden Observatory, Leiden University, Postbus 9513, 2300 RA, Leiden, The Netherlands \\
\email{cmalcolm@strw.leidenuniv.nl} \label{instLeiden}
\and
Department of Astronomy, California Institute of Technology, Pasadena, CA 91125, USA \label{instCalTech} \and
Department of Physics, University of Warwick, Coventry CV4 7AL, UK \label{instWarwick} \and
Centre for Exoplanets and Habitability, University of Warwick, Gibbet Hill Road, Coventry CV4 7AL, UK \label{instCEH} \and
Max-Planck-Institut für Astronomie, Königstuhl 17, 69117 Heidelberg, Germany \label{instMPIA}\and
School of Physics, Trinity College Dublin, University of Dublin, Dublin, Ireland \label{instTCD} \and
IPAC, Mail Code 100-22, Caltech, 1200 E. California Boulevard, Pasadena, CA 91125, USA \label{instIPAC}
}
\date{}

\abstract 
{The distinct formation pathways of brown dwarfs and giant exoplanets may be encoded in their atmospheric elemental and isotopic composition. The ESO SupJup Survey uses high-resolution CRIRES+ spectroscopy to systematically characterise the $^{12}$C/$^{13}$C isotope ratio, C/O ratio, and metallicity across a sample of 49 isolated brown dwarfs, hosts, and companions.}
{We present atmospheric retrievals for six isolated brown dwarfs of spectral types M7--L2.5, aiming to constrain their thermal structures, chemical compositions, and isotope ratios.}
{We analysed CRIRES+ $K$-band spectra using retrievals that couple the radiative transfer code \texttt{petitRADTRANS} with the nested sampling algorithm \texttt{PyMultiNest}. Both free- and equilibrium-chemistry frameworks were employed for each target.}
{The L0 dwarf 2MASS~J09532126$-$1014205 emerges as one of the fastest-rotating ultracool dwarfs characterised to date, with $v\sin i = 86.9\pm0.6$~km\,s$^{-1}$. H$_2^{16}$O and $^{12}$CO are strongly detected in all six targets. $^{13}$CO is significantly detected in DENIS~J060852.8$-$275358 (S/N\,$=$\,5.6) and tentatively in three further targets (S/N\,$=$\,2.7--4.1). Retrieved compositions are consistent with isolated brown dwarfs: near-solar C/O ratios ($0.51$--$0.63$), predominantly near-solar metallicities, and $^{12}$C/$^{13}$C ratios of $\sim$91--309, at or above the local interstellar medium (ISM) value, with constraints for the two fastest rotators resting on the spectral fit but not corroborated by a $^{13}$CO cross-correlation peak. Apparent H$_2^{18}$O constraints for two targets are found to be spurious, and their H$_2^{16}$O/H$_2^{18}$O ratios are presented as lower limits.}
{The near-solar C/O ratios and metallicities, with $^{12}$C/$^{13}$C ratios at or above the ISM value, support a molecular cloud fragmentation origin for the sample. The agreement of $^{12}$C/$^{13}$C between chemistry frameworks supports the robustness of these ratios. Spurious H$_2^{18}$O constraints demonstrate the importance of cross-correlation validation for minor species detections.}


   \maketitle

\section{Introduction}
\label{sec:introduction}

Brown dwarfs occupy the mass regime between the lowest-mass stars and the most massive planets, sharing atmospheric temperatures with directly imaged giant exoplanets while being unaffected by host star contamination \citep{Patience2012, Faherty2016}. This makes them ideal laboratories for developing and testing atmospheric characterisation techniques applicable to exoplanet science. High-resolution spectroscopy (HRS) has proven particularly powerful for constraining atmospheric properties, enabling precise measurements of molecular abundances \citep{Brogi2019, Line2021}, isotope ratios that may trace formation histories \citep{MolliereSnellen2019, Zhang2021_BD}, and other properties (e.g. \citealt{Snellen2014, Whiteford2023, Landman2024}).

The carbon isotope ratio, $^{12}$C/$^{13}$C, has emerged as a potential formation diagnostic, with isolated brown dwarfs consistently showing ratios at or above the local interstellar medium value of $68 \pm 15$ \citep{Milam2005}, while substellar companions exhibit greater diversity \citep{Zhang2021_BD, Zhang2021_SJ, deRegt2024, Picos2024, Picos2025, Gandhi2025, Grasser2025}. The ESO SupJup Survey (Programme ID: 1110.C-4264, PI: Snellen) aims to systematically characterise the $^{12}$C/$^{13}$C ratio, C/O ratio, and metallicity across a sample of 49 objects spanning isolated brown dwarfs, lower-mass companions, and host stars \citep{deRegt2024}, with spectra obtained from the upgraded CRyogenic InfraRed Echelle Spectrograph (CRIRES$+$) on the Very Large Telescope (VLT), located at Paranal Observatory in Chile's Atacama Desert \citep{Dorn2023}. Early results have demonstrated the sensitivity of CRIRES+ $K$-band spectroscopy for constraining these key parameters \citep{deRegt2024, Picos2024, Zhang2024, Picos2025, Gandhi2025, Mulder2025, deRegt2025, Grasser2025, deRegt_2026, Grasser2026}.

\begin{figure}
    \centering
    \includegraphics[width=\linewidth]{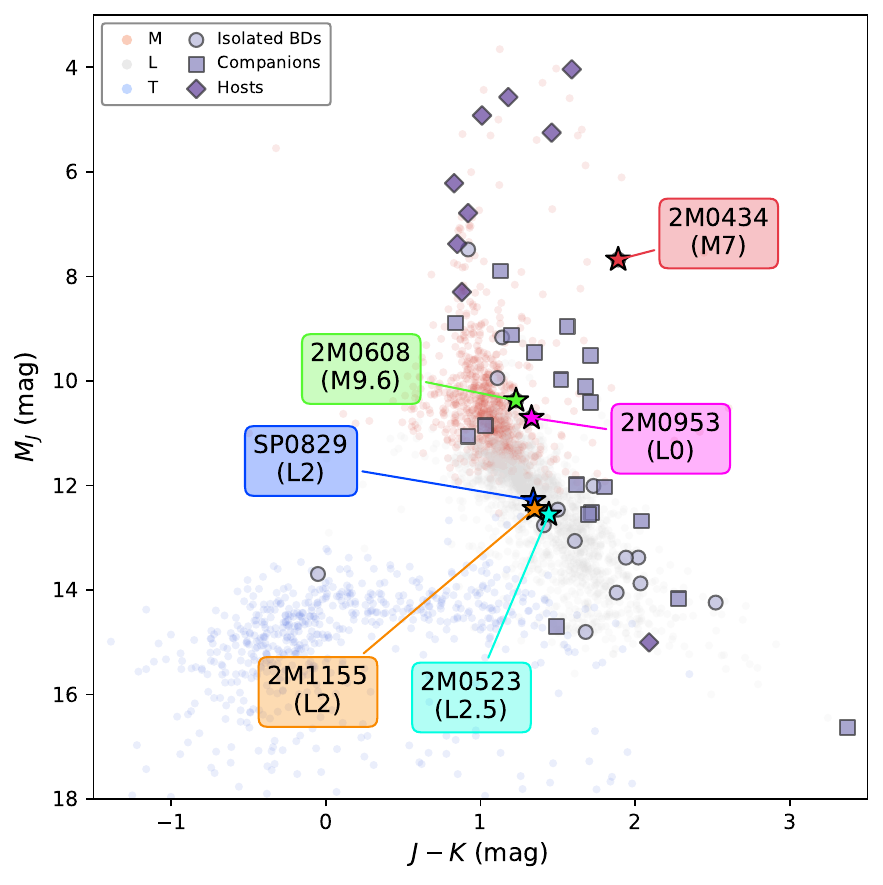}
    \caption{Colour-magnitude diagram displaying absolute J-band magnitudes vs J-K colours for our six-target sample in the context of the broader ultracool dwarf population. The background points show M, L, and T dwarfs from the UltracoolSheet database (\url{https://doi.org/10.5281/zenodo.13993077}), colour-coded by spectral type. The full SupJup survey sample is shown in purple, distinguished by system architecture: isolated brown dwarfs (circles), companions (squares), and host objects (diamonds). Our six targets are emphasised with star-shaped data markers to illustrate their placement within this population of ultracool dwarfs and substellar companions.}
    \label{fig:CMD_SupJup}
\end{figure}

In this work, we present atmospheric retrievals for six isolated ultracool dwarfs observed as part of the SupJup Survey, spanning spectral types M7 to L2.5 (Fig.~\ref{fig:CMD_SupJup}). These six targets are the isolated brown dwarfs not presented by earlier SupJup studies and have spectra of sufficient quality for our retrieval analysis. Two additional isolated targets were excluded, as their low S/N did not enable reliable model fits. This sample includes the candidate Taurus member 2MASS~J04341527+2250309 \citep{Gagne2018}, which may represent one of the youngest brown dwarfs characterised with this methodology, with a corresponding age of 1--2~Myr \citep{Kenyon1995}. The sample of six targets presented here represents one of the largest single contributions to the SupJup brown dwarf inventory, spanning a broad range of spectral types and including both candidate young objects and field dwarfs. 

Section~\ref{sec:observations} describes our observations and data reduction pipeline. Section~\ref{sec:methods} outlines the retrieval framework. Section~\ref{sec:results} presents the retrieved atmospheric properties and discusses their implications for formation pathways. Section~\ref{sec:conclusions} summarises our conclusions.

\section{Observations and reduction}
\label{sec:observations}

Our sample comprises six isolated brown dwarfs spanning spectral types M7 to L2.5, observed as part of the ESO SupJup Survey. The targets and their properties are summarised in Table~\ref{tab:target_properties}. The sample includes two M dwarfs and four early-L dwarfs. Notably, the M7 dwarf 2MASS~J04341527+2250309 (hereafter 2M0434) is a candidate member of the Taurus star-forming region with an 87.8\% probability from the BANYAN~$\Sigma$ classification \citep{Gagne2018}, suggesting an age of 1--2~Myr \citep{Kenyon1995}. The remaining five targets are classified as field objects \citep{Gagne2018}. Young objects with independently constrained ages are valuable from an evolutionary perspective, as evolutionary models are least certain at the youngest ages \citep{Baraffe2002}, where an object's early accretion history makes ages inferred strongly model-dependent \citep{Baraffe_2017}. Their low surface gravities also make them close analogues of young giant planets \citep{Faherty2016}. At this young age, 2M0434 should reflect the composition of the present-day interstellar medium (ISM), providing a comparison for the $^{12}$C/$^{13}$C ratios of older objects in our sample (Sect.~\ref{sec:carbon_isotope}). Throughout this work, we refer to targets by their abbreviations listed in Table~\ref{tab:target_properties}.

\begin{table*}
\caption{Properties of the target sample.}
\label{tab:target_properties}
\centering
\scriptsize
\begin{tabular}{llccccccc}
\hline\hline
Name & Abbrev. & SpT & $d$ & BANYAN~$\Sigma$ & $K_\mathrm{S}$ & $T_\mathrm{eff}$ & $v_\mathrm{rad}$ & $v\sin i$ \\
 & & & [pc]$^{(1)}$ & Classification$^{(2)}$ & [mag]$^{(3)}$ & [K] & [km\,s$^{-1}$] & [km\,s$^{-1}$] \\
\hline
\object{2MASS J04341527+2250309} & 2M0434 & M7$^{(4)}$ & $163 \pm 10$ & TAU (87.8\%) & $11.85 \pm 0.02$ & -- & $19.13 \pm 0.04^{(6)}$ & -- \\
\object{DENIS J060852.8-275358} & 2M0608 & M9.6$^{(9)}$ & $44.2 \pm 0.3$ & FIELD (99.9\%) & $12.37 \pm 0.03$ & $2147 \pm 228^{(10)}$ & $26.35 \pm 0.07^{(10)}$ & -- \\
\object{2MASS J05233822-1403022} & 2M0523 & L2.5$^{(7)}$ & $12.73 \pm 0.02$ & FIELD (99.9\%) & $11.638 \pm 0.03$ & $2100^{(8)}$ & $12.21 \pm 0.09^{(8)}$ & $15.98 \pm 0.31^{(8)}$ \\
\object{SSSPM J0829-1309} & SP0829 & L2$^{(11)}$ & $11.68 \pm 0.02$ & FIELD (99.9\%) & $11.30 \pm 0.02$ & $2300^{(8)}$ & $25.85 \pm 0.08^{(8)}$ & $29.13 \pm 5.00^{(8)}$ \\
\object{2MASS J09532126-1014205} & 2M0953 & L0$^{(9)}$ & $35.7 \pm 0.3$ & FIELD (78.3\%) & $12.14 \pm 0.02$ & -- & $47.6 \pm 11.3^{(12)}$ & -- \\
\object{2MASS J11553952-3727350} & 2M1155 & L2$^{(13)}$ & $11.80 \pm 0.02$ & FIELD (99.9\%) & $11.46 \pm 0.02$ & $2200^{(8)}$ & $45.57 \pm 0.11^{(8)}$ & $13.61 \pm 0.31^{(8)}$ \\
\hline
\end{tabular}
\tablefoot{The age of 2M0434 is estimated to be 1--2~Myr based on tentative Taurus membership$^{(2,5)}.$ {\textbf {References}: 
(1)~\citep{Gaia2020}; 
(2)~\citep{Gagne2018}; 
(3)~\citep{2MASS2003}; 
(4)~\citep{Monin2010}; 
(5)~\citep{Kenyon1995}; 
(6)~\citep{Jonsson2020}; 
(7)~\citep{Cruz2003}; 
(8)~\citep{Blake2010}; 
(9)~\citep{Bardalez2014}; 
(10)~\citep{Faherty2016}; 
(11)~\citep{Scholz_2002}; 
(12)~\citep{Gaia2024}; 
(13)~\citep{Gizis2002}.
}}
\end{table*}

All targets were observed with CRIRES+ \citep{Dorn2023} on the 8.2\,m Unit Telescope~3 (UT3) of the ESO Very Large Telescope (VLT) as part of the ESO SupJup Survey. All observations employed the K2166 wavelength setting, covering 1920--2480\,nm, and the observational setup for each target is detailed in Table~\ref{tab:obs_setup}. No adaptive optics were used; all targets were observed through the 0.4\arcsec\ slit. The cross-dispersed format of CRIRES+ splits the $K$-band exposure into seven echelle orders, with each order further divided across the three detectors, yielding 21 contiguous spectral segments per exposure. Telluric standard stars were observed at similar airmass to enable accurate corrections of telluric absorption features. The spectral resolving power of our observations was determined from the full width at half maximum (FWHM) of telluric absorption line profiles fit by \texttt{Molecfit}. As the telluric standard star line profiles were modelled with Gaussian functions, the resolving power $\mathcal{R} = \lambda/\Delta\lambda$ could be calculated directly from the fitted FWHM. Our six targets exhibit spectral resolutions ranging from $\mathcal{R} \sim 58{,}300$ to $62{,}900$, with a mean of $\mathcal{R} \sim 60{,}000$. The slight variation in resolving power between targets likely reflects differences in observational conditions and uncertainties in the telluric line fitting process. 

\begin{table*}
\caption{Observational setup for the target sample.}
\label{tab:obs_setup}
\scriptsize
\centering
\begin{tabular}{lccccccccc}
\hline\hline
Target & Observation Date & Slit width & $t_\mathrm{exp}$ & Airmass & Seeing & Integrated Water Vapour & Standard star & Std. star airmass & S/N$^a$\\
 & & [$"$] & [min.] & & [$"$] & [mm] & & & \\
\hline
2M0434 & 2023-02-01 & 0.4 & 110 & $1.71 \pm 0.21$ & $0.79 \pm0.09$ & $13.14\pm0.39$ & 72~Tau & $1.48\pm 0.00$ & 48 \\
2M0608 & 2023-02-26 & 0.4 & 120 & $1.03 \pm 0.03$ & $0.79\pm0.04$ & $3.76\pm0.24$ & $\zeta$~CMa & $1.01\pm 0.00$ & 52\\
2M0523 & 2023-01-03 & 0.4 & 60 & $1.02 \pm 0.01$ & $1.20\pm0.07$ & $7.74\pm0.34$ & 8~Lep & $1.02\pm 0.00$ & 37 \\
SP0829 & 2023-01-03 & 0.4 & 60 & $1.08 \pm 0.03$ & $1.68\pm0.15$ & $4.96 \pm 0.49$  & 3~Hya & $1.12 \pm 0.07$ & 24\\
2M0953 & 2023-03-04 & 0.4 & 90 & $1.10 \pm 0.04$ & $0.81\pm0.08$ & $9.11\pm0.24$ & $\iota$~Sco & $1.88 \pm 0.02$ & 45 \\
2M1155 & 2023-02-01 & 0.4 & 20 & $1.03 \pm 0.00$ & $0.65\pm0.04$ & $13.63\pm0.03$ & $\beta$~Hya & $1.03 \pm 0.00$ & 48 \\
\hline
\end{tabular}
\tablefoot{
\tablefoottext{a}{S/N per pixel at the central wavelength (2.345\,$\mu$m), computed using the uncertainty scaling factor $s^2$ (Eq.~\ref{eq:s2}) from the equilibrium-chemistry model and rounded to the nearest integer.}
}
\end{table*}

Data reduction was performed using \texttt{excalibuhr}\footnote{\url{https://github.com/yapenzhang/excalibuhr}} \citep{Zhang2024}, a Python-based pipeline specifically designed for CRIRES+ spectroscopic data. The reduction procedure largely followed those described in \citet{deRegt2024} and \citet{Grasser2025}. We utilised the open-source software \texttt{Molecfit} \citep{Smette2015} to correct for transmission effects of Earth's atmosphere. The software fits a telluric model of these transmission effects based on key observational variables during the time of observation. These variables include airmass, precipitable water vapour, and the thermal profile of Earth's atmosphere during the observation time, as well as the abundance of key opacity sources including H$_2$O, CO$_2$, CO, CH$_4$, and N$_2$O. A further description of this application of \texttt{Molecfit} is described in Appendix A of \citet{Picos2024}. Using this fitted telluric model, we identified any detector pixels with a telluric transmission $<70\%$. These pixels cannot be adequately corrected and were thus masked. We scaled these telluric-corrected spectra by dividing by the blackbody spectrum of the standard star. The spectra consist primarily of the seven spectral orders, where wavelength regions were further spread over three detectors with 2048 pixels each, and all sub-spectra were normalised by the median flux. After the standard pipeline reduction, telluric correction and normalisation, we excised any outlier pixels that survived the earlier data pre-processing steps. This outlier rejection followed a sliding-window $\sigma$-clipping procedure adapted from the SupJup workflow of \citet{deRegt2024}, where we computed a running median within an 8-pixel window. Using the uncertainties from \texttt{excalibuhr} ($\sigma$), any pixel $>3\sigma$ was removed. Because this approach compares each pixel only to its local 8-pixel neighbourhood, $\sigma$-clipping removes any artefacts while preserving the true structure of spectral features.

\section{Retrieval framework}
\label{sec:methods}

Atmospheric retrievals typically consist of two frameworks: one to forward-model a spectrum based on physical parameters, and one to iteratively sample the prior space of each parameter to obtain a best-fitting model of the observed data. Our retrieval framework utilised the nested sampling tool \texttt{PyMultiNest} \citep{Buchner2014}, a Python wrapper of the Bayesian inference algorithm \texttt{MultiNest} \citep{Feroz2009}, to sample the parameter space. Model spectra were generated with the radiative transfer code \texttt{petitRADTRANS} \citep[pRT; Version 2.7;][]{Molliere2019} from parameters that influence the atmospheric spectrum, including chemical composition, thermal profile, cloud properties, and surface gravity. The framework built upon the methodology developed for SupJup Survey brown dwarfs (e.g. \citealt{deRegt2024, Grasser2025, Picos2024}), with key adaptations for our sample spanning spectral types M7--L2.5. Table~\ref{tab:free_params_priors} lists all free parameters included in our retrievals, along with their descriptions and prior ranges. To avoid biasing the results, we adopted wide uniform priors. The retrievals were run in parallel on the Leiden Exoplanet Machine (\texttt{LEM}). Following \texttt{MultiNest} recommendations, we used the importance nested sampling mode with a sampling efficiency of 5\% \citep{Feroz2019} and applied the constant efficiency mode. We utilised an evidence tolerance of 0.1 and further discuss the sampler configuration and convergence validation, including the number of live points used, in Sect. \ref{sec:sample_config_convergence}.

\subsection{Spectral forward model}
\label{sec:forward_model}

The forward model generated synthetic emission spectra that were compared against observations. Following the SupJup framework, pRT computed line-by-line radiative transfer with opacities down-sampled by a factor of three from the native $\mathcal{R} = 10^6$ resolution. The resulting spectrum underwent a series of transformations: Doppler shifting to the barycentric frame using the retrieved radial velocity $v_{\mathrm{rad}}$, rotational broadening via \texttt{fastRotBroad} \citep{Gray2008, Czesla2019} with retrieved $v\sin i$ and limb-darkening coefficient $u_{\mathrm{LD}}$, convolution to match the CRIRES+ instrumental resolution ($\mathcal{R} \approx 60{,}000$), interpolation onto the observed wavelength grid, and median normalisation mirroring the data reduction.

For 2M0434, the only target in our sample with a high-probability youth assignment (Table~\ref{tab:target_properties}), following \citet{Picos2024} we included a veiling component with the wavelength-dependent veiling factor $r_k(\lambda) = r_0\,(\lambda/\lambda_0)^{\alpha_\mathrm{veil}}$ added to the median-normalised spectrum. Here, $r_0$ is the veiling factor at the reference wavelength $\lambda_0 = 1.9\,\mu$m and $\alpha_\mathrm{veil}$ is the power-law index, both retrieved as free parameters. This was motivated by 2M0434's red $J-K$ colour ($=1.89$ mag; Fig.~\ref{fig:CMD_SupJup}), the highest in our sample and well above the $J-K=$1.23--1.50 mag range of the remaining targets. Inconsistent surface gravity and metallicity constraints are recovered by retrievals without the veiling prescription (Appendix~\ref{appdx:2M0434_logg_FeH}), consistent with a near-infrared excess of uncertain origin.

We adopted a composition-agnostic grey cloud model \citep{Molliere2020} in which cloud opacity varies with pressure according to
\begin{equation}
\kappa_{\mathrm{cl}}(P) = \kappa_{\mathrm{cl},0} \cdot \left(\frac{P}{P_{\mathrm{base}}}\right)^{f_{\mathrm{sed}}},
\label{eq:cloud}
\end{equation}
for $P \leq P_{\mathrm{base}}$, and zero otherwise. Here, $\kappa_{\mathrm{cl},0}$ is the cloud-base opacity, $P_{\mathrm{base}}$ the cloud-base pressure, and $f_{\mathrm{sed}}$ the sedimentation parameter controlling how steeply opacity decreases above the cloud deck. 

\subsubsection{Thermal structure}

The pressure-temperature (P-T) profile followed the gradient-based parametrisation of \citet{Zhang2023}. However, following \citet{Grasser2025} we adopted wide uniform priors on the temperature gradients rather than the Gaussian priors of \citet{Zhang2023}, which are based on the gradients of self-consistent models, so that the retrieved profiles were not pulled towards those models. Temperature gradients $\nabla T_i = d\ln T / d\ln P$ were retrieved at five pressure nodes ($P = 10^{-6}, 10^{-4}, 10^{-2}, 10^{0}, 10^{2}$~bar), quadratically interpolated across layers, and integrated upwards from a retrieved base temperature $T_4$ at $10^2$~bar. This approach ensures smooth profiles while remaining flexible enough to capture the thermal structure across our range of photospheric temperatures.

\subsubsection{Chemical abundances}
\label{sec:chemical_abundances}

We performed retrievals under both equilibrium- and free-chemistry assumptions to probe the thermochemical state of each atmosphere. In equilibrium mode, volume mixing ratios (VMRs) were interpolated from pre-computed \texttt{FastChem} \citep{Stock2018, Kitzmann2024} chemical equilibrium tables as functions of pressure, temperature, the C/O ratio, and metallicity [Fe/H]. The tables form a grid spanning $P = 10^{-6}$--$10^{3}$~bar in (log-uniform) steps of 0.1~dex, $T = 150$--$6000$~K in steps of 50~K, C/O $= 0.1$--$1.6$ in steps of 0.1, and $[\mathrm{Fe/H}] = -1$ to $+1$ in steps of 0.1, and abundances were interpolated linearly within the grid. The tables were computed with the equilibrium condensation treatment of \texttt{FastChem}, where the fraction of each species that condenses (e.g. into silicate clouds) is determined locally at every $(P,T)$ point, rather than with the rainout treatment, which would additionally remove the condensed species from all layers above the condensation point \citep{Kitzmann2024}. Isotopologue ratios ($^{12}$CO/$^{13}$CO, C$^{16}$O/C$^{17}$O, C$^{16}$O/C$^{18}$O, H$_2^{16}$O/H$_2^{18}$O) were retrieved separately and used to redistribute mass among isotopologues. This configuration has fewer free parameters than free chemistry and enforces abundance profiles that vary with altitude to maintain thermochemical equilibrium across atmospheric layers.

In free-chemistry mode, each species VMR was retrieved independently as a vertically constant value. While this increases dimensionality, it avoids assumptions about chemical equilibrium and can reveal signatures of atmospheric mixing or photochemistry \citep{Line2015, deRegt2024}. As H$_2$ and He are expected as the main atmospheric constituents, we fixed $\mathrm{VMR}_{\mathrm{He}} = 0.15$ and computed $\mathrm{VMR}_{\mathrm{H}_2}$ to ensure all VMRs sum to unity: $\sum_k \mathrm{VMR}_k = 1$. Derived quantities including C/O, [C/H] as a metallicity proxy, and isotopologue ratios were computed from the retrieved VMRs. The atmospheric metallicity proxy [C/H] was approximated through the number fraction of carbon relative to hydrogen, scaled to the solar composition, utilising $\log_{10}(n_{\mathrm{C}}/n_{\mathrm{H}})_\odot = -3.54$ \citep{Asplund2021}:

\begin{equation}
    [\mathrm{Fe/H}] \approx [\mathrm{C/H}] = \log_{10}\left(\frac{n_{\mathrm{C}}}{n_{\mathrm{H}}}\right) - \log_{10}\left(\frac{n_{\mathrm{C}}}{n_{\mathrm{H}}}\right)_\odot.
\end{equation}

\subsubsection{Opacity Sources}
\label{sec:species_selection}

A key methodological decision in atmospheric retrievals concerns which opacity sources to include. Including too few species risks biasing abundance estimates and thermal profiles, while including too many increases computational cost and can introduce degeneracies with weakly constrained parameters. Prior SupJup retrievals (e.g. \citealt{deRegt2024, Grasser2025}) selected species based on expectations for L-type brown dwarfs, but our sample spans a broader range of spectral types (M7--L2.5) where chemical abundances can vary significantly.

Continuum opacities incorporated in the model spectra included collision-induced absorption from H$_2$--H$_2$ and H$_2$--He, as well as Rayleigh scattering of H$_2$ and He \citep{Dalgarno1962, Chan1965, Borysow1988, Richard2012}. For the two warmer M-type targets (2M0434 and 2M0608), we additionally included H$^-$ continuum opacity, following the approach of \citet{Picos2024} for late-M brown dwarfs. At the cooler photospheric temperatures of the L-type targets, the free-electron abundance becomes negligible (e.g. \citealt{Marley2015}), consistent with its exclusion from previous SupJup L dwarf retrievals (e.g. \citealt{Grasser2025, deRegt2024, Mulder2025}).

To determine the inclusion of line species, we adopted a data-driven approach to species selection. This approach was tailored to each target and is described in Sect. \ref{sec:species_selection_acf}, with the species list per-target given in Table \ref{tab:species_per_target}.

\subsection{Statistical framework}

\subsubsection{Likelihood evaluation}

We adopted the likelihood formalism of \citet{Ruffio2019}, computing log-likelihood contributions for each of the 21 order-detector segments independently before summing them. This approach analytically marginalises over a flux-scaling factor $\phi$ that absorbs calibration differences between model and data. As our spectra are median-normalised, we did not retrieve a radius parameter because the associated scaling of the model flux is absorbed by $\phi$, consistent with previous SupJup retrievals of normalised CRIRES+ spectra \citep{Picos2024, Mulder2025, Grasser2025}. The optimal scaling for each segment is
\begin{equation}
\phi = \frac{\vec{m}^T\boldsymbol{\Sigma}_0^{-1}\vec{d}}{\vec{m}^T\boldsymbol{\Sigma}_0^{-1}\vec{m}},
\label{eq:phi}
\end{equation}
where $\vec{d}$ and $\vec{m}$ are the data and model flux vectors, and $\boldsymbol{\Sigma}_0$ is the covariance matrix. The residuals are then $\vec{R} = \vec{d} - \phi\vec{m}$, and the log-likelihood is
\begin{equation}
{\small
\ln\mathcal{L} = -\frac{1}{2}\left[\ln|\boldsymbol{\Sigma}_0| + \ln(\vec{m}^T\boldsymbol{\Sigma}_0^{-1}\vec{m}) + (N_d - N_\phi + \alpha - 1)\ln\chi_0^2\right],}
\label{eq:lnL}
\end{equation}
with $\chi_0^2 = \vec{R}^T\boldsymbol{\Sigma}_0^{-1}\vec{R}$, valid pixel count $N_d$, $N_\phi = 1$ scaling parameter per segment, and $\alpha = 2$ \citep{Ruffio2019}. The second term penalises models with low overall flux (preventing degenerate solutions), while the third term quantifies the goodness-of-fit. To assess the pipeline flux uncertainties, we computed an uncertainty scaling factor for each segment following \citet{Ruffio2019}:
\begin{equation}
s^2 = \frac{\chi_0^2}{N_d} = \frac{\vec{R}^T\boldsymbol{\Sigma}_0^{-1}\vec{R}}{N_d}.
\label{eq:s2}
\end{equation}
Using $s^2$ we derived the uncertainty-scaled covariance matrix $\Sigma = s^2\Sigma_0$, with a reduced chi-squared equal to unity: $\chi_{\mathrm{red}}^2 = \vec{R}^T\boldsymbol{\Sigma}^{-1}\vec{R}/{N_d} = 1$.

\subsubsection{Correlated noise treatment}

High-resolution spectroscopy introduces pixel-to-pixel correlations from instrumental effects, incomplete telluric correction, and imperfections in molecular line lists \citep{Czekala2015}. Ignoring these correlations can lead to underestimated parameter uncertainties and can bias retrieved values \citep{deRegt2024}. To handle correlation between individual pixels, which correspond to the off-diagonal elements of our covariance matrix, we adopted Gaussian processes (GPs) following the approach of \citet{deRegt2024}, which is based on \citet{Kawahara2022}. Specifically, for every order-detector pair, we built the covariance as the sum of an uncorrelated (diagonal) term and a correlated GP term:

\begin{equation}
{\small
\boldsymbol{\Sigma}_{0,ij}=
\underbrace{\delta_{ij}\,\sigma_{i}^{2}}_{\text{uncorrelated}}
\;+\;
\underbrace{a^{2}\,\sigma_{\mathrm{eff},ij}^{2}\,
           \exp \textrm{ }\!\Bigl[-(\lambda_i-\lambda_j)^{2}/(2\ell^{2})\Bigr],}_{\text{correlated\;(GP)}}
\label{eq:Sigma0_full}}
\end{equation}
where $\sigma_{i}$ is the 1$\sigma$ flux uncertainty of pixel $i$, provided by the data, $\lambda_i$ and $\lambda_j$ are the wavelengths of pixels i and j, the effective $\sigma_{\mathrm{eff},ij}$ is the median flux uncertainty of the order-detector pair, $\delta_{ij}$ is the Kronecker delta, and $a$ and $\ell$ are parameters relating to GPs. These GP parameters were introduced as a Gaussian with amplitude $a$ scaling the uncertainty, and length-scale $\ell$. This GP length-scale $\ell$ sets the contribution of off-diagonal elements. Following \citet{deRegt2024}, we neglected the off-diagonal elements with $|\lambda_i - \lambda_j| > 5\ell$, yielding a banded covariance matrix, exploiting that the largest values are concentrated near the diagonal. This enables efficient Cholesky decomposition, reducing computation while preserving numerical accuracy in computing $\boldsymbol{\Sigma}_0^{-1}$ and evaluating the likelihood (Eq.~\ref{eq:lnL}).

\subsubsection{Species detection significance}\label{sec:ccf_methods}
Detection significance for individual species was assessed via cross-correlation analysis following \citet{Zhang2021_SJ}. For every order-detector pair $(i,j)$, we defined a residual $\vec{R}_{ij} = \vec{d}_{ij} - \vec{m}_{ij}^{\,(w/o\;X)}$ and a species template $\vec{T}_{ij} = \vec{m}_{ij}^{\,(\mathrm{all})} - \vec{m}_{ij}^{\,(w/o\;X)}$, where $\vec{m}_{ij}^{\,(\mathrm{all})}$ is the best-fit model and $\vec{m}_{ij}^{\,(w/o\;X)}$ is the same model with species $X$ removed. For free-chemistry retrievals, $\vec{m}_{ij}^{\,(w/o\;X)}$ was constructed by setting the abundance of species $X$ to a negligibly small value ($\log_{10}\mathrm{VMR} = -14$). For equilibrium-chemistry retrievals, $\vec{m}_{ij}^{\,(w/o\;X)}$ was constructed by directly excluding the opacity contribution of species $X$ in the radiative transfer calculations, while leaving the equilibrium abundance structure unchanged. Using the segment-specific covariance matrix $\boldsymbol{\Sigma}_{0,ij}$ from Eq.~\ref{eq:Sigma0_full}, the covariance-weighted cross-correlation function (CCF) is
\begin{equation}
\mathrm{CCF}(v) = \sum_{ij} \frac{1}{s_{ij}^{2}}\, \vec{T}_{ij}^{\,T}(v)\, \boldsymbol{\Sigma}_{0,ij}^{-1}\, \vec{R}_{ij},
\label{eq:ccf}
\end{equation}

\noindent where $\vec{T}_{ij}(v)$ is the template Doppler-shifted by velocity $v$ and $s_{ij}^{2}$ is the uncertainty scaling factor from Eq.~\ref{eq:s2}. The CCF was evaluated on a velocity grid $v \in [-500, +500]$~km\,s$^{-1}$ in steps of 1~km\,s$^{-1}$. The S/N was computed as $\mathrm{S/N}_{X} = \mathrm{CCF}(0)/\sigma_{\mathrm{noise}}$, where, following \citet{deRegt2024}, $\sigma_{\mathrm{noise}}$ is the standard deviation of $\mathrm{CCF}(v)$ after subtracting the auto-correlation function (ACF), obtained by replacing the term $\vec{R}_{ij}$ with $\vec{T}_{ij}$ in Eq.~\ref{eq:ccf}, at $|v| > 100$~km\,s$^{-1}$; for targets with high retrieved $v\sin i$, the velocity range and noise window were adjusted to accommodate significant rotational broadening, as noted in Appendix~\ref{apdx:CCF_overall}. Following \citet{Landman2024}, we adopted $\mathrm{S/N}_{X} \geq 5$ as evidence of a significant detection of species ${X}$.

\begin{table}
\centering
\tiny
\setlength{\tabcolsep}{3pt}
\caption{Free parameters and uniform prior ranges for the retrievals.}
\label{tab:free_params_priors}
\begin{tabularx}{\linewidth}{l X l}
\hline\hline
Parameter & Description & Prior range \\
\hline
\hspace{0.5em}$v_{\mathrm{rad}}$ [km\,s$^{-1}$]   & Radial velocity                  & Target-specific$^{a}$ \\
\hspace{0.5em}$v\sin i$ [km\,s$^{-1}$]           & Projected rotational velocity    & $\mathcal{U}(0, 40)$$^{b}$ \\
\hspace{0.5em}$\log g$ [cm\,s$^{-2}$]            & Surface gravity                  & $\mathcal{U}(3, 6)$$^{c}$ \\
\hspace{0.5em}$u_{\mathrm{LD}}$                  & Limb-darkening coefficient       & $\mathcal{U}(0, 1)$ \\
\hline
\hspace{0.5em}$T_4$ [K]                          & Temperature at base \newline ($P_4 = 10^{2}$~bar)  & $\mathcal{U}(0, 10000)$ \\
\hspace{0.5em}$\nabla T_0, \dots, \nabla T_4$         & Temperature gradients at  \newline 5 pressure nodes$^{(d)}$ & $\mathcal{U}(0, 0.4)$ \\
\hline
\hspace{0.5em}$\log P_{\mathrm{base}}$ [bar]     & Cloud base pressure              & $\mathcal{U}(-6, 3)$ \\
\hspace{0.5em}$\log\kappa_{\mathrm{cl},0}$ [cm$^2$\,g$^{-1}$] & Opacity at cloud base ($P_{\mathrm{base}}$) & $\mathcal{U}(-10, 3)$ \\
\hspace{0.5em}$f_{\mathrm{sed}}$                 & Cloud sedimentation parameter    & $\mathcal{U}(0, 20)$ \\
\hline
\hspace{0.5em}$\log a$                           & GP amplitude                     & $\mathcal{U}(-1, 1)$ \\
\hspace{0.5em}$\log \ell$ [nm]                   & GP length-scale                  & $\mathcal{U}(-3, 1)$ \\
\hline
\multicolumn{3}{@{}l}{\textit{Free chemistry}} \\
\hspace{0.5em}$\log\mathrm{VMR}_X$ & Log$_{10}$ VMR per species $X$$^{e}$ & $\mathcal{U}(-12, -1)$\\
\hline
\multicolumn{3}{@{}l}{\textit{Equilibrium chemistry}} \\
\hspace{0.5em}C/O & Carbon-to-oxygen ratio & $\mathcal{U}(0.1, 1)$ \\
\hspace{0.5em}$[\mathrm{Fe/H}]$ & Metallicity & $\mathcal{U}(-1, 1)$ \\
\hspace{0.5em}$\log(^{12}\mathrm{CO}/^{13}\mathrm{CO})$ & $^{12}$C/$^{13}$C isotope ratio & $\mathcal{U}(1, 6)$ \\
\hspace{0.5em}$\log(^{12}\mathrm{CO}/\mathrm{C}^{18}\mathrm{O})$ & $^{16}$O/$^{18}$O isotope ratio via CO & $\mathcal{U}(1, 6)$ \\
\hspace{0.5em}$\log(\mathrm{H}_2^{16}\mathrm{O}/\mathrm{H}_2^{18}\mathrm{O})$ & $^{16}$O/$^{18}$O isotope ratio via H$_2$O & $\mathcal{U}(1, 6)$ \\
\hspace{0.5em}$\log(^{12}\mathrm{CO}/\mathrm{C}^{17}\mathrm{O}) $ & Oxygen-17 isotope ratio & $\mathcal{U}(1, 6)$ \\
\hline
\multicolumn{3}{@{}l}{\textit{Veiling$^{f}$}} \\
\hspace{0.5em}$r_0$                           & Veiling factor at shortest  \newline wavelength ($\lambda_0=1.9\,\mu$m)                  & $\mathcal{U}(0, 2)$ \\
\hspace{0.5em}$\alpha_\mathrm{veil}$                           & Veiling power law exponent                 & $\mathcal{U}(0, 3)$ \\
\hline
\end{tabularx}

\tablefoot{$\mathcal{U}$ denotes a uniform distribution within the given range.\\
\tablefoottext{a}{Target-specific $v_{\mathrm{rad}}$ priors based on literature values (see Table~\ref{tab:target_properties}): 
2M0434: $\mathcal{U}(10, 30)$; 
2M0608: $\mathcal{U}(20, 35)$; 
2M0523: $\mathcal{U}(1, 20)$;
SP0829: $\mathcal{U}(20, 35)$; 
2M0953: $\mathcal{U}(0, 60)$; 
2M1155: $\mathcal{U}(30, 55)$.}
\tablefoottext{b}{2M0953 used $\mathcal{U}(50, 120)$ due to its rapid rotation.}
\tablefoottext{c}{2M0434 used $\mathcal{U}(2, 5.5)$ following the analysis of young late-type M dwarfs by \citet{Picos2024} due to its probable Taurus membership and associated age of 1--2~Myr \citep{Gagne2018,Kenyon1995}}.
\tablefoottext{d}{$P_0 =10^{-6}$ bar, $P_1 =10^{-4}$ bar, $P_2 =10^{-2}$ bar, $P_3 =10^{0}$ bar, and $P_4 =10^{2}$ bar.}
\tablefoottext{e}{The species included vary by target; see Table~\ref{tab:species_per_target}.}
\tablefoottext{f}{Our veiling approach was only applied to 2M0434.}}
\end{table}

\subsection{Species selection} \label{sec:species_selection_acf}

We selected the line species included for each target with a data-driven detectability screen, which tested whether a candidate species would produce a detectable signal in our observed spectra and included only those predicted to do so. The screen began with a preliminary equilibrium-chemistry retrieval per target to establish baseline atmospheric parameters, using the 11 species from the L4 dwarf retrievals of \citet{Grasser2025}: H$_2^{16}$O, $^{12}$CO, $^{13}$CO, C$^{18}$O, C$^{17}$O, CH$_4$, NH$_3$, HCN, HF, H$_2$S, and H$_2^{18}$O. From this preliminary retrieval, we took the retrieved model parameters and interpolated from the \texttt{FastChem} equilibrium-chemistry tables via the baseline thermal structure, C/O and [Fe/H], including 42 candidate species with high-resolution $K$-band line lists described in the notes of Table~\ref{tab:species_per_target} and the isotopologues set by the baseline retrieved ratios. 

We computed a model spectrum and added synthetic Gaussian noise at the level of the \texttt{excalibuhr} uncertainties scaled by the uncertainty-scaling factors $s_{ij}$ (Eq.~\ref{eq:s2}) of the baseline retrieval. This generated simulated observations of a known atmosphere. For every candidate species, the noiseless template $\vec{T}_{ij}$ (computed from the model spectra before the synthetic noise was added) and the simulated residuals were then constructed exactly as in the detection analysis of Sect.~\ref{sec:ccf_methods}. The screen was built on the ACF, obtained by replacing $\vec{R}_{ij}$ with $\vec{T}_{ij}$ in Eq.~\ref{eq:ccf}, where $\mathrm{ACF}(0)$ represents the noise-free expectation of the CCF peak. The predicted detectability given by this ACF screen is $\mathrm{S/N}^{\mathrm{pred}}_{X} = \mathrm{ACF}(0)/\sigma_{\mathrm{noise}}$, where $\sigma_{\mathrm{noise}}$ was calculated from the CCF of the template with simulated residuals evaluated over the off-peak velocity range described in Sect.~\ref{sec:ccf_methods}, adjusting for the individual targets with high retrieved rotational velocity (Appendix~\ref{apdx:CCF_overall}). Species with $\mathrm{S/N}^{\mathrm{pred}} \geq 3$ were included in the retrievals of that target, and species falling in the marginal band $1 \leq \mathrm{S/N}^{\mathrm{pred}} < 3$ were retained for the validation runs of Sect.~\ref{sec:sample_config_convergence}, where none yields a significant detection, supporting this adopted inclusion threshold of $\mathrm{S/N}^{\mathrm{pred}} \geq 3$. The minor isotopologues $^{13}$CO, C$^{18}$O, C$^{17}$O, and H$_2^{18}$O were exempt from the screen and included for all targets, as the isotope ratios they trace are a core science goal of the SupJup Survey.

The screen was thus performed only on the preliminary equilibrium-chemistry retrievals. The abundance interpolation is itself an equilibrium calculation, making this the self-consistent choice of anchor. A single screen guarantees identical per-target species lists across chemistry frameworks, ensuring that the Bayesian model comparison (Sect.~\ref{sec:model_comparison}) reflects differences in the chemistry treatment rather than additional differing model content.

The screen yields a target-specific species list (Table~\ref{tab:species_per_target}), with all targets sharing a common set of ten species including the isotopologues. NH$_3$ enters only for 2M0523 and 2M1155, the atomic species Na, Ti, and the hydride SH are strongest in the warmer targets, AlH is included only in 2M0608, 2M0953 and 2M0523, and Mg in all targets except 2M0434. We note that the ACF S/N values in Table~\ref{tab:species_per_target} are not directly comparable to the detection significances reported in Appendix~\ref{apdx:CCF_overall}, where the former compare a noise-free template expectation against synthetic uncertainties and are correspondingly optimistic and inclusive for the purposes of a detectability screen, and the latter are measured against the empirical noise of the ACF-subtracted real CCF.

\begin{table}
\centering
\scriptsize
\setlength{\tabcolsep}{3.5pt}
\caption{Line species selected by the ACF screen for each target with their associated S/N$^\mathrm{pred}$ values.}
\label{tab:species_per_target}
\begin{tabular}{l c c c c c c}
\hline\hline
Species & 2M0434 & 2M0608 & 2M0953 & SP0829 & 2M0523 & 2M1155 \\
        & (M7)   & (M9.6) & (L0)   & (L2)   & (L2.5) & (L2)   \\
\hline
H$_2^{16}$O & $\bullet$ (41.9) & $\bullet$ (43.1) & $\bullet$ (12.1) & $\bullet$ (20.6) & $\bullet$ (42.2) & $\bullet$ (46.7) \\
H$_2^{18}$O$^{\dagger}$ & $\bullet$ & $\bullet$ & $\bullet$ & $\bullet$ & $\bullet$ & $\bullet$ \\
$^{12}$CO     & $\bullet$ (23.5) & $\bullet$ (31.2) & $\bullet$ (4.6)  & $\bullet$ (12.7) & $\bullet$ (20.3) & $\bullet$ (22.6) \\
$^{13}$CO$^{\dagger}$ & $\bullet$ & $\bullet$ & $\bullet$ & $\bullet$ & $\bullet$ & $\bullet$ \\
C$^{18}$O$^{\dagger}$ & $\bullet$ & $\bullet$ & $\bullet$ & $\bullet$ & $\bullet$ & $\bullet$ \\
C$^{17}$O$^{\dagger}$ & $\bullet$ & $\bullet$ & $\bullet$ & $\bullet$ & $\bullet$ & $\bullet$ \\
HF            & $\bullet$ (17.3) & $\bullet$ (30.3) & $\bullet$ (9.3)  & $\bullet$ (29.0) & $\bullet$ (33.8) & $\bullet$ (48.3) \\
Ca            & $\bullet$ (19.4) & $\bullet$ (17.1) & $\bullet$ (3.1)  & $\bullet$ (5.3)  & $\bullet$ (11.6) & $\bullet$ (13.7) \\
Sc            & $\bullet$ (26.5) & $\bullet$ (13.1) & $\bullet$ (7.0)  & $\bullet$ (5.0)  & $\bullet$ (8.6)  & $\bullet$ (7.3)  \\
AlO           & $\bullet$ (6.0)  & $\bullet$ (6.4)  & $\bullet$ (8.2)  & $\bullet$ (3.4)  & $\bullet$ (3.1)  & $\bullet$ (4.0)  \\
\hline
Mg            & $\circ$ (1.2)    & $\bullet$ (4.1)  & $\bullet$ (8.2)  & $\bullet$ (14.9) & $\bullet$ (21.4) & $\bullet$ (29.4) \\
Na            & $\bullet$ (7.1)  & $\bullet$ (8.6)  & $\bullet$ (12.8) & $\circ$ (1.7)    & $\circ$ (2.3)    & $\circ$ (1.4)    \\
Ti            & $\bullet$ (11.7) & $\bullet$ (8.1)  & $\bullet$ (4.6)  & $\circ$ (2.7)    & $\circ$ (2.6)    & $\circ$ (1.1)    \\
SH            & $\bullet$ (8.9)  & $\bullet$ (7.3)  & $\bullet$ (4.5)  & $\circ$ (2.1)    & $\circ$ (1.5)    & $\circ$ (1.5)    \\
H$_2$S        & $\circ$ (1.8)    & $\bullet$ (6.5)  & $\bullet$ (3.2)  & $\bullet$ (4.2)  & $\bullet$ (4.8)  & $\bullet$ (5.9)  \\
AlH           & $\circ$ (1.7)    & $\bullet$ (4.3)  & $\bullet$ (3.1)  & $\circ$ (2.3)    & $\bullet$ (3.5)  & $\circ$ (2.0)    \\
NH$_3$        &                  &                  &                  & $\circ$ (1.7)    & $\bullet$ (3.0)  & $\bullet$ (3.7)  \\
\hline
CH$_4$        &                  &                  &                  & $\circ$ (1.5)    & $\circ$ (2.0)    & $\circ$ (2.0)    \\
Cr            & $\circ$ (1.9)    &                  &                  &                  &                  &                  \\
V             & $\circ$ (1.7)    &                  &                  &                  &                  &                  \\
Fe            & $\circ$ (1.3)    &                  &                  &                  &                  &                  \\
K             & $\circ$ (1.1)    &                  &                  &                  &                  &                  \\
\hline
Total included & 13 & 16 & 16 & 12 & 14 & 13 \\
\hline
\end{tabular}
\tablefoot{Filled circles: included in the retrievals ($\mathrm{S/N}^{\mathrm{pred}} \geq 3$, value in parentheses). Open circles: marginal band ($1 \leq \mathrm{S/N}^{\mathrm{pred}} < 3$), included only in the validation retrievals of Sect.~\ref{sec:sample_config_convergence}. Empty cells: below the marginal band in a given target. Minor isotopologues, indicated by the dagger (${\dagger}$), are exempt from the screen and are always included. The first block lists the species included for all targets, the second block the species whose status varies across the sample, and the third block the species that reach at most the marginal band. The full 42-species candidate screen additionally comprises OH, HCl, MgO, TiO, FeH, Rb, SiO, Al, CN, NH, MgH, VO, CO$_2$, HCN, Ni, Si, Mn, CH, Cs, Co, CaH, Li, CrH, and TiH, none of which reaches the marginal band in any target screen.}
\end{table}

Our retrieval framework included the molecular and atomic species listed in Table~\ref{tab:species_per_target}. We used the HITEMP line lists for $^{12}$CO, $^{13}$CO, C$^{18}$O, C$^{17}$O \citep{Li2015}, CH$_4$ \citep{Hargreaves2020}, and CO$_2$ \citep{Rothman2010}. From the ExoMol database, we used line lists for H$_2^{16}$O \citep[POKAZATEL;][]{Polyansky2018}, H$_2^{18}$O \citep{Polyansky2017}, NH$_3$ \citep{Alderzi2015, Coles2019}, HCN \citep{Harris2006, Barber2014}, HF \citep{Li2013, Coxon2015, Somogyi2021}, H$_2$S \citep{Azzam2016, Chubb2018}, FeH \citep{Dulick2003, Bernath2020}, CrH \citep{Bernath2020}, SiO \citep{Yurchenko_2021}, AlH \citep{Yurchenko2023_AlH}, AlO \citep{Patrascu2015}, TiO \citep{McKemmish2019}, VO \citep{McKemmish2016}, SH \citep{Gorman2019}, MgO \citep{Li2019_MgO}, CaH and MgH \citep{Owens2022}, NH \citep{Perri2024}, OH and CN \citep{Wang2020, Tennyson2020}, TiH \citep{Burrows_2005, Bernath2020}, CH \citep{Masseron2014, Bernath2020}, and HCl \citep{Gordon2017}. We also included the atomic species Na, Ca, Sc, Ti, Mg, Fe, Cr, V, Mn, Ni, Si, Al, Co, Li, K, Rb, and Cs \citep{Kurucz2018}.

We did not include any ionic species, beyond H$^-$ for the M dwarfs, in our retrievals. We verified this omission by interpolating equilibrium chemistry (Sect.~\ref{sec:chemical_abundances}) at the retrieved photospheres under both chemistry frameworks (given in Tables~\ref{tab:results_eqbm} and \ref{tab:results_free}) to find the abundances of free electrons and the ions Na$^+$, K$^+$, Ca$^+$, Mg$^+$, Al$^+$, Fe$^+$, Ti$^+$, Sc$^+$, Cr$^+$, V$^+$, Si$^+$, Li$^+$, Rb$^+$, Cs$^+$, H$^-$, Cl$^-$, F$^-$, OH$^-$, and CN$^-$ in addition to the per-target species lists (Table~\ref{tab:species_per_target}). We find the free electron abundance $e^-_\mathrm{VMR}\approx 10^{-6}$ for the warmest target (2M0434) down to $\approx10^{-8}$ for the coolest (2M1155). By charge neutrality the total cation budget is approximately equal to the free electron abundance, with the anions several dex below $e^-$, and is carried at the $>98\%$ level by Na$^+$ and K$^+$, both closed-shell ions with no $K$-band transitions \citep{NIST_ASD}, for all targets except 2M0434 where it is $\sim$91\% (equilibrium chemistry) and $\sim$94\% (free chemistry). In 2M0434, the main contributors to this cation budget remainder are Ca$^+$, with its strong transitions lying shortwards of the $K$ band, and Al$^+$ whose transitions arise only between highly excited levels ($\gtrsim15$ eV above ground; \citealt{NIST_ASD}) that are unpopulated at these photospheric temperatures (Boltzmann factors of $\lesssim 10^{-28}$). For the remaining targets the $\lesssim$1\% remainder is split between Ca$^+$ and the closed-shell Rb$^+$. These values are largely identical across chemistry frameworks, suggesting that ionic opacity beyond the H$^-$ continuum for the M dwarfs is negligible for our sample, motivating the exclusion of ionic species consistent with previous SupJup retrieval studies (e.g. \citealt{Grasser2025, deRegt2024, Mulder2025, Picos2024}).

\subsection{Sampler configuration and convergence validation}\label{sec:sample_config_convergence}

To verify that our retrievals yield converged and unbiased posteriors, we performed a live-point sensitivity test on 2M0523, which has the lowest S/N among our more slowly rotating targets (S/N$=37$; $v\sin i \approx 17.5$~km\,s$^{-1}$), repeating the full retrieval of this target at $N_\mathrm{live} = 400$, 1000, 1500, and 2000 for equilibrium chemistry, and at $N_\mathrm{live} = 400$, 2000, 3000, 4000, and 5000 for the higher-dimensional free-chemistry framework, with all other settings identical. These validation runs additionally included the marginal-band species of the ACF screen (Sect.~\ref{sec:species_selection_acf}), making the adopted sampler configurations conservative for the fiducial retrievals, which excluded these species and therefore carried fewer free parameters in free chemistry. The full results are reported in Table~\ref{tab:ladder_2M0523}. 

We adopted, for each chemistry framework, the smallest $N_\mathrm{live}$ for which all runs with more live points reached both a stable maximum log-likelihood, where $|\Delta\ln\mathcal{L}| < 1$ for higher live-point runs, and returned every parameter within the reported uncertainties. These criteria yield $N_\mathrm{live} = 1000$ for equilibrium chemistry and $N_\mathrm{live} = 4000$ for free chemistry, which we applied to all targets. Additionally, in all converged runs, we see no constraints suggested by the CCF detection significance for the included marginal-band species, supporting our adopted ACF species screen threshold of $\mathrm{S/N}^{\mathrm{pred}} \geq 3$ and suggesting species excluded by the screen are those the data cannot reliably constrain.

The equilibrium framework is stable essentially from $N_\mathrm{live} = 400$ upward, with the $N_\mathrm{live} = 400$ run only marginally exceeding our $|\Delta\ln\mathcal{L}| < 1$ criterion ($|\Delta\ln\mathcal{L}| = 1.1$). The higher-dimensional free-chemistry framework requires substantially more live points, consistent with the behaviour reported by \citet{Molliere2020}, who find $N_\mathrm{live} = 400$ insufficient and adopt 4000 for a comparably high-dimensional parameter space. The under-sampled free-chemistry runs illustrate the failure modes of using an insufficient $N_\mathrm{live}$. At $N_\mathrm{live} \leq 2000$ the H$_2$O/H$_2^{18}$O ratio constraint (termed apparent, as H$_2^{18}$O itself is not detected via the CCF; Sect.~\ref{sec:ccf_methods}) is missed entirely, emerging only for $N_\mathrm{live} \geq 3000$, where it is stable and in agreement with the value recovered by all equilibrium-chemistry runs. At $N_\mathrm{live} = 400$, the sampler returns a spurious apparent C$^{18}$O constraint ($\log\mathrm{VMR} = -6.28^{+0.20}_{-0.26}$ vs effectively unconstrained at every higher $N_\mathrm{live}$ run), and an inconsistent surface gravity, while reaching a log-likelihood 127 below the converged value. These cases suggest that an insufficient number of live points can both introduce constraints absent from the converged posteriors and suppress constraints within them.

\section{Results and discussion}
\label{sec:results}

We present the atmospheric retrieval results for our sample of six dwarfs spanning spectral types M7 to L2.5. For each target, we performed two independent retrievals assuming free chemistry and equilibrium chemistry, yielding 12 complete analyses of the physical parameters, thermal structure and cloud properties, species abundances, isotope ratios, and implications for formation history across the sample.

\subsection{Best-fit models and Bayesian model comparison}
\label{sec:model_comparison}

We compared the free- and equilibrium-chemistry models using the Bayesian evidence ($\ln \mathcal{Z}$) from nested sampling. The log Bayes factor was computed as $\ln B = \ln \mathcal{Z}_\mathrm{free} - \ln \mathcal{Z}_\mathrm{eqbm}$. Table~\ref{tab:bayes} summarises the model comparison for all six targets. We report $\Delta\ln\mathcal{L}$ alongside $\ln B$ to separate raw fit quality from the complexity penalty inherent in the Bayesian evidence, since $\ln B$ penalises free chemistry's larger parameter space regardless of fit quality: free chemistry allows each species abundance to vary independently ($\sim$12--16 free parameters), while equilibrium enforces thermochemical consistency through only six (C/O, [Fe/H], and the isotopologue ratios).

For all six targets, the more flexible free-chemistry retrieval produces a slightly better fit to the data than the equilibrium retrieval, with log-likelihoods higher by $\Delta\ln\mathcal{L} = 3.36$ (2M0608) to $15.92$ (2M0434), excluding 2M0953 with $\Delta\ln\mathcal{L} = 69.54$, which we discuss further as an exception below. Differences of this size are modest, with the reduced $\chi^2_0$ of the free-chemistry model (Table~\ref{tab:bayes}) lower than that of the equilibrium model by only 0.02--0.34\% across our sample, with the largest difference for 2M0953, so that both frameworks provide a similarly good fit with free chemistry only marginally better. \citet{Grasser2025} report similarly good fits to the spectra between chemistry frameworks for two L4 dwarfs, with comparable or larger likelihood differences ($\Delta\ln\mathcal{L} = 11.88$ and $33.01$) and $\chi^2$ differences ($\sim$0.1--0.6\%). According to the Bayes factor, equilibrium chemistry is decisively preferred for five of six targets (excluding 2M0953) ranging from $\ln B = -17.10$ (SP0829) to $-41.31$ (2M0608). This difference reflects the smaller number of free parameters required by the equilibrium retrieval to fit the data nearly as well. Following \citet{Grasser2025}, we therefore consider both approaches valid, with the equilibrium setup requiring a simpler parametrisation, and the free-chemistry approach able to retrieve the abundances of individual species directly. 

\begin{figure*}[t!]
    \centering
    \includegraphics[width={0.9\textwidth}]{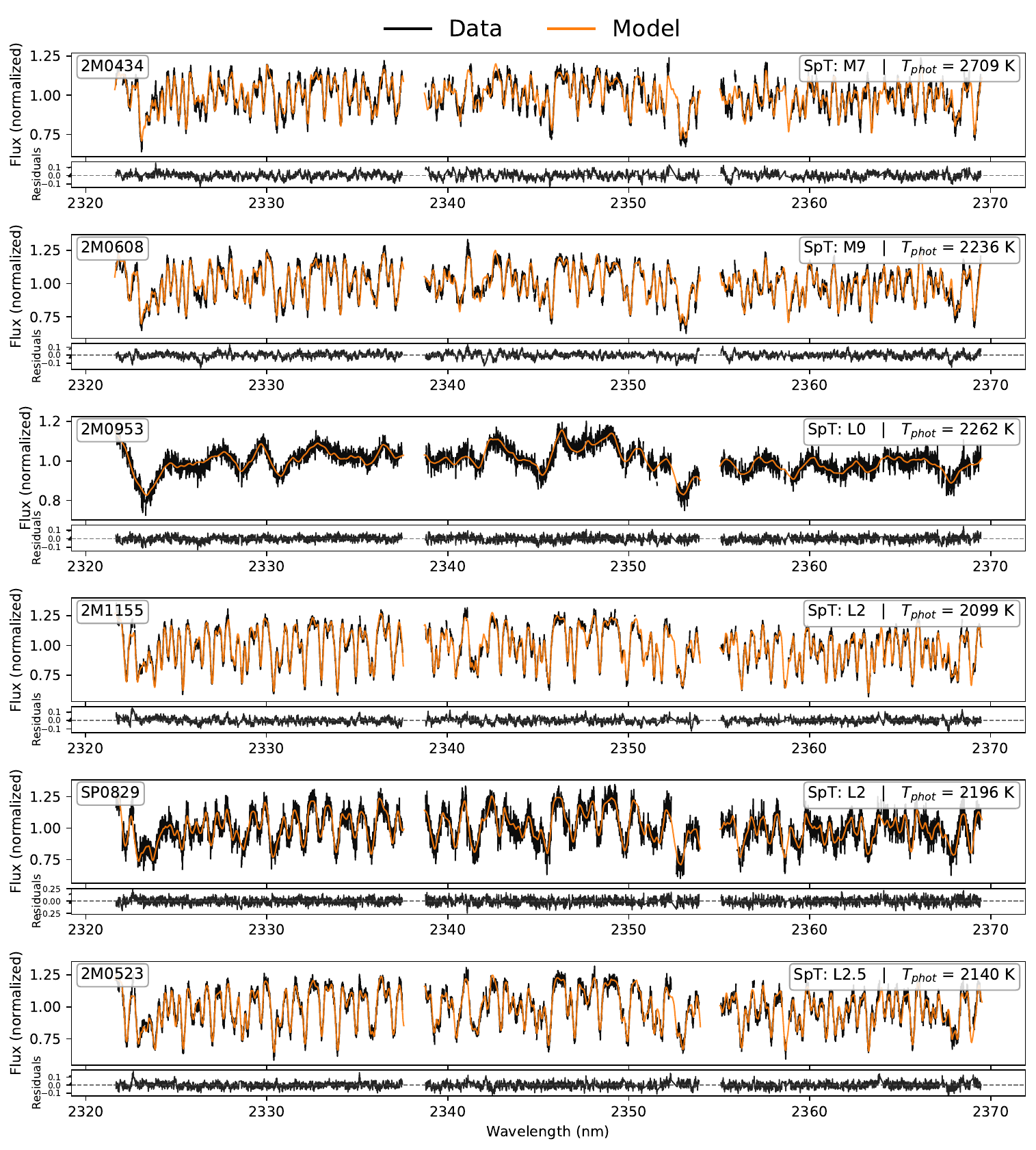}
    \caption{Best-fit equilibrium-chemistry model spectra (orange) vs the corrected observational data (black) for all six targets in spectral order 6 ($K$ band). The targets are ordered by spectral type from M7 (top) to L2.5 (bottom). The photospheric temperature ($T_\mathrm{phot}$) at the pressure of maximal emission contribution is indicated for each target and was derived from the retrieval results. The residuals are shown below each spectrum.}
    \label{fig:order6_comparison}
\end{figure*}

2M0953 is the exception in our sample, with a likelihood improvement large enough to overcome the complexity penalty ($\ln B = +38.17$). However, this target is by far the fastest rotator in the sample ($v\sin i \approx 86$--$87$~km\,s$^{-1}$ between chemistry frameworks; Tables \ref{tab:results_eqbm} and \ref{tab:results_free}), and its rotational broadening severely blends the individual lines, as seen in Fig. \ref{fig:order6_comparison}. In such a spectrum, the freedom of the free-chemistry model to adjust each abundance independently may provide additional leverage on residual structure that the equilibrium model, with its abundances tied to the bulk C/O and [Fe/H], cannot exploit. 2M0953's free-chemistry retrieval returns narrow posteriors for several species that are not supported by the cross-correlation analysis (Tables~\ref{tab:results_free} and \ref{tab:ccf_snr}), which we discuss further in Sect.~\ref{sec:2m0953_rotation} and Appendix~\ref{appdx:H218O_appendix}. The absence of any dominant peak structure at the systemic velocity indicates these constraints are not driven by coherent signal, and we therefore interpret the likelihood gain for 2M0953 as the flexibility of the free model in fitting a severely rotationally broadened spectrum, rather than as evidence of departures from equilibrium chemistry.

As noted above, the direction of trends in $\ln B$ and $\Delta\ln\mathcal{L}$ matches those found by \citet{Grasser2025} for two L4 dwarfs, but the balance differs: excluding 2M0953, our free-chemistry likelihood gains are comparable or slightly weaker, while Bayesian preference for equilibrium is markedly stronger (compared to $\ln B = -6.66$ and $-2.67$; \citealt{Grasser2025}), motivating our focus on the equilibrium-chemistry results. The similarly good fits between chemistry frameworks are consistent with theoretical expectations that chemical equilibrium is well-maintained at the photospheric temperatures of our sample ($T_\mathrm{phot} \approx 2100$--$2710$~K; \citealt{LoddersFegley2002, Visscher2010}). The equilibrium framework also directly retrieves constraints on bulk atmospheric properties of interest, including C/O, [Fe/H], and the isotopologue ratios, making it a scientifically informative framework as the focus of our analysis. We therefore present the equilibrium-chemistry retrieval results in this work unless otherwise stated, for consistency across target comparisons. 
 
\begin{figure}
    \centering
    \includegraphics[width=1.0\linewidth]{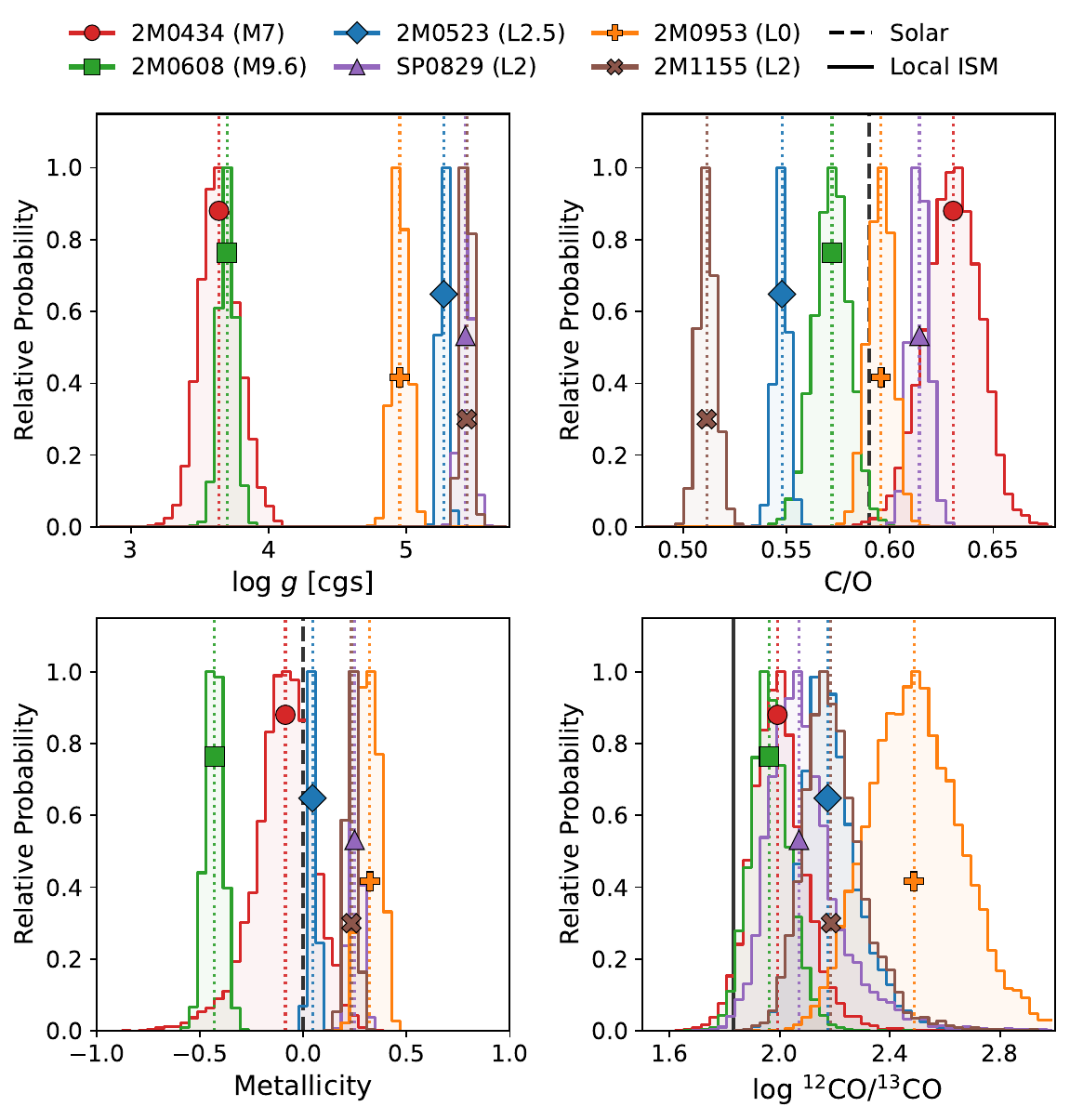}
    \caption{Marginalised posterior distributions for four key atmospheric parameters across the six-target sample: surface gravity (log~$g$), carbon-to-oxygen ratio (C/O), metallicity ([Fe/H]), and carbon isotope ratio (log~$^{12}$CO/$^{13}$CO). Each target is shown in a distinct colour, with the vertical dotted lines indicating the retrieved median. A target-specific symbol is overplotted at a staggered height to aid identification where medians overlap (see legend). The dashed black lines mark the solar reference values for C/O ($= 0.59$; \citealt{Asplund2021}) and the metallicity ($= 0.0$). The solid black line indicates the local ISM value of $^{12}$C/$^{13}$C~$= 68 \pm 15$ \citep{Milam2005}. Results are shown for the equilibrium-chemistry models, where metallicity was retrieved as [Fe/H].} 
    \label{fig:posterior_4panel}
\end{figure}

Figure~\ref{fig:order6_comparison} presents the best-fit equilibrium-chemistry model spectra for all six targets. All spectra display only the sixth spectral order (of seven total): a wavelength interval from roughly 2320--2370 nm within the full infrared $K$-band coverage. This region captures prominent CO and H$_2$O absorption features that strongly respond to temperature and carbon-oxygen chemistry. Selected parameter posterior distributions from the equilibrium retrieval models are displayed in Fig.~\ref{fig:posterior_4panel}.

\begin{table}
\centering
\caption{Bayesian model comparison between the free- and equilibrium-chemistry retrievals.}
\label{tab:bayes}
\small
\begin{tabular}{lcccc}
\hline\hline
Target & $\Delta\ln\mathcal{L}$ & $\ln B$ & $\chi^2_{0,\mathrm{red,free}}$ & $\chi^2_{0,\mathrm{red,eqbm}}$ \\
\hline
2M0434 & $15.92$ & $-25.45$ & 6.164 & 6.172 \\
2M0523 & $13.07$ & $-26.31$ & 7.301 & 7.310 \\
2M0608 & $3.36$ & $-41.31$ & 4.911 & 4.912 \\
SP0829 & $11.11$ & $-17.10$ & 17.792 & 17.804 \\
2M0953 & $69.54$ & $38.17$ & 5.569 & 5.588 \\
2M1155 & $15.88$ & $-20.75$ & 5.212 & 5.228 \\
\hline
\end{tabular}
\tablefoot{$\Delta\ln\mathcal{L} = \ln\mathcal{L}_\mathrm{free} - \ln\mathcal{L}_\mathrm{eqbm}$, and $\ln B = \ln \mathcal{Z}_\mathrm{free} - \ln \mathcal{Z}_\mathrm{eqbm}$ is the log Bayes factor. The positive values favour free chemistry, and the negative values equilibrium chemistry for these first two columns. $\chi^2_{0,\mathrm{red}} = \chi_0^2/N_d$ is the reduced chi-squared of each model prior to uncertainty scaling (Eq.~\ref{eq:s2}), which sets $\chi^2_\mathrm{red} = 1$ by construction. It was computed for both frameworks using the covariance (Eq.~\ref{eq:Sigma0_full}) of the equilibrium retrieval (i.e. the same $a$ and $\ell$) for consistent comparison.}
\end{table}

\subsection{Consistency trends between chemistry frameworks}
\label{sec:parameter_tables}

The retrieved parameters from both equilibrium- and free-chemistry retrievals are presented in Tables~\ref{tab:results_eqbm} and \ref{tab:results_free}. These comprehensive tables include retrieved physical parameters, temperature structure, cloud properties, species abundances, isotopologue ratios, and GP hyperparameters. Following \citet{Grasser2025}, we note that the small formal uncertainties on several of the retrieved parameters may not be fully representative, as overly narrow posteriors are a reported issue of \texttt{MultiNest} retrievals \citep[e.g.][]{Buchner2016, Ardevol2022}. Several key trends emerge from these retrieved parameters.

The physical parameters: radial velocities ($v_\mathrm{rad}$), rotational velocities ($v\sin i$), surface gravities ($\log g$), and limb darkening coefficients ($u_\mathrm{LD}$), show good consistency between free- and equilibrium-chemistry retrievals, with typical differences within 1--2$\sigma$ uncertainties. The one exception is 2M0953's $v_\mathrm{rad}$ of $21.37^{+0.22}_{-0.24}$~km\,s$^{-1}$ (equilibrium) versus $20.03^{+0.17}_{-0.18}$~km\,s$^{-1}$ (free). This broad consistency demonstrates the robustness of these fundamental parameters regardless of the chemistry framework applied, including for the M7 dwarf 2M0434, for which the two frameworks agree on surface gravity ($\log g = 3.64^{+0.15}_{-0.13}$ in equilibrium vs $3.69^{+0.17}_{-0.14}$ in free chemistry) once the veiling component is included; the $K$-band gravity--metallicity degeneracy that separates the frameworks without this veiling is briefly discussed in Appendix~\ref{appdx:2M0434_logg_FeH}.

The cloud parameters are unconstrained, with large uncertainties ($\gtrsim$1.5~dex) reflecting the limited sensitivity of $K$-band spectroscopy to grey continuum opacity. Regarding bulk composition, the equilibrium-chemistry C/O ratios range from 0.51 to 0.63, while the free-chemistry values are systematically slightly higher ($0.59$--$0.69$), an offset of $\sim$0.04--0.08 (Sect.~\ref{sec:co_ratio}). Equilibrium [Fe/H] spans $-0.43\pm0.04$ (2M0608) to $+0.32^{+0.05}_{-0.04}$ (2M0953) and free [C/H] spans $-0.34^{+0.03}_{-0.04}$ (2M0608) to $+0.30^{+0.02}_{-0.04}$ (SP0829), with broadly consistent target-to-target trends despite the different elemental tracers.

\subsection{Radial and rotational velocities}
\label{sec:velocities}

The high spectral resolution of CRIRES+ enables a precise determination of radial ($v_\mathrm{rad}$) and projected rotational ($v\sin i$) velocities through the Doppler shift and rotational broadening of molecular lines. For targets with prior velocity measurements from HRS presented in \citet{Blake2010} ($\mathcal{R}=25{,}000$), we find reasonable agreement: the $v_\mathrm{rad}$ values for 2M0523, SP0829, and 2M1155 agree within $\sim$1.3~km\,s$^{-1}$, and the rotational velocities of SP0829 and 2M1155 show excellent agreement, while 2M0523 exhibits a $\sim$1.6~km\,s$^{-1}$ offset in $v\sin i$ ($17.54\pm0.21$~km\,s$^{-1}$ vs $15.98 \pm 0.31$~km\,s$^{-1}$ from \citealt{Blake2010}). Other targets show similar minor discrepancies with literature values. For 2M0434, our retrieved $v_\mathrm{rad} = 17.76 \pm 0.05$~km\,s$^{-1}$ differs by $\sim$1.4~km\,s$^{-1}$ from the APOGEE value of $19.13 \pm 0.04$~km\,s$^{-1}$ \citep{Jonsson2020}; this discrepancy may arise from systematic differences between instruments, including different wavelength ranges ($H$ band for APOGEE vs $K$ band for CRIRES+) or spectral resolution ($\mathcal{R}_{\mathrm{APOGEE}}=22{,}500$; \citealt{Wilson2019}). Similarly, we find $v_\mathrm{rad} = 24.79 \pm 0.03$~km\,s$^{-1}$ for 2M0608, differing by $\sim$1.6~km\,s$^{-1}$ from the value of $26.35 \pm 0.07$~km\,s$^{-1}$ reported by \citet{Faherty2016}, based on measurements from Magellan Clay MIKE and Gemini South Phoenix. The most significant discrepancy is for 2M0953, where our $v_\mathrm{rad} = 21.37^{+0.22}_{-0.24}$~km\,s$^{-1}$ differs dramatically from the $47.6 \pm 11.3$~km\,s$^{-1}$ measured from lower resolution ($\mathcal{R}\approx2500$) GTC/OSIRIS optical spectra by \citet{Gaia2024}; however, this value is the weighted mean of two discrepant methods (line centre fitting and cross correlation) and has very large uncertainties, reflecting the difficulty of measuring radial velocities for faint, rapidly rotating ultracool dwarfs at low spectral resolution. Our typical $v_\mathrm{rad}$ precision of $\sigma \lesssim 0.1$~km\,s$^{-1}$ ($\sim$0.2~km\,s$^{-1}$ for 2M0953) represents improvement for most measurements, demonstrating the power of CRIRES+ for precision radial velocity work.

\subsection{Surface gravity}
\label{sec:logg}

The retrieved surface gravities span $\log g = 3.64^{+0.15}_{-0.13}$ (2M0434) to $5.44^{+0.03}_{-0.04}$ (2M1155) (cgs), with significant variation across the sample that likely reflects differences in age and mass. The lowest gravities are found for the two M dwarfs, 2M0434 ($\log g = 3.64^{+0.15}_{-0.13}$), and 2M0608 ($3.70 \pm 0.06$). Low surface gravity brown dwarfs have been shown to be younger objects \citep{Baraffe2002, Allers2007, Bonnefoy2014, Picos2024}, and although 2M0608 is classified as a field object (Table~\ref{tab:target_properties}), it is spectroscopically flagged as a low-gravity, likely young, object in the sample of \citet{Faherty2016}, so that its retrieved gravity is consistent with those of other young late-M dwarfs ($\log g \approx 3.3$--$3.9$; \citealt{Picos2024}). 2M0434, classified as a probable Taurus member (Table~\ref{tab:target_properties}), returns consistent surface gravities between chemistry frameworks, $\log g = 3.64^{+0.15}_{-0.13}$ (equilibrium) versus $3.69^{+0.17}_{-0.14}$ (free), in line with the age of 1--2~Myr that a confirmed Taurus membership would imply \citep{Kenyon1995}. This agreement depends on the veiling component: both frameworks retrieve a non-zero veiling continuum for this target, whereas without it the two frameworks separate to opposite ends of the well-documented degeneracy between metallicity and surface gravity in the $K$ band, a consequence of the limited wavelength coverage and the lack of reliable gravity-sensitive features \citep{Zhang2021_BD}, and this degeneracy is present in all of our retrievals (Appendix~\ref{appdx:2M0434_logg_FeH}). We defer a full analysis to Appendix~\ref{appdx:2M0434_logg_FeH}, where we compare both frameworks with and without veiling and show that the $^{12}$CO/$^{13}$CO and C/O ratios remain robust regardless. However, we emphasise that the $\log g$ of 2M0434 should be interpreted with more caution than its formal uncertainty suggests. 

2M0953 shows an intermediate gravity of $\log g = 4.95^{+0.07}_{-0.05}$, which we discuss further in Sect.~\ref{sec:2m0953_rotation}. In contrast, 2M0523, SP0829, and 2M1155 all exhibit high gravities ($\log g > 5.2$), consistent with field-age L dwarfs. These values align with similar retrievals of isolated L dwarfs: \citet{deRegt2024} report $\log g \sim 5.27$ for DENIS~J0255$-$4700, while for the young L dwarfs 2M1425 and 2M0355 \citet{Grasser2025} find $\log g \sim 4.98$ and $\log g \sim 4.75$, respectively.

\subsection{2M0953: An ultra-fast rotator}
\label{sec:2m0953_rotation}

The L0 dwarf 2M0953 is the fastest rotator in our sample by a substantial margin, with projected rotational velocity $v\sin i = 86.89^{+0.48}_{-0.61}$~km\,s$^{-1}$, exceeding the fastest benchmark substellar companion in the recent KPIC spin survey of \citet{Hsu_2026} (LP349-25~B at $v\sin i = 79.6^{+2.4}_{-2.6}$~km\,s$^{-1}$). Assuming $R \approx 1\,R_\mathrm{Jup}$, our retrieved $v\sin i$ implies a period $P \leq1.436^{+0.010}_{-0.008}$ hours (with equality for an equator-on geometry), comparable to the shortest periods measured for any brown dwarf \citep[1.08--1.23~h;][]{Tannock2021}. Brown dwarfs spin up as they cool and contract, conserving angular momentum after disk dissipation \citep{Bouvier2014, Bryan2020}. The extreme spin of 2M0953 fits naturally into this picture for a mature, isolated brown dwarf at the fast end of the rotation distribution. Despite the exceptional rotation, the bulk atmospheric composition is unremarkable within our sample: C/O~$= 0.60 \pm 0.01$ is consistent with the other targets, and $[\mathrm{Fe/H}] = +0.32^{+0.05}_{-0.04}$ is comparable to the mildly super-solar metallicities of the field L dwarfs SP0829 and 2M1155 (Sect.~\ref{sec:metallicity}).

The retrieved surface gravity of $\log g = 4.95^{+0.07}_{-0.05}$ is intermediate compared to the higher-gravity objects in our sample and might naively suggest a younger age, though as discussed in Sect.~\ref{sec:logg} and Appendix~\ref{appdx:2M0434_logg_FeH}, $K$-band retrievals have limited capacity to constrain gravity due to a lack of gravity-sensitive features \citep{Zhang2021_BD, Picos2024}. BANYAN~$\Sigma$ assigns a 78.3\% field probability for 2M0953, with the remaining probability attributed almost entirely to Columba (COL: 20.4\%) \citep{Gagne2018}. These constraints are ambiguous enough to leave open the possibility that it may be part of the Columba association, which, if confirmed, would assign an age of 28.4$^{+1.8}_{-2.9}$ Myr \citep{Couture_2026}. We consider this scenario unlikely given that brown dwarfs spin up with age, and 2M0953's extreme rotation is difficult to reconcile with an age of only $\sim28$~Myr. Marginal independent support for an older age comes from the retrieved ratio of $\log_{10}\!^{12}$CO/$^{13}$CO = $2.49^{+0.17}_{-0.16}$ ($\approx 309^{+148}_{-95}$), which is the highest in our sample and comparable to the elevated ratio of 184$^{+61}_{-40}$ found for the field L dwarf DENIS~J0255$-$4700 \citep{deRegt2024}, pointing towards a field-age rather than young origin (Sect.~\ref{sec:carbon_isotope}). This ratio should be treated with caution, as no significant $^{13}$CO S/N peak is detected in the cross-correlation for 2M0953 (S/N~$= 1.0$; Table~\ref{tab:ccf_snr}). The ratio is instead constrained by the full spectral likelihood, with the tight posterior reflecting a response to the $^{13}$CO opacity while the lack of a direct $^{13}$CO cross-correlation peak is likely due to extreme rotational broadening suppressing the CCF amplitude (Sect.~\ref{sec:abundances}). The free-chemistry retrieval also responds to the same opacity, with its $^{13}$CO posterior showing a sharp peak at the upper limit $\log_{10}\mathrm{VMR}\sim-5.3$, with an extended tail towards lower abundances, therefore yielding a lower limit of $\log_{10}\!^{12}$CO/$^{13}$CO~$\gtrsim 2.3$ ($\gtrsim 200$; Table~\ref{tab:results_free}) that is consistent with the equilibrium value. The high $v\sin i$ and the tentatively elevated $^{12}$CO/$^{13}$CO ratio together favour an evolved, field-age object, suggesting that the intermediate surface gravity may reflect the limited sensitivity of $K$-band observations rather than a genuinely lower gravity.

2M0953's extreme rotation also appears to affect the retrieval itself. As noted in Sect.~\ref{sec:model_comparison}, it is the only target for which the Bayesian evidence prefers the free-chemistry model, and, unlike $^{13}$CO, which both frameworks constrain consistently, certain free-chemistry abundances appear tightly constrained without support from the cross-correlation analysis: H$_2$S ($\log_{10}\mathrm{VMR} = -3.93^{+0.06}_{-0.08}$) and Ti ($-7.01^{+0.05}_{-0.06}$), lying 0.6 and 0.5~dex above the equilibrium predictions (Table~\ref{tab:results_eqbm}), well beyond framework uncertainties and with no CCF peak at the systemic velocity in either framework. For an ultra-fast rotator, the CCF S/N itself is not a decisive diagnostic, since the rotational broadening spreads CCF power and suppresses the peak amplitude (Appendix~\ref{appdx:H218O_appendix}), but it is the complete absence of peak structure at the systemic velocity that indicates these constraints do not arise from coherent signal. We therefore interpret them as the free-chemistry model absorbing residual structure in the blended spectrum rather than as credible abundance constraints. 

\subsection{Thermal structure and cloud properties}
\label{sec:teff}

We report the photospheric temperature ($T_\mathrm{phot}$) at the pressure level of maximal emission contribution rather than a bolometric effective temperature ($T_\mathrm{eff}$), because we are not sensitive to the latter due to limited wavelength coverage. This distinction is important: $T_\mathrm{phot}$ characterises the temperature of the atmospheric layers contributing most strongly to the observed $K$-band flux, while $T_\mathrm{eff}$ is defined by the total bolometric luminosity. The retrieved $T_\mathrm{phot}$ values range from $2099$~K (2M1155) to $2709$~K (2M0434), correlating with spectral type as expected. Comparison with literature effective temperatures (Table~\ref{tab:target_properties}, and for 2M0434 and 2M0953 the spectral-type based estimates of \citealt{Sanghi_2023}, Appendix~\ref{apdx:PT}) shows general agreement, with $T_\mathrm{phot}$ within $\sim$100~K of $T_\mathrm{eff}$. This offset likely reflects the difference between the $K$-band photospheric temperature and the bolometric effective temperature.

The retrieved P-T profiles follow the expected decreasing temperature with decreasing pressure characteristic of radiatively stable dwarf atmospheres. The profiles for equilibrium- and free-chemistry retrievals are displayed in Appendix~\ref{apdx:PT}. All targets show photospheric temperatures in the observed range of $T_\mathrm{phot} \approx 2100$--$2710$~K at pressures of $P \approx 0.1$--$0.4$~bar for the M dwarfs and $\sim 1$--$3$~bar for the L-types, declining to $T \lesssim 1000$~K at $P \sim 10^{-6}$~bar. The temperature gradients ($\nabla T_i$) at the five pressure nodes show generally consistent values between free- and equilibrium-chemistry retrievals, with very small differences at photospheric pressures. All targets retrieve an isothermal region above the photosphere, with $\nabla T_2 = 0.01$--$0.03$ at $10^{-2}$ bar for the L dwarfs, and the same structure $\nabla T_1  \sim0.01$--$0.03$ higher up at $10^{-4}$ bar for the M dwarfs. 2M0434's profile appears as an exception, showing a second low-gradient region below the photosphere, with $\nabla T_3 = 0.02\pm0.01$ (free chemistry) and $0.06\pm0.01$ (equilibrium) at 1~bar, compared to $0.11$--$0.12$ for all other targets. For 2M0434, however, this node ($P=1$~bar) lies $\sim1$~dex below the peak of the emission contribution function, whereas for the other targets it lies within $\sim$0.5~dex of this peak (Fig.~\ref{fig:Pt_profs}). This lower $\nabla T_3$ therefore reflects the poorly constrained region below the $K$-band photosphere, rather than a property of 2M0434's atmosphere.

Compared with the self-consistent cloudy and cloudless Sonora Diamondback \citep{SonoraDiamondback_Morley2024} and, for 2M0434, cloudless SPHINX\footnote{\url{https://doi.org/10.5281/zenodo.11392341}} \citep{SPHINX_Iyer_2023}  profiles in Fig.~\ref{fig:Pt_profs}, the retrieved profiles, excluding 2M0434 and least pronounced for 2M0523, show shallower temperature gradients over the photospheric pressure than in the cloudy ($f_\mathrm{sed} = 8$) Sonora Diamondback models, with the cloudless counterparts even steeper. A similar behaviour is found by \citet{Burningham2021}, whose thermal profile retrieved from 1--2.5~$\mu$m coverage alone deviates at all pressures from a self-consistent grid profile that is matched closely at pressures deeper than $\sim$1~bar with the profile retrieved from 1--15~$\mu$m coverage, illustrating how a narrow wavelength range may poorly constrain the thermal profile. At the contribution peak the retrieved and cloudy Diamondback model temperatures agree closely for 2M0608 and 2M0523, whereas for SP0829, 2M1155, and 2M0953 the retrieved profiles cross the model profiles above the contribution peak and are cooler than the models at the peak itself. The cloud-free models are only slightly cooler at this peak, so the cloud treatment does not account for the offset, suggesting the offset may be driven instead by the different profile shape noted above, and for SP0829 and 2M1155 possibly also the literature $T_\mathrm{eff}$ used for model selection, which exceeds their $T_\mathrm{phot}$ by $\sim$100~K (Table~\ref{tab:target_properties}). For 2M0434, the SPHINX profile does not show the same offset in slope through the photosphere, but is $\sim$100~K hotter through the region. We note that the deep gradient $\nabla T_4$ is strongly degenerate with the base temperature $T_4$, with Pearson correlation coefficients $r = +0.77$ to $+0.99$ across all retrievals, and corresponding to a region where the profile is poorly constrained (Fig.~\ref{fig:Pt_profs}), differing between chemistry frameworks beyond the formal uncertainties for two targets, so that our data cannot reliably test for sub-adiabatic deep profiles suggested for late-L and cooler brown dwarfs \citep[e.g.][]{Tremblin_2016, Tremblin_2019, Leggett_2021, Leggett2023}.

The grey cloud model parameters, namely base pressure ($\log P_\mathrm{cl}$), base opacity ($\log \kappa_\mathrm{cl}$), and sedimentation efficiency ($f_\mathrm{sed}$), remain unconstrained by the $K$-band data alone. Large uncertainties ($\sigma \gtrsim1.5$~dex for the pressure and opacity parameters) reflect the limited sensitivity of high-resolution $K$-band spectroscopy to grey cloud opacity, which primarily affects the continuum level rather than individual line profiles. The retrievals show no evidence of optically thick cloud decks in the $K$-band photosphere, consistent with similar findings from $K$-band SupJup retrievals of isolated L dwarfs \citep{deRegt2024, Mulder2025, Grasser2025}.

The condensation curves for Al$_2$O$_3$, CaTiO$_3$, Fe, Mg$_2$SiO$_4$ and MgSiO$_3$ in Fig.~\ref{fig:Pt_profs}, evaluated at the retrieved equilibrium-chemistry metallicities, show that the photospheric pressure layers of the two M dwarfs are largely hotter than the condensation temperatures of all five species considered. The three L2--L2.5 dwarfs show a contrast, with the profiles crossing the Al$_2$O$_3$, CaTiO$_3$, and Fe curves near the peak of the emission contribution function, and Mg$_2$SiO$_4$ and MgSiO$_3$ near the lowest pressures probed by the observations. The L0 2M0953 shows its profiles crossing the Al$_2$O$_3$ and CaTiO$_3$ curves near the lowest photospheric pressures. These intersections indicate that corundum, perovskite, and iron may condense at pressures probed by the observations, as well as potentially forsterite and enstatite in the case of the L2--L2.5 targets. As \citet{Grasser2025} also note for their cooler L4 target 2M1425, where silicate clouds may likewise form at pressures probed by the observations, the possible presence of condensates at photospheric pressures is not reflected in a cloud opacity that is constrained: the opacity of these condensates may be negligible or fail to be captured by our grey cloud model. Cloud-free solutions are consistent with previous retrievals of high-resolution $K$-band spectra (e.g. \citealt{Zhang2021_BD, Xuan2022, deRegt2024, Landman2024}), and are expected, since the opacity of gaseous absorbers rises towards longer wavelengths, so that the $K$ band probes higher altitudes than the $J$ and $H$ bands, whereas the opacity of small cloud particles falls with wavelength \citep{Marley2015, deRegt2024}, leaving the $K$ band comparatively insensitive to clouds \citep{Landman2023}. Broader wavelength coverage, such as with JWST, is better able to detect them \citep{Miles2023}.

\subsection{Carbon-to-oxygen ratio}
\label{sec:co_ratio}

The equilibrium-chemistry retrievals yield C/O ratios ranging from $0.51^{+0.005}_{-0.004}$ (2M1155) to $0.63\pm0.01$ (2M0434), clustering around the solar value of C/O$_\odot = 0.59 \pm 0.08$ \citep{Asplund2021}. The free-chemistry retrievals give systematically slightly higher C/O values, ranging from $0.59\pm0.004$ (2M1155) to $0.69\pm0.01$ (2M0434), offset from the equilibrium values by $0.04$--$0.08$ for every target, but remain broadly consistent with solar composition. Such an offset has been attributed to oxygen sequestration in refractory condensates: free-chemistry retrievals measure only the gaseous C/O ratio, whereas equilibrium-chemistry retrievals account for the total atmospheric C/O including oxygen possibly locked in condensates such as silicate clouds (e.g. MgSiO$_3$) \citep{Line2021, Grasser2025}. 

In our sample however, condensation should not be the sole cause of this offset, as the photospheric temperature(s) of the two M dwarfs lie largely above the condensation temperatures (Sect.~\ref{sec:teff}; Fig.~\ref{fig:Pt_profs}). Instead, the offset arises largely from the bookkeeping of the gas-phase oxygen. The free-chemistry C/O was computed from the retrieved species alone (Sect.~\ref{sec:chemical_abundances}), whereas the equilibrium-chemistry treatment partitions the oxygen among additional gas-phase species and condensates at each pressure and temperature. At the sample photospheric temperatures $T_\mathrm{phot} \approx 2100$--$2700$~K the tables place $\sim5$--$7\%$ of the gas-phase oxygen in SiO, which is not among the retrieved species (Sect.~\ref{sec:species_selection_acf}). However, including SiO as a free parameter does not recover this oxygen: in an earlier iteration of our retrievals in which SiO was retrieved for all six targets, its free-chemistry abundance remained unconstrained and $\gtrsim1$--$2$~dex below the equilibrium value, in line with its exclusion by our species screen (Sect.~\ref{sec:species_selection_acf}), so that the offset in C/O remains consistent with the trend presented here. Evaluating the C/O ratio of the retrieved species alone at the peak of the contribution function according to our \texttt{FastChem} equilibrium interpolation and retrieved equilibrium parameters and uncertainties reproduces the retrieved free-chemistry values for 2M0434, 2M0608, and 2M0953 ($0.69 \pm 0.01$, $0.61 \pm 0.01$, and $0.64 \pm 0.01$ predicted vs $0.69 \pm 0.01$, $0.61 \pm 0.01$, and $0.65 \pm 0.01$ retrieved). This retrieved-species-alone approach also accounts for roughly half to two-thirds of the offset in the L2--L2.5 dwarfs ($0.59$, $0.66$, and $0.55$ predicted vs $0.61$, $0.68$, and $0.59$ retrieved, all with uncertainties $\leq 0.01$), with the remainder consistent in direction with oxygen sequestration by condensates, as predicted theoretically \citep{Calamari_2024} and discussed for high-resolution retrievals of later L-type dwarfs \citep{Grasser2025, deRegt2024}. This remaining offset is smaller than for those later L-type targets, likely because our warmer photospheres exceed the silicate condensation temperatures at the pressures of peak contribution, reducing gas-phase oxygen depletion relative to the cooler targets of \citet{Grasser2025} and \citet{deRegt2024} \citep{Visscher2010, Line2021}.

\subsection{Metallicity}
\label{sec:metallicity}

The retrieved metallicities for the six targets span $[\mathrm{Fe/H}] = -0.43\pm0.04$ (2M0608) to $+0.32^{+0.05}_{-0.04}$ (2M0953). 2M0523, SP0829, 2M0953, and 2M1155 are mildly super-solar 
($[\mathrm{Fe/H}] = +0.05^{+0.02}_{-0.01}$ to $+0.32^{+0.05}_{-0.04}$), 2M0434 is 
consistent with solar ($-0.09^{+0.13}_{-0.14}$) and 2M0608 is sub-solar. These predominantly near-solar values are consistent with previous retrieval studies of ultracool dwarfs (e.g. \citealt{Xuan_2024_GLIESE229BAB, Zalesky_2022, deRegt2024}).

Despite the different metallicity measurements between frameworks, with [Fe/H] sampled in equilibrium and [C/H] derived a posteriori in free chemistry, the two measurements agree to within $0.1$~dex for five targets ($[\mathrm{C/H}] = -0.34^{+0.03}_{-0.04}$ to $+0.30^{+0.02}_{-0.04}$; Table~\ref{tab:results_free}) and within the larger uncertainties for 2M0434 ($[\mathrm{C/H}] = +0.20^{+0.17}_{-0.19}$), with similar target-to-target ordering in both frameworks according to the different measurements. The broader 2M0434 posteriors reflect the well-documented $K$-band degeneracy between metallicity and surface gravity \citep{Zhang2021_BD, Picos2024}, which is present in all of our retrievals but appears more loosely constrained in 2M0434, which we explore further in Appendix~\ref{appdx:2M0434_logg_FeH}.

\subsection{Molecular abundances and detection significance}
\label{sec:abundances}

Cross-correlation analysis (Sect.~\ref{sec:ccf_methods}) quantifies the detection significance of each species across all six targets for both chemistry frameworks. Table~\ref{tab:ccf_snr} summarises the CCF S/N values and Fig.~\ref{fig:ccf_2m0434} presents the CCF curves for 2M0434 with its 13 included species. Our discussion quotes the equilibrium-chemistry CCF S/N values unless stated otherwise.

H$_2^{16}$O and $^{12}$CO dominate the $K$-band opacity and are detected for all six targets, with H$_2^{16}$O at S/N\,$=$\,70.0--145.7 and $^{12}$CO at S/N\,$=$\,36.9--92.8 (Table~\ref{tab:ccf_snr}), with the lowest values in both cases belonging to the rapidly rotating 2M0953. These abundances are consistent with previous studies of L \& M dwarfs (e.g. \citealt{Picos2024, Grasser2025}) and theoretical predictions for photospheres where H$_2$O and CO contain the bulk of atmospheric oxygen and carbon \citep{LoddersFegley2002}. $^{13}$CO is significantly detected only in 2M0608 (S/N\,$=$\,5.6), but all remaining targets except SP0829 and 2M0953 show clear peaks centred at the systemic velocity (S/N\,$\sim$\,2.7--4.1; see Fig.~\ref{fig:ccf_2m0434}), providing tentative evidence of $^{13}$CO in these objects. SP0829 and 2M0953 show no discernible $^{13}$CO peak (S/N\,$=$\,1.9 and 1.0, respectively). For 2M0953, the extreme rotation ($v\sin i = 86.89^{+0.48}_{-0.61}$~km\,s$^{-1}$) is expected to suppress the CCF peak, consistent with the trend that detection significance decreases with increased rotational broadening \citep{Grasser2025}. For SP0829, with its more modest rotation ($v\sin i = 29.48^{+0.40}_{-0.38}$~km\,s$^{-1}$), the absence of a significant $^{13}$CO cross-correlation peak is likely driven by its low S/N ($=24$; the lowest in our sample), with rotational broadening a contributing factor. For comparison, \citet{deRegt2024} recover a marginal $^{13}$CO cross-correlation peak (S/N = 3.7) for the L dwarf DENIS~J0255$-$4700 ($v\sin i = 41.05\pm0.19$~km\,s$^{-1}$) with higher-quality observations ($\mathcal{R}\approx100{,}000$; S/N $\sim40$), suggesting that for the more slowly rotating SP0829, peak recovery depends on a combination of resolution, S/N, and rotation rather than rotation alone.

Atomic calcium is detected in four of six targets (2M0434, 2M0608, 2M0523, and SP0829; S/N\,$=$\,17.3--33.6), with a weaker peak seen for 2M1155 (S/N\,$=$\,4.1; 2.9 in free chemistry). In 2M0953, the CCF shows a peak at the systemic velocity (S/N\,$=$\,14.8), but a higher secondary peak (S/N\,$\approx$\,20) near $v \approx +800$~km\,s$^{-1}$, likely attributable to telluric contamination in the Ca spectral region for this object, preventing a formal detection. As noted by \citet{Picos2024}, Ca features originate primarily from a small spectral region ($\sim1.9$--$2\,\mu$m; the bluest spectral order) where telluric contamination is prevalent, so interpretation of Ca abundances requires caution. Among the other atomic species, Na and Ti are detected in the M7 dwarf 2M0434 (S/N\,$=$\,6.2 and 5.0), whereas in 2M0608 the Na peak at the systemic velocity (S/N\,$=$\,4.1) is exceeded by a feature near $v \approx 400$~km\,s$^{-1}$, possibly reflecting the same contamination or template artefact effects, and Ti is not recovered. HF shows a clear peak above the noise in most targets, with a significant detection only in 2M0608 (S/N\,$=$\,5.4), 2M1155 approaching the threshold (S/N\,$=$\,4.7), and the other targets showing weaker peaks (Fig.~\ref{fig:ccf_2m0434}). These patterns are qualitatively consistent with \citet{Grasser2025}, who report $^{13}$CO and HF detections in both L4 dwarf retrievals, but note that detection significance decreases with increasing rotational broadening.

Mg appears detected in the equilibrium-chemistry CCFs for 2M0523 and 2M1155 (S/N\,$=$\,5.6 and 7.3), and tentatively in SP0829 (S/N\,$=$\,4.7), whereas the free-chemistry retrievals return only upper limits on its abundances ($\log_{10}\mathrm{VMR} 
\lesssim -7$, compared with $\approx -4.0$ to $-4.2$ in equilibrium) and show no CCF signal (S/N\,$\leq$\,0.8). To test whether Mg lines present in the data are simply left unfitted by the free-chemistry models, we cross-correlated their residuals with the equilibrium Mg templates for these targets and again find no signal (S/N\,$=$\,$-1.0$ to $0.4$). This indicates that the more flexible free-chemistry models account for the absorption at the positions of the Mg lines without Mg, and since this absorption is attributed to Mg in just one chemistry framework, we regard Mg as only tentative in these three targets.

While equilibrium chemistry predicts H$_2$S at $\log_{10}\mathrm{VMR} \approx -4.5$ to $-5.4$ for the five targets where it is included, the CCF yields non-detections across these targets, indicating that the weak $K$-band H$_2$S features are not constraining at the S/N and spectral resolution of our data. This contrasts with \citet{Grasser2025}, who report a robust H$_2$S detection (S/N\,$=$\,6.2) in the L4 dwarf 2M0355. Although the free-chemistry retrievals return constrained AlO abundances for the two M dwarfs (Table~\ref{tab:results_free}), the corresponding CCF peaks are marginal (S/N\,$=$\,3.3 and 3.1), and for 2M0608 comparable to a feature elsewhere in the velocity range, so we do not consider AlO as detected. The 2M0434 peak is nonetheless the highest point of its CCF (Fig.~\ref{fig:ccf_2m0434}), and its top per-pixel contributions (computed as in Appendix~\ref{appdx:H218O_appendix}) are spread across the orders where AlO has opacity, in contrast to the single-order concentration found for spurious H$_2^{18}$O signals. A very weak AlO signal would also be consistent with its profile reaching the corundum (Al$_2$O$_3$) condensation curve only well above the peak of the emission contribution function (Sect.~\ref{sec:teff}; Fig.~\ref{fig:Pt_profs}), which leaves gas-phase AlO largely undepleted in the probed layers. However, at S/N $=3.3$, this is at most a weak hint. The remaining minor species that passed the ACF screen (Sc, SH, AlH, and NH$_3$) and the minor CO isotopologues C$^{18}$O and C$^{17}$O show no significant CCF signal (S/N\,$\leq$\,2.9). The broad posteriors for most of these species in the free-chemistry retrievals (Table~\ref{tab:results_free}) indicate that they are unconstrained by the $K$-band data alone, consistent with the treatment of unconstrained species by \citet{Grasser2025}, who note that $\log_{10}\mathrm{VMR} \lesssim -8$ values are largely compatible with a non-detection.

\subsection{Carbon isotope ratio}
\label{sec:carbon_isotope}
The $^{12}$C/$^{13}$C isotope ratio provides a key diagnostic of formation environment and has emerged as a potentially powerful tracer for distinguishing formation pathways \citep{Zhang2021_BD, deRegt2024}. Following \citet{Picos2024}, we treated the CO isotopologue ratio $^{12}$CO/$^{13}$CO as representative of the atmospheric $^{12}$C/$^{13}$C ratio. The equilibrium-chemistry framework sampled this ratio directly as a free parameter, yielding well-constrained posteriors for all targets:

\vspace{3pt}

\noindent\hspace{2em}\textbf{2M0434:} $\log_{10}\!^{12}$CO/$^{13}$CO $= 1.99^{+0.10}_{-0.09}$ ($\approx 98^{+25}_{-18}$);\\[6pt]
\hspace*{2em}\textbf{2M0608:} $\log_{10}\!^{12}$CO/$^{13}$CO $= 1.96^{+0.08}_{-0.06}$ ($\approx 91^{+18}_{-12}$);\\[6pt]
\hspace*{2em}\textbf{2M0523:} $\log_{10}\!^{12}$CO/$^{13}$CO $= 2.17^{+0.11}_{-0.08}$ ($\approx 148^{+43}_{-25}$);\\[6pt]
\hspace*{2em}\textbf{SP0829:} $\log_{10}\!^{12}$CO/$^{13}$CO $= 2.07^{+0.14}_{-0.10}$ ($\approx 117^{+45}_{-24}$);\\[6pt]
\hspace*{2em}\textbf{2M0953:} $\log_{10}\!^{12}$CO/$^{13}$CO $= 2.49^{+0.17}_{-0.16}$ ($\approx 309^{+148}_{-95}$);\\[6pt]
\hspace*{2em}\textbf{2M1155:} $\log_{10}\!^{12}$CO/$^{13}$CO $= 2.19^{+0.12}_{-0.09}$ ($\approx 155^{+49}_{-29}$).
\vspace{3pt}

\noindent All targets return ratios above the local ISM value of $^{12}$C/$^{13}$C\,$=$\,$68 \pm 15$ \citep{Milam2005}. The M dwarfs are both within $\sim1.3\sigma$ of the ISM, while the L dwarfs are higher, echoing the result seen for DENIS J0255 \citep{deRegt2024}.

Cross-correlation analysis (Sect.~\ref{sec:abundances}) supports $^{13}$CO detection with at least marginal significance in four targets (2M0434, 2M0608, 2M0523, 2M1155), with clear CCF S/N peaks centred at the systemic velocity. For SP0829 and 2M0953, the equilibrium framework produces well-constrained $^{12}$CO/$^{13}$CO posteriors, with uncertainties comparable to the other targets for SP0829 and marginally higher for 2M0953, despite the absence of $^{13}$CO cross-correlation peaks (Sect.~\ref{sec:abundances}). Notably, the SP0829 ratio ($2.07^{+0.14}_{-0.10}$) is nearly identical to the free-chemistry value ($2.06^{+0.13}_{-0.10}$), suggesting cross-framework consistency despite the CCF non-detection in both cases. For 2M0953, by contrast, the free-chemistry $^{12}$CO/$^{13}$CO posterior peaks sharply at $\approx$\,2.3, its $-1\sigma$ bound, with a long tail towards higher ratios ($2.96^{+3.57}_{-0.69}$), therefore providing only a lower limit ($\gtrsim 2.3$, i.e. $\gtrsim 200$) that is consistent with the equilibrium value (Sect.~\ref{sec:2m0953_rotation}). For the remaining four targets, the free-chemistry ratios, as for SP0829, agree with their equilibrium values to within $1\sigma$ (Table~\ref{tab:results_free}).

Our $^{12}$C/$^{13}$C values align with the growing consensus from high-resolution SupJup studies of isolated L- and M-type dwarfs. \citet{Grasser2025} report $^{12}$C/$^{13}$C\,$=\,95.5^{+6.8}_{-6.4}$--$109.6^{+10.6}_{-9.6}$ for the two L4 dwarfs, while \citet{Picos2024} find ratios of $79^{+20}_{-14}$--$114^{+69}_{-33}$ for three young late-M brown dwarfs. As confirmed by recent observations of M dwarf stars \citep{Dario_Nature2025}, the ISM $^{12}$C/$^{13}$C ratio has decreased over Galactic history due to $^{13}$C enrichment \citep{Romano2017, Milam2005}, so older objects may preserve higher ratios from their natal environment. The elevated ratios in some targets (2M0953, 2M1155, and 2M0523) echo the findings of \citet{deRegt2024}, who measure $^{12}$C/$^{13}$C\,$=$\,$184^{+61}_{-40}$ for the older L dwarf DENIS~J0255$-$4700, suggesting this elevated value may reflect the $^{13}$C-poor ISM at the time of formation. A comparable $^{12}$C/$^{13}$C-based formation-age argument was recently applied to the interstellar comet 3I/ATLAS \citep{Cordiner2026}, whose elevated carbon isotope ratios imply formation in the early Galaxy.

\subsection{Oxygen isotope ratio}
\label{sec:oxygen_isotopes}

The oxygen isotope ratio $^{16}$O/$^{18}$O was probed through the H$_2^{16}$O/H$_2^{18}$O abundance ratio, following \citet{Picos2024}. For four of our six targets, the equilibrium retrieval yields largely unconstrained (excluding 2M0523) log\,H$_2^{16}$O/H$_2^{18}$O~$= 2.79^{+1.00}_{-0.39}$ (2M0434), $3.06^{+1.05}_{-0.39}$ (2M0608), $2.91^{+1.11}_{-0.39}$ (2M1155), and $2.44^{+0.20}_{-0.12}$ (2M0523), with H$_2^{18}$O CCF S/N (Table~\ref{tab:ccf_snr}) all significantly below any detection threshold. For SP0829 and 2M0953, the equilibrium retrieval returns anomalously low constraints: log\,H$_2^{16}$O/H$_2^{18}$O~$=2.10^{+0.11}_{-0.09}$ 
($\sim 126^{+36}_{-24}$) and $1.86^{+0.08}_{-0.07}$ 
($\sim 72^{+15}_{-11}$). If confirmed, these would be the lowest $^{16}$O/$^{18}$O ratio(s) measured for any substellar atmosphere, by a significant margin in the case of 2M0953. They would fall below the solar ($^{16}$O/$^{18}$O $= 525 \pm 21$; \citealt{Lyons2018}) and ISM ($557 \pm 30$; \citealt{Wilson1999}) values, the sub-solar ratios reported by \citet{Picos2024} for young M dwarfs ($\sim 141$--$205$), and the tentative constraint for HIP~55507~B ($240^{+145}_{-80}$; \citealt{Xuan2024}). However, there is observational precedent for caution: \citet{Picos2024} report $^{16}$O/$^{18}$O $\approx 205^{+140}_{-62}$ for the young brown dwarf TWA~28 from $K$-band CRIRES+ spectroscopy, which \citet{Picos_2025_JWST_NIRSPEC} subsequently revise to $681^{+53}_{-50}$ with JWST/NIRSpec spectroscopy spanning 0.97--5.27~$\mu$m. This suggests that $K$-band H$_2^{18}$O constraints may be systematically unreliable without broader wavelength coverage, or challenging at the S/N of our current datasets. \citet{deRegt_2026} similarly caution that retrieved H$_2^{18}$O abundances for Luhman~16AB are likely biased by poorly fitted telluric or atmospheric H$_2$O lines, with rotational broadening further blending the weak isotopologue features, which is an effect directly relevant to SP0829 and 2M0953 as the two fastest rotators in our sample. Our H$_2^{18}$O S/N is only 3.1 for both 2M0953 and SP0829 (Table~\ref{tab:ccf_snr}), below our detection threshold, motivating CCF robustness tests described in Appendix~\ref{appdx:H218O_appendix}. Following these diagnostic tests, we treat H$_2^{18}$O in SP0829 and 2M0953 as non-detections. The retrieved log\,H$_2^{16}$O/H$_2^{18}$O values are thus lower limits.

The CO-based oxygen isotope ratios (C$^{16}$O/C$^{18}$O and C$^{16}$O/C$^{17}$O) are only weakly constrained across all targets, with CCF S/N~$< 1$ for both C$^{18}$O and C$^{17}$O in all cases (Table~\ref{tab:ccf_snr}), consistent with the narrower wavelength coverage of CO features in the $K$ band \citep{Grasser2025}. The narrowest of these posteriors is the equilibrium $\log\,^{12}$CO/C$^{18}$O = $3.15^{+0.39}_{-0.35}$ (C$^{16}$O/C$^{18}$O $\sim 1400$) of 2M0953, which would exceed its H$_2$O-based ratio by more than an order of magnitude if taken at face value. This inconsistency between the two $^{18}$O tracers further supports treating the H$_2^{18}$O constraint for 2M0953 as spurious.

\subsection{Implications for formation pathways}
\label{sec:formation}

The combination of C/O ratios, metallicities, and carbon isotope ratios provides complementary constraints on formation history, though no single tracer can definitively distinguish formation pathways \citep{Zhang2021_SJ}. For isolated targets where comparison to a host is not possible, unlike companion systems such as ROXs12 \citep{Grasser2026}, GQ~Lup \citep{Picos2025}, or HIP~55507 \citep{Xuan2024}, the $^{12}$C/$^{13}$C ratio is one of few direct probes of the natal environment's composition.

For isolated brown dwarfs that formed via gravitational collapse of a molecular cloud, near-solar C/O and metallicity are expected: the object directly inherits the bulk elemental composition of its parent cloud without disk-mediated fractionation processes that can alter these ratios in planetary systems. Unlike planet formation, where volatile ice lines and disk chemistry can significantly enrich or deplete carbon and oxygen relative to each other \citep{Oberg2011}, gravitational collapse preserves the bulk elemental ratios of the natal cloud. Our sample shows near-solar C/O ratios ($0.51$--$0.63$; Sect.~\ref{sec:co_ratio}), predominantly near-solar metallicities ($[\mathrm{Fe/H}] = -0.43\pm0.04$ to $+0.32^{+0.05}_{-0.04}$; Sect.~\ref{sec:metallicity}; Appendix~\ref{appdx:2M0434_logg_FeH}), and $^{12}$C/$^{13}$C ratios at or above the local ISM value of $68 \pm 15$ \citep{Milam2005} (Sect.~\ref{sec:carbon_isotope}). This compositional picture is consistent with expectations for isolated brown dwarf formation via molecular cloud fragmentation.

The $^{12}$C/$^{13}$C ratios show excellent consistency between chemistry frameworks for all targets where both are well constrained, in contrast to the systematic C/O offsets of $\sim$0.04--0.08 between frameworks (Sect.~\ref{sec:co_ratio}). These offsets arise largely from the equilibrium framework's accounting of oxygen in species that are not retrieved, with a smaller additional contribution in the three L2--L2.5 dwarfs consistent with oxygen sequestration by condensates, yet the $^{12}$C/$^{13}$C ratios of these three targets agree between frameworks to within $0.4\sigma$. This supports the argument of \citet{deRegt2024} that isotope ratios may be largely insensitive to cloud condensation and oxygen sequestration effects. The ISM $^{12}$C/$^{13}$C ratio has generally decreased over Galactic history due to $^{13}$C enrichment \citep{Dario_Nature2025, Romano2017, Milam2005}, so older objects may preserve higher ratios from their natal environment. As such, elevated $^{12}$C/$^{13}$C ratios in 2M0953, 2M1155, and 2M0523 may reflect the $^{13}$C-poor ISM at the time of formation. 

Of our targets, only 2M0434 retains comparatively loosely constrained fundamental parameters: both chemistry frameworks agree on its surface gravity and metallicity with veiling included, but with comparatively broad posteriors along the $K$-band $\log g$--metallicity degeneracy, along which they separate without veiling (Sect.~\ref{sec:logg}; Appendix~\ref{appdx:2M0434_logg_FeH}). Crucially, 2M0434's $^{12}$C/$^{13}$C ratio is consistent between frameworks in all of these retrievals and the C/O shows the same small systematic offset seen across the full sample, explained entirely by oxygen bookkeeping effects for this target (Sect.~\ref{sec:co_ratio}), suggesting both parameters are insensitive to the gravity--metallicity degeneracy and the relative isotopic composition of 2M0434 is reliably constrained despite its uncertain fundamental parameters. 

Most of the limitations encountered in this work trace back to the restriction of the $K$ band's narrow wavelength coverage. The degeneracy between surface gravity and metallicity calls for gravity-sensitive features at shorter wavelengths (Appendix~\ref{appdx:2M0434_logg_FeH}), and the thermal profiles are only weakly constrained by the $K$ band alone (Sect.~\ref{sec:teff}). The oxygen isotope ratios are similarly affected: the $K$-band H$_2^{18}$O constraint for TWA~28 is substantially revised by JWST spectroscopy covering 0.97--5.27~$\mu$m \citep{Picos_2025_JWST_NIRSPEC}. Combining $K$-band spectra with broader wavelength coverage is therefore the most direct way to address these limitations. Analysis of the growing SupJup sample may reveal population-level trends with age or birth environment. The elevated $^{12}$C/$^{13}$C ratios in our sample provide measurements that can test predictions of chemical evolution models \citep{Romano2017, Romano2022}, while future observations combining broader wavelength coverage with kinematic constraints will further establish the compositional diversity of the isolated brown dwarf population.

\section{Conclusions}
\label{sec:conclusions}

We presented atmospheric retrievals for six isolated ultracool dwarfs spanning spectral types M7 to L2.5, observed with CRIRES+ as part of the ESO SupJup Survey. Equilibrium chemistry is statistically preferred for five of the six targets ($\ln B = -17.10$ to $-41.31$), although both frameworks fit the data similarly well (with the reduced $\chi^2_0$ differing by only $0.02$--$0.34\%$), and was adopted as the model of focus in this work for consistent comparison. H$_2^{16}$O and $^{12}$CO are robustly detected in all targets (S/N\,$=$\,70.0--145.7 and 36.9--92.8). $^{13}$CO is significantly detected in 2M0608 (S/N\,$=$\,5.6) and tentatively in 2M0434, 2M0523, and 2M1155 (S/N\,$=$\,2.7--4.1). SP0829 and 2M0953 show no detectable $^{13}$CO signal, attributable to low S/N for SP0829 and to extreme rotational broadening for 2M0953 (Sect.~\ref{sec:abundances}). Atomic Ca is detected in four targets, while Na, Ti, and HF are detected in individual M dwarfs. Mg detection in two of the equilibrium-chemistry retrieval CCFs remains tentative, since the free-chemistry models account for the same absorption without Mg. The remaining minor species are undetected. The retrieved compositions are consistent with expectations for formation via molecular cloud fragmentation: near-solar C/O ratios ($0.51$--$0.63$), largely near-solar metallicities ($[\mathrm{Fe/H}] = -0.43 \pm 0.04$ to $+0.32^{+0.05}_{-0.04}$), and $^{12}$C/$^{13}$C ratios of $\sim91$--155 ($\sim309$ for 2M0953) at or above the local ISM value of $68 \pm 15$ \citep{Milam2005}, with the values for the two fastest rotators resting on the spectral fit but not corroborated by a $^{13}$CO cross-correlation peak. The free-chemistry C/O ratios are higher by $0.04$--$0.08$, an offset that arises largely from the bookkeeping of oxygen in species that are not directly retrieved (chiefly SiO), with a smaller contribution in the L2--L2.5 dwarfs consistent with oxygen sequestration in potential condensates (Sect.~\ref{sec:co_ratio}). The M7 dwarf 2M0434, the youngest object in our sample with an associated age of 1--2~Myr \citep{Gagne2018,Kenyon1995}, retrieves a non-zero veiling continuum in both frameworks, which agree on its gravity and metallicity only with veiling included. Without this continuum, the two frameworks diverge along the $K$-band gravity--metallicity degeneracy ridge (Sect.~\ref{sec:logg}; Appendix~\ref{appdx:2M0434_logg_FeH}). The excellent agreement of $^{12}$C/$^{13}$C between chemistry frameworks supports the argument of \citet{deRegt2024} that carbon isotope ratios may be largely insensitive to cloud condensation and oxygen sequestration, which can bias C/O. The case of 2M0434 additionally suggests that these ratios are also insensitive to the $K$-band gravity--metallicity degeneracy. The L0 dwarf 2M0953 emerges as the fastest rotator in our sample ($v\sin i=86.89^{+0.48}_{-0.61}$~km\,s$^{-1}$) and among the fastest-known rotating ultracool dwarfs \citep{Tannock2021}, with its rapid spin and tentatively elevated $^{12}$C/$^{13}$C ($\approx309^{+148}_{-95}$) consistent with a mature, field-age object. Apparent H$_2^{18}$O constraints for 2M0953 and SP0829 are shown to be spurious via CCF robustness tests (Appendix~\ref{appdx:H218O_appendix}): the signals originate from anomalous residuals in the last spectral order rather than coherent isotopologue detections, and for 2M0953 the implied $^{16}$O/$^{18}$O is also inconsistent with its retrieved CO-based ratio, so that the retrieved $\log_{10}$H$_2^{16}$O/H$_2^{18}$O values of $1.86^{+0.08}_{-0.07}$ and $2.10^{+0.11}_{-0.09}$ serve as lower limits. The JWST/NIRSpec upward revision of TWA~28's $^{16}$O/$^{18}$O ratio by a factor of $\sim$3 \citep{Picos2024, Picos_2025_JWST_NIRSPEC} further highlights the limitations of oxygen isotope constraints from relatively low S/N $K$-band spectroscopy. This work adds six objects to the growing SupJup sample. Population-level trends in $^{12}$C/$^{13}$C across isolated brown dwarfs, companions, and hosts will further test isotope ratios as potential tracers of substellar formation pathways.

\begin{acknowledgements}
We are grateful to the anonymous referee for their constructive comments and suggestions, which helped to improve the manuscript. This work was supported by NL-NWO Spinoza SPI.2022.004 and NWO grant OCENW.M.21.010 (D.G.P., S.d.R., I.S.). This work is based on observations collected at the European Organisation for Astronomical Research in the Southern Hemisphere under ESO programme 1110.C-4264. This research has used NASA Astrophysics Data System (ADS) and the Python packages NumPy \citep{Numpy2020}, SciPy \citep{Scipy2020}, Matplotlib \citep{Hunter2007}, petitRADTRANS \citep{Molliere2019}, PyAstronomy \citep{Czesla2019}, Astropy \citep{Astropy2022}, and corner \citep{Foreman2016}.
\end{acknowledgements}

\bibliographystyle{aa}
\bibliography{bib.bib}

\begin{appendix}
\onecolumn
\section{Comprehensive retrieval results} \label{apdx:results_tab_eqbm}

\begin{table}[ht!]
\centering
\scriptsize
\caption{Retrieved parameters from equilibrium-chemistry retrievals.}
\label{tab:results_eqbm}
\setlength{\tabcolsep}{16pt}
\begin{tabular}{lcccccc}
\hline\hline
Parameter & 2M0434 & 2M0608 & 2M0523 & SP0829 & 2M0953 & 2M1155 \\
 & (M7) & (M9.6) & (L2.5) & (L2) & (L0) & (L2) \\
\hline
\multicolumn{7}{c}{\textit{Physical Parameters}} \\
\hline
$v_\mathrm{rad}$ [km\,s$^{-1}$] & $17.76^{+0.05}_{-0.05}$ & $24.79^{+0.03}_{-0.03}$ & $13.25^{+0.03}_{-0.03}$ & $27.14^{+0.08}_{-0.08}$ & $21.37^{+0.22}_{-0.24}$ & $45.98^{+0.03}_{-0.03}$ \\
$v\sin i$ [km\,s$^{-1}$] & $15.69^{+0.13}_{-0.09}$ & $16.64^{+0.11}_{-0.09}$ & $17.54^{+0.21}_{-0.21}$ & $29.48^{+0.40}_{-0.38}$ & $86.89^{+0.48}_{-0.61}$ & $13.81^{+0.06}_{-0.07}$ \\
$\log g$ [cgs] & $3.64^{+0.15}_{-0.13}$ & $3.70^{+0.06}_{-0.06}$ & $5.27^{+0.03}_{-0.03}$ & $5.43^{+0.04}_{-0.04}$ & $4.95^{+0.07}_{-0.05}$ & $5.44^{+0.03}_{-0.04}$ \\
$u_\mathrm{LD}$ & $0.53^{+0.05}_{-0.05}$ & $0.74^{+0.04}_{-0.04}$ & $0.88^{+0.06}_{-0.07}$ & $0.55^{+0.09}_{-0.09}$ & $0.97^{+0.02}_{-0.03}$ & $0.97^{+0.02}_{-0.02}$ \\
\hline
\multicolumn{7}{c}{\textit{Temperature Structure}} \\
\hline
$\nabla T_0$ & $0.23^{+0.08}_{-0.09}$ & $0.31^{+0.05}_{-0.06}$ & $0.22^{+0.09}_{-0.09}$ & $0.17^{+0.09}_{-0.08}$ & $0.17^{+0.06}_{-0.07}$ & $0.21^{+0.10}_{-0.10}$ \\
$\nabla T_1$ & $0.03^{+0.03}_{-0.02}$ & $0.01^{+0.01}_{-0.01}$ & $0.25^{+0.07}_{-0.08}$ & $0.30^{+0.05}_{-0.06}$ & $0.04^{+0.02}_{-0.02}$ & $0.28^{+0.06}_{-0.10}$ \\
$\nabla T_2$ & $0.17^{+0.02}_{-0.01}$ & $0.06^{+0.00}_{-0.00}$ & $0.02^{+0.01}_{-0.01}$ & $0.01^{+0.01}_{-0.01}$ & $0.01^{+0.00}_{-0.00}$ & $0.01^{+0.01}_{-0.01}$ \\
$\nabla T_3$ & $0.06^{+0.01}_{-0.01}$ & $0.12^{+0.01}_{-0.01}$ & $0.12^{+0.00}_{-0.00}$ & $0.12^{+0.00}_{-0.00}$ & $0.11^{+0.00}_{-0.00}$ & $0.12^{+0.00}_{-0.00}$ \\
$\nabla T_4$ & $0.23^{+0.06}_{-0.07}$ & $0.05^{+0.04}_{-0.03}$ & $0.33^{+0.03}_{-0.04}$ & $0.21^{+0.04}_{-0.03}$ & $0.06^{+0.04}_{-0.03}$ & $0.22^{+0.05}_{-0.03}$ \\
$T_4$ [K] & $5221^{+547}_{-538}$ & $4051^{+374}_{-230}$ & $4804^{+202}_{-254}$ & $4140^{+248}_{-172}$ & $3583^{+261}_{-177}$ & $3979^{+313}_{-171}$ \\
\textcolor{gray}{$T_\mathrm{phot}$ [K]\space \tablefootmark{(a)}} & \textcolor{gray}{\textit{$2709$}} & \textcolor{gray}{\textit{$2236$}} & \textcolor{gray}{\textit{$2140$}} & \textcolor{gray}{\textit{$2196$}} & \textcolor{gray}{\textit{$2262$}} & \textcolor{gray}{\textit{$2099$}} \\
\hline
\multicolumn{7}{c}{\textit{Cloud Properties}} \\
\hline
$\log\kappa_\mathrm{cl}$ [cm$^2$\,g$^{-1}$] & $-4.95^{+2.06}_{-2.03}$ & $-5.24^{+2.55}_{-2.31}$ & $-4.98^{+2.32}_{-2.39}$ & $-5.72^{+2.22}_{-2.00}$ & $-6.32^{+1.68}_{-1.47}$ & $-4.59^{+2.64}_{-2.46}$ \\
$\log P_\mathrm{cl}$ [bar] & $-1.00^{+2.04}_{-2.12}$ & $-0.49^{+2.01}_{-3.08}$ & $-1.26^{+2.18}_{-2.25}$ & $-1.10^{+2.05}_{-2.34}$ & $-0.46^{+1.61}_{-1.71}$ & $-2.67^{+2.05}_{-1.69}$ \\
$f_\mathrm{sed}$ & $9.30^{+4.63}_{-4.24}$ & $12.31^{+3.77}_{-4.49}$ & $8.55^{+5.03}_{-4.49}$ & $13.51^{+3.10}_{-3.97}$ & $16.34^{+1.90}_{-2.52}$ & $9.92^{+4.45}_{-4.40}$ \\
\hline
\multicolumn{7}{c}{\textit{Composition (Sampled)}} \\
\hline
C/O & $0.63^{+0.01}_{-0.01}$ & $0.57^{+0.01}_{-0.01}$ & $0.55^{+0.00}_{-0.00}$ & $0.61^{+0.00}_{-0.00}$ & $0.60^{+0.01}_{-0.01}$ & $0.51^{+0.00}_{-0.00}$ \\
$[$Fe/H$]$ [dex] & $-0.09^{+0.13}_{-0.14}$ & $-0.43^{+0.04}_{-0.04}$ & $0.05^{+0.02}_{-0.01}$ & $0.25^{+0.03}_{-0.02}$ & $0.32^{+0.05}_{-0.04}$ & $0.23^{+0.02}_{-0.03}$ \\
$\log\,^{12}$CO/$^{13}$CO & $1.99^{+0.10}_{-0.09}$ & $1.96^{+0.08}_{-0.06}$ & $2.17^{+0.11}_{-0.08}$ & $2.07^{+0.14}_{-0.10}$ & $2.49^{+0.17}_{-0.16}$ & $2.19^{+0.12}_{-0.09}$ \\
$\log\,^{12}$CO/C$^{18}$O & $4.09^{+0.63}_{-0.55}$ & $4.39^{+0.80}_{-0.80}$ & $4.57^{+0.69}_{-0.69}$ & $4.40^{+0.81}_{-0.92}$ & $3.15^{+0.39}_{-0.35}$ & $4.27^{+0.84}_{-0.80}$ \\
$\log$ H$_2$O/H$_2^{18}$O & $2.79^{+1.00}_{-0.39}$ & $3.06^{+1.05}_{-0.39}$ & $2.44^{+0.20}_{-0.12}$ & $2.10^{+0.11}_{-0.09}$ & $1.86^{+0.08}_{-0.07}$ & $2.91^{+1.11}_{-0.39}$ \\
$\log$ C$^{16}$O/C$^{17}$O & $3.49^{+0.62}_{-0.49}$ & $4.23^{+0.62}_{-0.56}$ & $4.31^{+0.80}_{-0.76}$ & $4.16^{+0.66}_{-0.64}$ & $4.75^{+0.62}_{-0.79}$ & $4.17^{+0.68}_{-0.61}$ \\
\hline
\multicolumn{7}{c}{\textit{Derived Species Abundances } \tablefootmark{(b)}} \\
\hline
\textcolor{gray}{$\log$ H$_2$O} & \textcolor{gray}{\textit{$-3.98^{+0.34}_{-0.14}$}} & \textcolor{gray}{\textit{$-3.97^{+0.05}_{-0.05}$}} & \textcolor{gray}{\textit{$-3.42^{+0.03}_{-0.04}$}} & \textcolor{gray}{\textit{$-3.32^{+0.02}_{-0.05}$}} & \textcolor{gray}{\textit{$-3.26^{+0.06}_{-0.06}$}} & \textcolor{gray}{\textit{$-3.20^{+0.04}_{-0.03}$}} \\
\textcolor{gray}{$\log$ H$_2^{18}$O} & \textcolor{gray}{\textit{$-7.69^{+0.66}_{-0.40}$}} & \textcolor{gray}{\textit{$-7.54^{+0.98}_{-1.01}$}} & \textcolor{gray}{\textit{$-6.33^{+0.45}_{-0.50}$}} & \textcolor{gray}{\textit{$-5.62^{+0.24}_{-0.15}$}} & \textcolor{gray}{\textit{$-5.23^{+0.09}_{-0.14}$}} & \textcolor{gray}{\textit{$-6.54^{+0.40}_{-1.47}$}} \\
\textcolor{gray}{$\log\,^{12}$CO} & \textcolor{gray}{\textit{$-3.64^{+0.31}_{-0.19}$}} & \textcolor{gray}{\textit{$-3.77^{+0.02}_{-0.03}$}} & \textcolor{gray}{\textit{$-3.28^{+0.01}_{-0.03}$}} & \textcolor{gray}{\textit{$-3.05^{+0.03}_{-0.03}$}} & \textcolor{gray}{\textit{$-3.03^{+0.07}_{-0.04}$}} & \textcolor{gray}{\textit{$-3.13^{+0.04}_{-0.03}$}} \\
\textcolor{gray}{$\log\,^{13}$CO} & \textcolor{gray}{\textit{$-5.74^{+0.33}_{-0.36}$}} & \textcolor{gray}{\textit{$-5.82^{+0.20}_{-0.13}$}} & \textcolor{gray}{\textit{$-5.69^{+0.31}_{-0.26}$}} & \textcolor{gray}{\textit{$-5.99^{+0.81}_{-1.55}$}} & \textcolor{gray}{\textit{$-5.44^{+0.15}_{-0.09}$}} & \textcolor{gray}{\textit{$-5.94^{+0.61}_{-0.39}$}} \\
\textcolor{gray}{$\log$ C$^{18}$O} & \textcolor{gray}{\textit{$-7.87^{+0.30}_{-0.84}$}} & \textcolor{gray}{\textit{$-8.03^{+0.55}_{-1.30}$}} & \textcolor{gray}{\textit{$-7.16^{+0.13}_{-0.99}$}} & \textcolor{gray}{\textit{$-6.88^{+0.97}_{-1.06}$}} & \textcolor{gray}{\textit{$-5.84^{+0.20}_{-0.97}$}} & \textcolor{gray}{\textit{$-6.80^{+0.66}_{-0.89}$}} \\
\textcolor{gray}{$\log$ C$^{17}$O} & \textcolor{gray}{\textit{$-6.80^{+0.52}_{-0.93}$}} & \textcolor{gray}{\textit{$-7.69^{+0.54}_{-0.40}$}} & \textcolor{gray}{\textit{$-7.75^{+0.90}_{-0.58}$}} & \textcolor{gray}{\textit{$-7.10^{+0.75}_{-0.80}$}} & \textcolor{gray}{\textit{$-8.03^{+0.76}_{-0.29}$}} & \textcolor{gray}{\textit{$-7.21^{+0.44}_{-0.56}$}} \\
\textcolor{gray}{$\log$ H$_2$S} & -- & \textcolor{gray}{\textit{$-5.38^{+0.05}_{-0.03}$}} & \textcolor{gray}{\textit{$-4.72^{+0.03}_{-0.03}$}} & \textcolor{gray}{\textit{$-4.53^{+0.03}_{-0.04}$}} & \textcolor{gray}{\textit{$-4.57^{+0.07}_{-0.04}$}} & \textcolor{gray}{\textit{$-4.53^{+0.04}_{-0.03}$}} \\
\textcolor{gray}{$\log$ HF} & \textcolor{gray}{\textit{$-7.72^{+0.32}_{-0.18}$}} & \textcolor{gray}{\textit{$-7.84^{+0.03}_{-0.03}$}} & \textcolor{gray}{\textit{$-7.45^{+0.02}_{-0.02}$}} & \textcolor{gray}{\textit{$-7.25^{+0.03}_{-0.02}$}} & \textcolor{gray}{\textit{$-7.16^{+0.05}_{-0.04}$}} & \textcolor{gray}{\textit{$-7.32^{+0.03}_{-0.03}$}} \\
\textcolor{gray}{$\log$ NH$_3$} & -- & -- & \textcolor{gray}{\textit{$-6.45^{+0.02}_{-0.02}$}} & -- & -- & \textcolor{gray}{\textit{$-6.32^{+0.02}_{-0.02}$}} \\
\textcolor{gray}{$\log$ Na} & \textcolor{gray}{\textit{$-6.08^{+0.37}_{-0.21}$}} & \textcolor{gray}{\textit{$-6.01^{+0.03}_{-0.03}$}} & -- & -- & \textcolor{gray}{\textit{$-5.26^{+0.06}_{-0.05}$}} & -- \\
\textcolor{gray}{$\log$ Sc} & \textcolor{gray}{\textit{$-8.98^{+0.33}_{-0.17}$}} & \textcolor{gray}{\textit{$-9.08^{+0.03}_{-0.03}$}} & \textcolor{gray}{\textit{$-8.58^{+0.03}_{-0.03}$}} & \textcolor{gray}{\textit{$-8.37^{+0.03}_{-0.03}$}} & \textcolor{gray}{\textit{$-8.34^{+0.06}_{-0.05}$}} & \textcolor{gray}{\textit{$-8.41^{+0.04}_{-0.03}$}} \\
\textcolor{gray}{$\log$ Ca} & \textcolor{gray}{\textit{$-5.86^{+0.35}_{-0.15}$}} & \textcolor{gray}{\textit{$-5.95^{+0.03}_{-0.03}$}} & \textcolor{gray}{\textit{$-5.50^{+0.03}_{-0.02}$}} & \textcolor{gray}{\textit{$-5.30^{+0.03}_{-0.03}$}} & \textcolor{gray}{\textit{$-5.23^{+0.06}_{-0.05}$}} & \textcolor{gray}{\textit{$-5.33^{+0.04}_{-0.03}$}} \\
\textcolor{gray}{$\log$ AlO} & \textcolor{gray}{\textit{$-8.93^{+0.66}_{-0.34}$}} & \textcolor{gray}{\textit{$-9.11^{+0.08}_{-0.09}$}} & \textcolor{gray}{\textit{$-8.29^{+0.05}_{-0.04}$}} & \textcolor{gray}{\textit{$-7.95^{+0.03}_{-0.07}$}} & \textcolor{gray}{\textit{$-7.74^{+0.10}_{-0.09}$}} & \textcolor{gray}{\textit{$-8.04^{+0.03}_{-0.06}$}} \\
\textcolor{gray}{$\log$ Mg} & -- & \textcolor{gray}{\textit{$-4.67^{+0.03}_{-0.03}$}} & \textcolor{gray}{\textit{$-4.17^{+0.03}_{-0.03}$}} & \textcolor{gray}{\textit{$-3.97^{+0.03}_{-0.03}$}} & \textcolor{gray}{\textit{$-3.93^{+0.06}_{-0.05}$}} & \textcolor{gray}{\textit{$-4.00^{+0.04}_{-0.03}$}} \\
\textcolor{gray}{$\log$ Ti} & \textcolor{gray}{\textit{$-7.31^{+0.17}_{-0.21}$}} & \textcolor{gray}{\textit{$-7.74^{+0.04}_{-0.04}$}} & -- & -- & \textcolor{gray}{\textit{$-7.55^{+0.04}_{-0.02}$}} & -- \\
\textcolor{gray}{$\log$ SH} & \textcolor{gray}{\textit{$-5.48^{+0.31}_{-0.17}$}} & \textcolor{gray}{\textit{$-5.65^{+0.06}_{-0.05}$}} & -- & -- & \textcolor{gray}{\textit{$-5.05^{+0.06}_{-0.06}$}} & -- \\
\textcolor{gray}{$\log$ AlH} & -- & \textcolor{gray}{\textit{$-6.87^{+0.04}_{-0.03}$}} & \textcolor{gray}{\textit{$-6.01^{+0.02}_{-0.01}$}} & -- & \textcolor{gray}{\textit{$-5.97^{+0.06}_{-0.04}$}} & -- \\
\hline
\multicolumn{7}{c}{\textit{Veiling Parameters (2M0434)}} \\
\hline
$r_0$ & $0.21^{+0.04}_{-0.03}$ & -- & -- & -- & -- & -- \\
$\alpha_\mathrm{veil}$ & $2.40^{+0.27}_{-0.35}$ & -- & -- & -- & -- & -- \\
\hline
\multicolumn{7}{c}{\textit{GP Hyperparameters}} \\
\hline
$\log a$ & $0.22^{+0.00}_{-0.00}$ & $0.24^{+0.00}_{-0.00}$ & $0.04^{+0.01}_{-0.00}$ & $0.05^{+0.00}_{-0.00}$ & $0.03^{+0.00}_{-0.00}$ & $0.14^{+0.00}_{-0.00}$ \\
$\log \ell$ [nm] & $-1.44^{+0.01}_{-0.01}$ & $-1.35^{+0.00}_{-0.00}$ & $-1.72^{+0.01}_{-0.01}$ & $-2.06^{+0.01}_{-0.01}$ & $-1.90^{+0.01}_{-0.01}$ & $-1.48^{+0.01}_{-0.01}$ \\
\hline
\end{tabular}
\tablefoot{The uncertainties are 1$\sigma$ (68\% credible interval) from the posterior distributions. The grey entries indicate parameters derived a posteriori from retrieved free parameters rather than directly sampled. The species marked with an em dash (--) were not included in the retrieval for that target following the ACF species screen (Sect.~\ref{sec:species_selection_acf}).\\ \tablefoottext{a}{$T_\mathrm{phot}$ is the photospheric temperature at the pressure of maximal emission contribution, not the bolometric effective temperature. The derived photospheric temperatures are rounded to the nearest degree for clarity.} \tablefoottext{b}{$\log$ denotes $\log_{10}$ of the VMR, calculated at the pressure of maximum emission contribution.}}
\end{table}

\clearpage

\begin{table*}
\centering
\caption{Retrieved parameters from free-chemistry retrievals.}
\label{tab:results_free}
\scriptsize
\setlength{\tabcolsep}{16pt}
\begin{tabular}{l cccccc}
\hline\hline
Parameter & 2M0434 & 2M0608 & 2M0523 & SP0829 & 2M0953 & 2M1155 \\
 & (M7) & (M9.6) & (L2.5) & (L2) & (L0) & (L2) \\
\hline
\multicolumn{7}{c}{\textit{Physical Parameters}} \\
\hline
$v_\mathrm{rad}$ [km\,s$^{-1}$] & $17.76^{+0.04}_{-0.04}$ & $24.78^{+0.03}_{-0.03}$ & $13.25^{+0.02}_{-0.02}$ & $27.14^{+0.06}_{-0.06}$ & $20.03^{+0.17}_{-0.18}$ & $46.00^{+0.02}_{-0.02}$ \\
$v\sin i$ [km\,s$^{-1}$] & $15.74^{+0.13}_{-0.10}$ & $16.53^{+0.06}_{-0.05}$ & $17.43^{+0.18}_{-0.15}$ & $29.68^{+0.43}_{-0.35}$ & $85.90^{+0.33}_{-0.32}$ & $13.89^{+0.08}_{-0.10}$ \\
$\log g$ [cgs] & $3.69^{+0.17}_{-0.14}$ & $3.77^{+0.05}_{-0.06}$ & $5.27^{+0.04}_{-0.03}$ & $5.49^{+0.03}_{-0.05}$ & $4.85^{+0.04}_{-0.04}$ & $5.36^{+0.03}_{-0.03}$ \\
$u_\mathrm{LD}$ & $0.56^{+0.06}_{-0.05}$ & $0.69^{+0.03}_{-0.03}$ & $0.83^{+0.05}_{-0.05}$ & $0.60^{+0.10}_{-0.08}$ & $0.99^{+0.01}_{-0.01}$ & $0.96^{+0.02}_{-0.03}$ \\
\hline
\multicolumn{7}{c}{\textit{Temperature Structure}} \\
\hline
$\nabla T_0$ & $0.24^{+0.07}_{-0.07}$ & $0.28^{+0.06}_{-0.07}$ & $0.23^{+0.08}_{-0.09}$ & $0.17^{+0.07}_{-0.07}$ & $0.24^{+0.07}_{-0.08}$ & $0.24^{+0.06}_{-0.07}$ \\
$\nabla T_1$ & $0.01^{+0.01}_{-0.01}$ & $0.01^{+0.01}_{-0.01}$ & $0.17^{+0.08}_{-0.07}$ & $0.32^{+0.04}_{-0.05}$ & $0.10^{+0.04}_{-0.04}$ & $0.17^{+0.08}_{-0.08}$ \\
$\nabla T_2$ & $0.14^{+0.01}_{-0.01}$ & $0.07^{+0.00}_{-0.00}$ & $0.02^{+0.01}_{-0.01}$ & $0.02^{+0.01}_{-0.01}$ & $0.01^{+0.00}_{-0.00}$ & $0.03^{+0.01}_{-0.01}$ \\
$\nabla T_3$ & $0.02^{+0.01}_{-0.01}$ & $0.11^{+0.00}_{-0.00}$ & $0.11^{+0.00}_{-0.00}$ & $0.12^{+0.00}_{-0.00}$ & $0.11^{+0.00}_{-0.00}$ & $0.11^{+0.00}_{-0.00}$ \\
$\nabla T_4$ & $0.31^{+0.04}_{-0.05}$ & $0.05^{+0.03}_{-0.02}$ & $0.27^{+0.03}_{-0.03}$ & $0.12^{+0.04}_{-0.03}$ & $0.04^{+0.03}_{-0.02}$ & $0.26^{+0.05}_{-0.05}$ \\
$T_4$ [K] & $4988^{+448}_{-393}$ & $3747^{+225}_{-171}$ & $4215^{+201}_{-176}$ & $3547^{+198}_{-130}$ & $3553^{+170}_{-113}$ & $4091^{+275}_{-250}$ \\
\textcolor{gray}{$T_\mathrm{phot}$ [K]\space \tablefootmark{(a)}} & \textcolor{gray}{\textit{$2635$}} & \textcolor{gray}{\textit{$2233$}} & \textcolor{gray}{\textit{$2118$}} & \textcolor{gray}{\textit{$2192$}} & \textcolor{gray}{\textit{$2291$}} & \textcolor{gray}{\textit{$2111$}} \\
\hline
\multicolumn{7}{c}{\textit{Cloud Properties}} \\
\hline
$\log\kappa_\mathrm{cl}$ [cm$^2$\,g$^{-1}$] & $-6.20^{+1.74}_{-1.60}$ & $-4.67^{+1.68}_{-1.77}$ & $-5.77^{+2.09}_{-1.86}$ & $-4.50^{+2.21}_{-2.13}$ & $-5.32^{+2.11}_{-2.00}$ & $-5.31^{+1.94}_{-1.87}$ \\
$\log P_\mathrm{cl}$ [bar] & $-0.52^{+1.45}_{-1.75}$ & $-0.64^{+1.64}_{-2.04}$ & $-1.90^{+1.66}_{-1.63}$ & $-2.65^{+1.82}_{-1.48}$ & $-1.83^{+1.86}_{-1.74}$ & $-1.21^{+1.88}_{-2.11}$ \\
$f_\mathrm{sed}$ & $7.60^{+4.37}_{-3.36}$ & $11.15^{+3.64}_{-3.89}$ & $8.46^{+3.98}_{-3.64}$ & $10.21^{+3.66}_{-3.89}$ & $8.40^{+4.02}_{-3.71}$ & $9.15^{+4.49}_{-3.85}$ \\
\hline
\multicolumn{7}{c}{\textit{Species Abundances (Sampled)}\space \tablefootmark{(b)}} \\
\hline
$\log$ H$_2$O & $-3.50^{+0.17}_{-0.18}$ & $-3.85^{+0.04}_{-0.04}$ & $-3.44^{+0.03}_{-0.02}$ & $-3.33^{+0.02}_{-0.04}$ & $-3.35^{+0.03}_{-0.03}$ & $-3.29^{+0.02}_{-0.02}$ \\
$\log$ H$_2^{18}$O & $-8.10^{+1.64}_{-1.64}$ & $-9.17^{+2.11}_{-1.44}$ & $-5.95^{+0.14}_{-0.33}$ & $-5.44^{+0.09}_{-0.11}$ & $-5.18^{+0.05}_{-0.05}$ & $-8.44^{+2.03}_{-1.72}$ \\
$\log\,^{12}$CO & $-3.15^{+0.17}_{-0.19}$ & $-3.65^{+0.03}_{-0.04}$ & $-3.25^{+0.03}_{-0.02}$ & $-3.01^{+0.02}_{-0.04}$ & $-3.08^{+0.03}_{-0.03}$ & $-3.14^{+0.02}_{-0.02}$ \\
$\log\,^{13}$CO & $-5.23^{+0.18}_{-0.20}$ & $-5.69^{+0.08}_{-0.08}$ & $-5.38^{+0.07}_{-0.08}$ & $-5.08^{+0.10}_{-0.14}$ & $-6.03^{+0.70}_{-3.58}$ & $-5.35^{+0.09}_{-0.11}$ \\
$\log$ C$^{18}$O & $-8.82^{+1.22}_{-1.32}$ & $-8.40^{+0.92}_{-1.12}$ & $-9.60^{+1.59}_{-1.19}$ & $-9.47^{+1.47}_{-1.20}$ & $-8.32^{+1.16}_{-1.25}$ & $-7.71^{+0.86}_{-1.13}$ \\
$\log$ C$^{17}$O & $-8.33^{+1.11}_{-1.15}$ & $-9.35^{+1.00}_{-1.04}$ & $-8.72^{+1.15}_{-1.24}$ & $-8.15^{+1.02}_{-1.13}$ & $-9.16^{+1.41}_{-1.29}$ & $-8.34^{+0.91}_{-1.01}$ \\
$\log$ H$_2$S & -- & $-9.28^{+1.51}_{-1.23}$ & $-7.80^{+1.31}_{-1.61}$ & $-8.73^{+1.70}_{-1.44}$ & $-3.93^{+0.06}_{-0.08}$ & $-8.56^{+1.31}_{-1.35}$ \\
$\log$ HF & $-7.47^{+0.21}_{-0.22}$ & $-7.94^{+0.09}_{-0.10}$ & $-7.85^{+0.09}_{-0.10}$ & $-7.25^{+0.06}_{-0.07}$ & $-7.42^{+0.07}_{-0.08}$ & $-7.59^{+0.07}_{-0.07}$ \\
$\log$ NH$_3$ & -- & -- & $-9.51^{+1.04}_{-0.99}$ & -- & -- & $-9.61^{+1.08}_{-1.01}$ \\
$\log$ Na & $-5.63^{+0.17}_{-0.25}$ & $-5.89^{+0.11}_{-0.11}$ & -- & -- & $-6.09^{+0.37}_{-2.95}$ & -- \\
$\log$ Sc & $-9.17^{+0.29}_{-0.34}$ & $-10.25^{+0.44}_{-0.53}$ & $-10.48^{+0.55}_{-0.57}$ & $-9.95^{+0.51}_{-0.58}$ & $-10.90^{+0.47}_{-0.46}$ & $-10.62^{+0.78}_{-0.61}$ \\
$\log$ Ca & $-4.87^{+0.18}_{-0.23}$ & $-5.77^{+0.06}_{-0.07}$ & $-5.76^{+0.05}_{-0.04}$ & $-5.74^{+0.04}_{-0.06}$ & $-5.32^{+0.04}_{-0.05}$ & $-7.68^{+0.46}_{-0.97}$ \\
$\log$ AlO & $-8.11^{+0.18}_{-0.20}$ & $-8.81^{+0.11}_{-0.13}$ & $-8.95^{+0.28}_{-0.70}$ & $-9.13^{+0.56}_{-1.03}$ & $-10.05^{+0.70}_{-0.75}$ & $-8.81^{+0.33}_{-1.50}$ \\
$\log$ Mg & -- & $-8.88^{+1.60}_{-1.38}$ & $-8.23^{+1.54}_{-1.61}$ & $-8.54^{+1.58}_{-1.49}$ & $-8.48^{+1.81}_{-1.57}$ & $-8.71^{+1.63}_{-1.45}$ \\
$\log$ Ti & $-6.46^{+0.18}_{-0.22}$ & $-7.83^{+0.29}_{-0.45}$ & -- & -- & $-7.01^{+0.05}_{-0.06}$ & -- \\
$\log$ SH & $-4.97^{+0.20}_{-0.24}$ & $-5.47^{+0.10}_{-0.10}$ & -- & -- & $-8.97^{+2.58}_{-1.45}$ & -- \\
$\log$ AlH & -- & $-10.12^{+0.88}_{-0.82}$ & $-8.86^{+0.79}_{-0.90}$ & -- & $-9.35^{+0.99}_{-1.04}$ & -- \\
\hline
\multicolumn{7}{c}{\textit{Derived Composition}} \\
\hline
\textcolor{gray}{C/O} & \textcolor{gray}{\textit{$0.69^{+0.01}_{-0.01}$}} & \textcolor{gray}{\textit{$0.61^{+0.01}_{-0.01}$}} & \textcolor{gray}{\textit{$0.61^{+0.00}_{-0.00}$}} & \textcolor{gray}{\textit{$0.68^{+0.00}_{-0.00}$}} & \textcolor{gray}{\textit{$0.65^{+0.01}_{-0.01}$}} & \textcolor{gray}{\textit{$0.59^{+0.00}_{-0.00}$}} \\
\textcolor{gray}{$[$C/H$]$ [dex]} & \textcolor{gray}{\textit{$0.20^{+0.17}_{-0.19}$}} & \textcolor{gray}{\textit{$-0.34^{+0.03}_{-0.04}$}} & \textcolor{gray}{\textit{$0.06^{+0.03}_{-0.02}$}} & \textcolor{gray}{\textit{$0.30^{+0.02}_{-0.04}$}} & \textcolor{gray}{\textit{$0.24^{+0.03}_{-0.03}$}} & \textcolor{gray}{\textit{$0.17^{+0.02}_{-0.02}$}} \\
\textcolor{gray}{$\log\,^{12}$CO/$^{13}$CO} & \textcolor{gray}{\textit{$2.09^{+0.09}_{-0.08}$}} & \textcolor{gray}{\textit{$2.04^{+0.07}_{-0.06}$}} & \textcolor{gray}{\textit{$2.13^{+0.08}_{-0.07}$}} & \textcolor{gray}{\textit{$2.06^{+0.13}_{-0.10}$}} & \textcolor{gray}{\textit{$2.96^{+3.57}_{-0.69}$}} & \textcolor{gray}{\textit{$2.21^{+0.11}_{-0.09}$}} \\
\textcolor{gray}{$\log\,^{12}$CO/C$^{18}$O} & \textcolor{gray}{\textit{$5.67^{+1.29}_{-1.19}$}} & \textcolor{gray}{\textit{$4.75^{+1.12}_{-0.92}$}} & \textcolor{gray}{\textit{$6.35^{+1.19}_{-1.60}$}} & \textcolor{gray}{\textit{$6.46^{+1.20}_{-1.48}$}} & \textcolor{gray}{\textit{$5.25^{+1.26}_{-1.15}$}} & \textcolor{gray}{\textit{$4.57^{+1.13}_{-0.86}$}} \\
\textcolor{gray}{$\log$ H$_2$O/H$_2^{18}$O} & \textcolor{gray}{\textit{$4.59^{+1.65}_{-1.64}$}} & \textcolor{gray}{\textit{$5.32^{+1.44}_{-2.13}$}} & \textcolor{gray}{\textit{$2.51^{+0.32}_{-0.14}$}} & \textcolor{gray}{\textit{$2.11^{+0.11}_{-0.08}$}} & \textcolor{gray}{\textit{$1.84^{+0.04}_{-0.04}$}} & \textcolor{gray}{\textit{$5.15^{+1.73}_{-2.03}$}} \\
\textcolor{gray}{$\log$ C$^{16}$O/C$^{17}$O} & \textcolor{gray}{\textit{$5.18^{+1.17}_{-1.14}$}} & \textcolor{gray}{\textit{$5.70^{+1.04}_{-1.01}$}} & \textcolor{gray}{\textit{$5.47^{+1.23}_{-1.14}$}} & \textcolor{gray}{\textit{$5.14^{+1.12}_{-1.02}$}} & \textcolor{gray}{\textit{$6.08^{+1.29}_{-1.42}$}} & \textcolor{gray}{\textit{$5.20^{+1.02}_{-0.90}$}} \\
\hline
\multicolumn{7}{c}{\textit{Veiling Parameters (2M0434)}} \\
\hline
$r_0$ & $0.13^{+0.02}_{-0.02}$ & -- & -- & -- & -- & -- \\
$\alpha_\mathrm{veil}$ & $2.53^{+0.20}_{-0.23}$ & -- & -- & -- & -- & -- \\
\hline
\multicolumn{7}{c}{\textit{GP Hyperparameters}} \\
\hline
$\log a$ & $0.22^{+0.00}_{-0.00}$ & $0.24^{+0.00}_{-0.00}$ & $0.05^{+0.00}_{-0.00}$ & $0.05^{+0.00}_{-0.00}$ & $0.05^{+0.00}_{-0.00}$ & $0.16^{+0.00}_{-0.00}$ \\
$\log \ell$ [nm] & $-1.44^{+0.00}_{-0.00}$ & $-1.34^{+0.00}_{-0.00}$ & $-1.72^{+0.01}_{-0.01}$ & $-2.05^{+0.00}_{-0.00}$ & $-1.95^{+0.01}_{-0.01}$ & $-1.48^{+0.01}_{-0.01}$ \\
\hline
\end{tabular}
\tablefoot{The uncertainties are 1$\sigma$ (68\% credible interval) from the posterior distributions. The grey entries indicate parameters derived a posteriori from retrieved free parameters rather than directly sampled. The free-chemistry model treats each molecular species as an independent free parameter. The species marked with an em dash (--) were not included in the retrieval for that target following the ACF species screen (Sect.~\ref{sec:species_selection_acf}).\\ \tablefoottext{a}{$T_\mathrm{phot}$ is the photospheric temperature at the pressure of maximal emission contribution, not the bolometric effective temperature. The derived photospheric temperatures are rounded to the nearest degree for clarity.} \tablefoottext{b}{log denotes $\log_{10}$ of the VMR, calculated at the pressure of maximum emission contribution.}}
\end{table*}
\clearpage

\section{Sampler convergence and sensitivity test - 2M0523}

\begin{table*}[ht!]
\centering
\caption{Live-point sensitivity test for 2M0523, with the retrieved parameters as functions of the number of live points.}
\label{tab:ladder_2M0523}
\setlength{\tabcolsep}{11pt}
\scalebox{0.70}{
\begin{tabular}{l cccc|ccccc}
\hline\hline
 & \multicolumn{4}{c|}{\textit{Equilibrium chemistry}} & \multicolumn{5}{c}{\textit{Free chemistry}} \\
$N_\mathrm{live}$ & 400 & 1000 & 1500 & 2000 & 400 & 2000 & 3000 & 4000 & 5000 \\
\hline
\multicolumn{10}{c}{\textit{Physical Parameters}} \\
\hline
$v_\mathrm{rad}$ [km\,s$^{-1}$] & $13.25^{+0.03}_{-0.03}$ & $13.25^{+0.03}_{-0.03}$ & $13.25^{+0.03}_{-0.03}$ & $13.25^{+0.03}_{-0.03}$ & $13.20^{+0.02}_{-0.03}$ & $13.31^{+0.02}_{-0.02}$ & $13.23^{+0.02}_{-0.02}$ & $13.25^{+0.03}_{-0.02}$ & $13.25^{+0.03}_{-0.03}$ \\
$v\sin i$ [km\,s$^{-1}$] & $17.35^{+0.17}_{-0.18}$ & $17.49^{+0.21}_{-0.20}$ & $17.50^{+0.21}_{-0.19}$ & $17.50^{+0.22}_{-0.20}$ & $17.35^{+0.12}_{-0.11}$ & $17.19^{+0.11}_{-0.28}$ & $17.39^{+0.15}_{-0.14}$ & $17.49^{+0.18}_{-0.17}$ & $17.52^{+0.18}_{-0.18}$ \\
$\log g$ [cgs] & $5.30^{+0.03}_{-0.03}$ & $5.29^{+0.03}_{-0.03}$ & $5.28^{+0.03}_{-0.03}$ & $5.28^{+0.03}_{-0.03}$ & $5.13^{+0.05}_{-0.02}$ & $5.35^{+0.02}_{-0.02}$ & $5.34^{+0.02}_{-0.03}$ & $5.34^{+0.02}_{-0.03}$ & $5.33^{+0.02}_{-0.02}$ \\
$u_\mathrm{LD}$ & $0.81^{+0.06}_{-0.07}$ & $0.86^{+0.06}_{-0.06}$ & $0.87^{+0.06}_{-0.06}$ & $0.87^{+0.06}_{-0.07}$ & $0.70^{+0.05}_{-0.04}$ & $0.75^{+0.05}_{-0.09}$ & $0.83^{+0.05}_{-0.05}$ & $0.86^{+0.05}_{-0.06}$ & $0.88^{+0.05}_{-0.06}$ \\
\hline
\multicolumn{10}{c}{\textit{Temperature Structure}} \\
\hline
$\nabla T_0$ & $0.10^{+0.06}_{-0.05}$ & $0.18^{+0.09}_{-0.09}$ & $0.18^{+0.10}_{-0.09}$ & $0.19^{+0.10}_{-0.09}$ & $0.14^{+0.05}_{-0.05}$ & $0.23^{+0.06}_{-0.06}$ & $0.14^{+0.06}_{-0.06}$ & $0.20^{+0.08}_{-0.08}$ & $0.19^{+0.09}_{-0.09}$ \\
$\nabla T_1$ & $0.29^{+0.06}_{-0.08}$ & $0.33^{+0.04}_{-0.07}$ & $0.31^{+0.05}_{-0.10}$ & $0.33^{+0.04}_{-0.07}$ & $0.12^{+0.09}_{-0.06}$ & $0.18^{+0.06}_{-0.05}$ & $0.17^{+0.09}_{-0.07}$ & $0.25^{+0.07}_{-0.07}$ & $0.21^{+0.08}_{-0.09}$ \\
$\nabla T_2$ & $0.03^{+0.01}_{-0.01}$ & $0.03^{+0.01}_{-0.01}$ & $0.02^{+0.01}_{-0.01}$ & $0.02^{+0.01}_{-0.01}$ & $0.05^{+0.02}_{-0.01}$ & $0.01^{+0.01}_{-0.01}$ & $0.01^{+0.01}_{-0.01}$ & $0.02^{+0.01}_{-0.01}$ & $0.02^{+0.01}_{-0.01}$ \\
$\nabla T_3$ & $0.12^{+0.00}_{-0.00}$ & $0.12^{+0.00}_{-0.00}$ & $0.12^{+0.00}_{-0.00}$ & $0.12^{+0.00}_{-0.00}$ & $0.12^{+0.00}_{-0.00}$ & $0.11^{+0.00}_{-0.00}$ & $0.11^{+0.00}_{-0.00}$ & $0.12^{+0.00}_{-0.00}$ & $0.11^{+0.00}_{-0.00}$ \\
$\nabla T_4$ & $0.32^{+0.03}_{-0.03}$ & $0.31^{+0.03}_{-0.04}$ & $0.32^{+0.04}_{-0.04}$ & $0.31^{+0.04}_{-0.05}$ & $0.29^{+0.03}_{-0.03}$ & $0.18^{+0.02}_{-0.02}$ & $0.22^{+0.03}_{-0.03}$ & $0.19^{+0.03}_{-0.03}$ & $0.23^{+0.03}_{-0.03}$ \\
$T_4$ [K] & $4728^{+195}_{-204}$ & $4681^{+222}_{-230}$ & $4741^{+244}_{-267}$ & $4679^{+234}_{-301}$ & $4399^{+198}_{-190}$ & $3670^{+95}_{-85}$ & $3858^{+141}_{-159}$ & $3730^{+157}_{-131}$ & $3945^{+173}_{-185}$ \\
\textcolor{gray}{$T_\mathrm{phot}$ [K]\space\tablefootmark{(a)}} & \textcolor{gray}{\textit{$2136$}} & \textcolor{gray}{\textit{$2133$}} & \textcolor{gray}{\textit{$2131$}} & \textcolor{gray}{\textit{$2136$}} & \textcolor{gray}{\textit{$2043$}} & \textcolor{gray}{\textit{$2106$}} & \textcolor{gray}{\textit{$2102$}} & \textcolor{gray}{\textit{$2106$}} & \textcolor{gray}{\textit{$2105$}} \\
\hline
\multicolumn{10}{c}{\textit{Cloud Properties}} \\
\hline
$\log\kappa_\mathrm{cl}$ [cm$^2$\,g$^{-1}$] & $-4.37^{+2.05}_{-2.03}$ & $-5.20^{+2.31}_{-2.28}$ & $-4.47^{+2.92}_{-2.66}$ & $-4.38^{+2.74}_{-2.62}$ & $-6.05^{+2.12}_{-1.77}$ & $-6.64^{+1.32}_{-1.29}$ & $-4.51^{+2.17}_{-2.16}$ & $-4.92^{+2.18}_{-2.15}$ & $-5.11^{+2.44}_{-2.27}$ \\
$\log P_\mathrm{cl}$ [bar] & $-2.15^{+2.39}_{-1.95}$ & $-1.52^{+2.33}_{-2.25}$ & $-1.83^{+2.55}_{-2.27}$ & $-1.94^{+2.67}_{-2.07}$ & $-1.97^{+1.76}_{-1.67}$ & $0.38^{+1.03}_{-1.20}$ & $-2.69^{+1.80}_{-1.46}$ & $-1.78^{+1.94}_{-1.82}$ & $-1.68^{+2.13}_{-1.99}$ \\
$f_\mathrm{sed}$ & $7.17^{+4.32}_{-3.66}$ & $9.88^{+4.97}_{-4.88}$ & $9.83^{+4.77}_{-4.87}$ & $9.75^{+5.16}_{-4.77}$ & $11.06^{+4.00}_{-4.38}$ & $12.77^{+2.52}_{-2.60}$ & $10.59^{+3.39}_{-3.55}$ & $9.13^{+4.50}_{-4.15}$ & $10.48^{+4.27}_{-4.48}$ \\
\hline
\multicolumn{10}{c}{\textit{Composition}} \\
\hline
C/O & $0.54^{+0.00}_{-0.00}$ & $0.55^{+0.00}_{-0.00}$ & $0.55^{+0.00}_{-0.00}$ & $0.55^{+0.00}_{-0.00}$ & \textcolor{gray}{\textit{$0.62^{+0.00}_{-0.00}$}} & \textcolor{gray}{\textit{$0.61^{+0.00}_{-0.00}$}} & \textcolor{gray}{\textit{$0.60^{+0.00}_{-0.00}$}} & \textcolor{gray}{\textit{$0.60^{+0.00}_{-0.00}$}} & \textcolor{gray}{\textit{$0.60^{+0.00}_{-0.00}$}} \\
$[$Fe/H$]$ [dex]\space\tablefootmark{(c)} & $0.06^{+0.01}_{-0.02}$ & $0.06^{+0.02}_{-0.02}$ & $0.05^{+0.02}_{-0.02}$ & $0.05^{+0.02}_{-0.02}$ & \textcolor{gray}{\textit{$0.02^{+0.02}_{-0.01}$}} & \textcolor{gray}{\textit{$0.12^{+0.01}_{-0.01}$}} & \textcolor{gray}{\textit{$0.11^{+0.02}_{-0.02}$}} & \textcolor{gray}{\textit{$0.11^{+0.02}_{-0.02}$}} & \textcolor{gray}{\textit{$0.09^{+0.02}_{-0.02}$}} \\
$\log\,^{12}$CO/$^{13}$CO & $2.22^{+0.11}_{-0.09}$ & $2.17^{+0.10}_{-0.09}$ & $2.17^{+0.11}_{-0.08}$ & $2.16^{+0.11}_{-0.09}$ & \textcolor{gray}{\textit{$2.27^{+3.38}_{-0.13}$}} & \textcolor{gray}{\textit{$2.10^{+0.07}_{-0.07}$}} & \textcolor{gray}{\textit{$2.10^{+0.06}_{-0.06}$}} & \textcolor{gray}{\textit{$2.09^{+0.06}_{-0.05}$}} & \textcolor{gray}{\textit{$2.08^{+0.07}_{-0.06}$}} \\
$\log\,^{12}$CO/C$^{18}$O & $4.05^{+0.58}_{-0.54}$ & $4.59^{+0.71}_{-0.77}$ & $4.30^{+0.72}_{-0.70}$ & $4.52^{+0.71}_{-0.76}$ & \textcolor{gray}{\textit{$2.99^{+0.27}_{-0.21}$}} & \textcolor{gray}{\textit{$5.62^{+1.00}_{-1.02}$}} & \textcolor{gray}{\textit{$5.82^{+1.29}_{-1.21}$}} & \textcolor{gray}{\textit{$5.95^{+1.22}_{-1.26}$}} & \textcolor{gray}{\textit{$5.66^{+1.34}_{-1.30}$}} \\
$\log$ H$_2$O/H$_2^{18}$O & $2.47^{+0.18}_{-0.12}$ & $2.43^{+0.19}_{-0.11}$ & $2.41^{+0.17}_{-0.11}$ & $2.42^{+0.16}_{-0.11}$ & \textcolor{gray}{\textit{$6.45^{+0.95}_{-1.11}$}} & \textcolor{gray}{\textit{$6.80^{+0.84}_{-0.98}$}} & \textcolor{gray}{\textit{$2.47^{+0.22}_{-0.12}$}} & \textcolor{gray}{\textit{$2.46^{+0.28}_{-0.12}$}} & \textcolor{gray}{\textit{$2.42^{+0.17}_{-0.11}$}} \\
$\log$ C$^{16}$O/C$^{17}$O & $4.24^{+0.72}_{-0.66}$ & $4.47^{+0.71}_{-0.73}$ & $4.13^{+0.75}_{-0.65}$ & $4.35^{+0.76}_{-0.74}$ & \textcolor{gray}{\textit{$5.76^{+0.81}_{-0.83}$}} & \textcolor{gray}{\textit{$6.35^{+0.83}_{-0.87}$}} & \textcolor{gray}{\textit{$6.38^{+1.07}_{-1.29}$}} & \textcolor{gray}{\textit{$5.62^{+1.22}_{-1.17}$}} & \textcolor{gray}{\textit{$5.97^{+1.26}_{-1.28}$}} \\
\hline
\multicolumn{10}{c}{\textit{Species Abundances}\space\tablefootmark{(b)}} \\
\hline
$\log$ H$_2$O & \textcolor{gray}{\textit{$-3.41^{+0.02}_{-0.05}$}} & \textcolor{gray}{\textit{$-3.42^{+0.03}_{-0.03}$}} & \textcolor{gray}{\textit{$-3.43^{+0.03}_{-0.02}$}} & \textcolor{gray}{\textit{$-3.42^{+0.02}_{-0.03}$}} & $-3.50^{+0.03}_{-0.01}$ & $-3.37^{+0.01}_{-0.02}$ & $-3.39^{+0.02}_{-0.02}$ & $-3.39^{+0.02}_{-0.02}$ & $-3.41^{+0.02}_{-0.02}$ \\
$\log$ H$_2^{18}$O & \textcolor{gray}{\textit{$-6.38^{+0.66}_{-0.58}$}} & \textcolor{gray}{\textit{$-6.28^{+0.32}_{-0.76}$}} & \textcolor{gray}{\textit{$-6.97^{+1.06}_{-1.29}$}} & \textcolor{gray}{\textit{$-6.35^{+0.46}_{-1.12}$}} & $-9.96^{+1.11}_{-0.94}$ & $-10.18^{+0.98}_{-0.84}$ & $-5.86^{+0.12}_{-0.22}$ & $-5.85^{+0.12}_{-0.28}$ & $-5.83^{+0.11}_{-0.18}$ \\
$\log\,^{12}$CO & \textcolor{gray}{\textit{$-3.27^{+0.03}_{-0.03}$}} & \textcolor{gray}{\textit{$-3.27^{+0.02}_{-0.04}$}} & \textcolor{gray}{\textit{$-3.28^{+0.02}_{-0.03}$}} & \textcolor{gray}{\textit{$-3.27^{+0.03}_{-0.03}$}} & $-3.29^{+0.02}_{-0.01}$ & $-3.19^{+0.01}_{-0.01}$ & $-3.21^{+0.02}_{-0.02}$ & $-3.21^{+0.02}_{-0.02}$ & $-3.22^{+0.02}_{-0.02}$ \\
$\log\,^{13}$CO & \textcolor{gray}{\textit{$-5.75^{+0.24}_{-1.16}$}} & \textcolor{gray}{\textit{$-5.82^{+0.44}_{-0.74}$}} & \textcolor{gray}{\textit{$-5.82^{+0.24}_{-0.69}$}} & \textcolor{gray}{\textit{$-6.00^{+0.54}_{-0.39}$}} & $-5.56^{+0.13}_{-3.35}$ & $-5.30^{+0.07}_{-0.07}$ & $-5.31^{+0.06}_{-0.07}$ & $-5.29^{+0.06}_{-0.06}$ & $-5.31^{+0.06}_{-0.07}$ \\
$\log$ C$^{18}$O & \textcolor{gray}{\textit{$-6.98^{+0.43}_{-0.66}$}} & \textcolor{gray}{\textit{$-8.13^{+1.20}_{-0.48}$}} & \textcolor{gray}{\textit{$-7.62^{+0.45}_{-0.83}$}} & \textcolor{gray}{\textit{$-7.58^{+1.31}_{-0.84}$}} & $-6.28^{+0.20}_{-0.26}$ & $-8.82^{+1.01}_{-1.01}$ & $-9.03^{+1.21}_{-1.29}$ & $-9.15^{+1.26}_{-1.22}$ & $-8.89^{+1.30}_{-1.33}$ \\
$\log$ C$^{17}$O & \textcolor{gray}{\textit{$-7.76^{+0.71}_{-0.64}$}} & \textcolor{gray}{\textit{$-8.05^{+0.98}_{-0.69}$}} & \textcolor{gray}{\textit{$-7.52^{+0.66}_{-0.92}$}} & \textcolor{gray}{\textit{$-7.21^{+0.51}_{-1.57}$}} & $-9.04^{+0.83}_{-0.81}$ & $-9.54^{+0.88}_{-0.84}$ & $-9.59^{+1.28}_{-1.07}$ & $-8.83^{+1.17}_{-1.22}$ & $-9.19^{+1.28}_{-1.26}$ \\
$\log$ HF & \textcolor{gray}{\textit{$-7.44^{+0.02}_{-0.02}$}} & \textcolor{gray}{\textit{$-7.44^{+0.02}_{-0.03}$}} & \textcolor{gray}{\textit{$-7.45^{+0.01}_{-0.03}$}} & \textcolor{gray}{\textit{$-7.44^{+0.01}_{-0.03}$}} & $-7.98^{+0.10}_{-0.09}$ & $-7.82^{+0.10}_{-0.12}$ & $-7.79^{+0.07}_{-0.07}$ & $-7.76^{+0.07}_{-0.08}$ & $-7.77^{+0.07}_{-0.08}$ \\
$\log$ Ca & \textcolor{gray}{\textit{$-5.48^{+0.03}_{-0.04}$}} & \textcolor{gray}{\textit{$-5.49^{+0.02}_{-0.04}$}} & \textcolor{gray}{\textit{$-5.50^{+0.03}_{-0.03}$}} & \textcolor{gray}{\textit{$-5.49^{+0.02}_{-0.03}$}} & $-5.92^{+0.03}_{-0.03}$ & $-5.70^{+0.03}_{-0.03}$ & $-5.68^{+0.03}_{-0.03}$ & $-5.67^{+0.03}_{-0.03}$ & $-5.69^{+0.03}_{-0.03}$ \\
$\log$ Sc & \textcolor{gray}{\textit{$-8.56^{+0.03}_{-0.04}$}} & \textcolor{gray}{\textit{$-8.56^{+0.02}_{-0.04}$}} & \textcolor{gray}{\textit{$-8.58^{+0.03}_{-0.03}$}} & \textcolor{gray}{\textit{$-8.57^{+0.02}_{-0.03}$}} & $-10.45^{+0.50}_{-0.64}$ & $-10.21^{+0.42}_{-0.42}$ & $-10.37^{+0.52}_{-0.54}$ & $-10.64^{+0.58}_{-0.58}$ & $-10.60^{+0.62}_{-0.62}$ \\
$\log$ AlO & \textcolor{gray}{\textit{$-8.27^{+0.04}_{-0.04}$}} & \textcolor{gray}{\textit{$-8.27^{+0.04}_{-0.07}$}} & \textcolor{gray}{\textit{$-8.28^{+0.04}_{-0.04}$}} & \textcolor{gray}{\textit{$-8.26^{+0.02}_{-0.06}$}} & $-9.17^{+0.30}_{-0.45}$ & $-9.67^{+0.79}_{-1.10}$ & $-9.02^{+0.38}_{-1.26}$ & $-8.78^{+0.21}_{-0.38}$ & $-8.80^{+0.23}_{-0.56}$ \\
$\log$ Mg & \textcolor{gray}{\textit{$-4.15^{+0.03}_{-0.04}$}} & \textcolor{gray}{\textit{$-4.15^{+0.02}_{-0.04}$}} & \textcolor{gray}{\textit{$-4.17^{+0.03}_{-0.03}$}} & \textcolor{gray}{\textit{$-4.16^{+0.02}_{-0.03}$}} & $-7.71^{+1.14}_{-1.31}$ & $-8.65^{+1.10}_{-1.06}$ & $-9.15^{+1.37}_{-1.25}$ & $-8.67^{+1.56}_{-1.48}$ & $-8.28^{+1.65}_{-1.69}$ \\
$\log$ H$_2$S & \textcolor{gray}{\textit{$-4.70^{+0.03}_{-0.04}$}} & \textcolor{gray}{\textit{$-4.70^{+0.02}_{-0.04}$}} & \textcolor{gray}{\textit{$-4.71^{+0.03}_{-0.02}$}} & \textcolor{gray}{\textit{$-4.71^{+0.02}_{-0.03}$}} & $-8.02^{+0.82}_{-0.95}$ & $-8.09^{+1.02}_{-1.09}$ & $-8.16^{+1.27}_{-1.41}$ & $-9.04^{+1.37}_{-1.29}$ & $-8.58^{+1.42}_{-1.46}$ \\
$\log$ AlH & \textcolor{gray}{\textit{$-6.00^{+0.02}_{-0.02}$}} & \textcolor{gray}{\textit{$-6.00^{+0.02}_{-0.02}$}} & \textcolor{gray}{\textit{$-6.01^{+0.01}_{-0.02}$}} & \textcolor{gray}{\textit{$-6.00^{+0.01}_{-0.02}$}} & $-10.04^{+1.00}_{-0.87}$ & $-9.05^{+1.06}_{-1.12}$ & $-9.36^{+0.96}_{-0.99}$ & $-8.95^{+0.91}_{-1.01}$ & $-9.33^{+1.08}_{-1.12}$ \\
$\log$ NH$_3$ & \textcolor{gray}{\textit{$-6.43^{+0.02}_{-0.03}$}} & \textcolor{gray}{\textit{$-6.44^{+0.01}_{-0.01}$}} & \textcolor{gray}{\textit{$-6.45^{+0.02}_{-0.01}$}} & \textcolor{gray}{\textit{$-6.44^{+0.01}_{-0.02}$}} & $-10.35^{+0.85}_{-0.78}$ & $-9.58^{+1.06}_{-0.96}$ & $-9.41^{+0.99}_{-0.99}$ & $-8.72^{+0.88}_{-1.01}$ & $-9.57^{+1.16}_{-1.11}$ \\
\hline
\multicolumn{10}{c}{\textit{Marginal-band species}\space\tablefootmark{(d)}} \\
\hline
$\log$ Na & \textcolor{gray}{\textit{$-5.48^{+0.03}_{-0.04}$}} & \textcolor{gray}{\textit{$-5.49^{+0.02}_{-0.04}$}} & \textcolor{gray}{\textit{$-5.50^{+0.03}_{-0.03}$}} & \textcolor{gray}{\textit{$-5.49^{+0.02}_{-0.03}$}} & $-9.17^{+1.19}_{-1.10}$ & $-8.93^{+1.14}_{-1.10}$ & $-8.68^{+1.43}_{-1.38}$ & $-8.46^{+1.43}_{-1.47}$ & $-8.66^{+1.61}_{-1.57}$ \\
$\log$ Ti & \textcolor{gray}{\textit{$-7.84^{+0.03}_{-0.02}$}} & \textcolor{gray}{\textit{$-7.82^{+0.02}_{-0.02}$}} & \textcolor{gray}{\textit{$-7.83^{+0.02}_{-0.02}$}} & \textcolor{gray}{\textit{$-7.81^{+0.01}_{-0.03}$}} & $-8.99^{+0.58}_{-0.65}$ & $-8.99^{+0.69}_{-0.97}$ & $-9.73^{+0.81}_{-0.85}$ & $-9.69^{+0.86}_{-0.89}$ & $-9.88^{+0.94}_{-0.95}$ \\
$\log$ SH & \textcolor{gray}{\textit{$-5.67^{+0.02}_{-0.01}$}} & \textcolor{gray}{\textit{$-5.66^{+0.04}_{-0.04}$}} & \textcolor{gray}{\textit{$-5.67^{+0.02}_{-0.02}$}} & \textcolor{gray}{\textit{$-5.65^{+0.01}_{-0.04}$}} & $-7.99^{+1.91}_{-1.37}$ & $-5.50^{+0.26}_{-0.83}$ & $-6.30^{+1.03}_{-3.46}$ & $-5.48^{+0.25}_{-3.86}$ & $-5.45^{+0.22}_{-3.84}$ \\
$\log$ CH$_4$ & \textcolor{gray}{\textit{$-6.72^{+0.04}_{-0.04}$}} & \textcolor{gray}{\textit{$-6.74^{+0.03}_{-0.02}$}} & \textcolor{gray}{\textit{$-6.73^{+0.01}_{-0.02}$}} & \textcolor{gray}{\textit{$-6.73^{+0.02}_{-0.04}$}} & $-7.86^{+0.67}_{-0.72}$ & $-7.54^{+0.89}_{-0.97}$ & $-8.67^{+1.30}_{-1.39}$ & $-9.17^{+1.43}_{-1.27}$ & $-8.82^{+1.39}_{-1.38}$ \\
\hline
\multicolumn{10}{c}{\textit{GP Hyperparameters \& Fit Quality}} \\
\hline
$\log a$ & $0.04^{+0.00}_{-0.00}$ & $0.04^{+0.00}_{-0.00}$ & $0.04^{+0.00}_{-0.00}$ & $0.04^{+0.00}_{-0.00}$ & $0.12^{+0.00}_{-0.00}$ & $0.05^{+0.00}_{-0.00}$ & $0.05^{+0.00}_{-0.00}$ & $0.04^{+0.00}_{-0.00}$ & $0.04^{+0.00}_{-0.00}$ \\
$\log \ell$ [nm] & $-1.71^{+0.01}_{-0.01}$ & $-1.72^{+0.01}_{-0.01}$ & $-1.72^{+0.01}_{-0.01}$ & $-1.72^{+0.01}_{-0.01}$ & $-1.71^{+0.01}_{-0.01}$ & $-1.70^{+0.01}_{-0.01}$ & $-1.72^{+0.01}_{-0.01}$ & $-1.72^{+0.01}_{-0.01}$ & $-1.73^{+0.01}_{-0.01}$ \\
$\Delta\ln\mathcal{L}$\space\tablefootmark{(e)} & $-1.1$ & $+0.0$ & $+0.2$ & $-0.0$ & $-126.9$ & $-10.0$ & $-1.3$ & $+0.0$ & $+0.5$ \\
\hline
\end{tabular}}
\tablefoot{The uncertainties are 1$\sigma$ (68\% credible interval) from the posterior distributions. The column headers mark the $N_\mathrm{live}$ configurations (Sect.~\ref{sec:sample_config_convergence}; these runs included the marginal-band species). The grey entries were derived a posteriori rather than directly sampled.\\
\tablefoottext{a}{$T_\mathrm{phot}$ is the photospheric temperature at the pressure of maximal emission contribution, not the bolometric effective temperature. The derived photospheric temperatures are rounded to the nearest degree for clarity.} \tablefoottext{b}{Values are $\log_{10}$ of the VMR.} \tablefoottext{c}{For the free-chemistry columns, this row reports the [C/H] metallicity proxy (Sect.~\ref{sec:chemical_abundances}).} \tablefoottext{d}{Species from the marginal band of the ACF screen ($1 \leq \mathrm{S/N}^{\mathrm{pred}} < 3$; Table~\ref{tab:species_per_target}).} \tablefoottext{e}{Maximum log-likelihood relative to the adopted configuration of each framework (1000 live points for equilibrium and 4000 for free chemistry).}}
\end{table*}

\clearpage
\section{The \textit{K}-band gravity--metallicity degeneracy: 2M0434 as a case study} \label{appdx:2M0434_logg_FeH}
\begingroup
\setcaptype{figure}
\noindent\sidecaption
\begin{overpic}[width={12cm}]{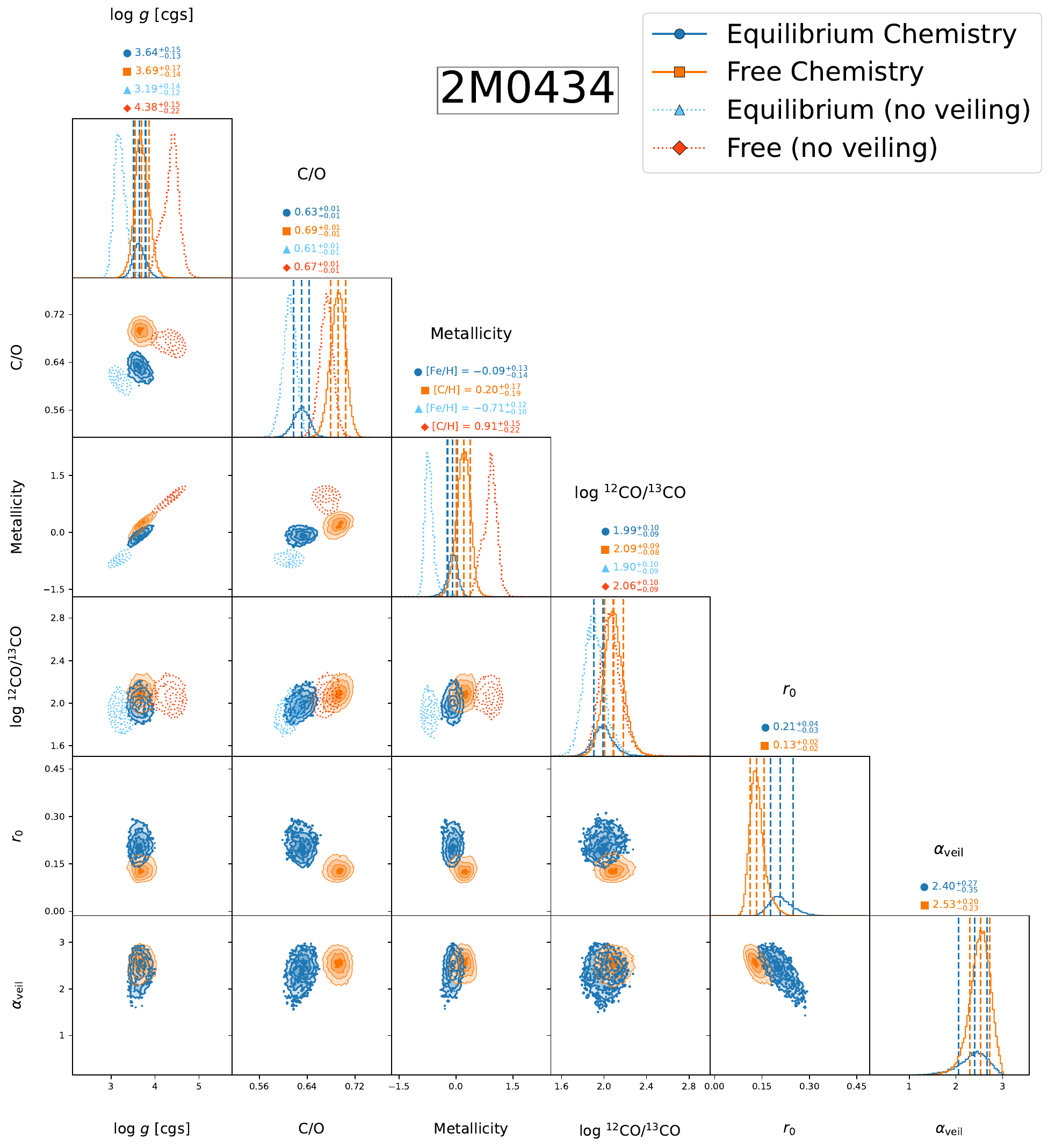}
\put(70,78){\makebox(0,0){%
    \setlength{\tabcolsep}{3pt}%
    \scriptsize
    \begin{tabular}{lcccc}
    \hline\hline
    Framework & $\ln\mathcal{L}_\mathrm{veil}$ & $\ln\mathcal{L}_\mathrm{no\,veil}$ & $\Delta\ln\mathcal{L}$ & $\ln B$ \\
    \hline
    Equilibrium & 57096.40 & 57085.64 & $+10.76$ & $+2.05$ \\
    Free & 57112.32 & 57106.48 & $+5.83$ & $-0.98$ \\
    \hline
    \end{tabular}}}
\end{overpic}
\caption{Joint posterior distributions for 2M0434 in $\log g$, metallicity, C/O ratio, $\log\,^{12}$CO/$^{13}$CO, and veiling parameters $r_0$ and $\alpha_\mathrm{veil}$, for the fiducial equilibrium- (blue, filled contours) and free-chemistry (orange, filled contours) retrievals, and for the corresponding retrievals without the veiling prescription (dotted contours; light blue for equilibrium, red for free). Metallicity is [Fe/H] for the equilibrium runs and [C/H] proxy for the free runs. Retrieved median values with 1$\sigma$ uncertainties are quoted above each diagonal panel with a symbol for the corresponding run (see legend), and the vertical dashed lines mark the 16th, 50th, and 84th percentiles of the fiducial posteriors. The retrievals without veiling do not sample $r_0$ and $\alpha_\mathrm{veil}$ and are absent from the two lowest rows. The inset table below the legend provides a Bayesian comparison of the retrievals with and without veiling. Positive $\ln B$ and $\Delta\ln\mathcal{L}$ favour the inclusion of the veiling component.}
\label{fig:2M0434_corner}
\par\endgroup

\begin{multicols}{2}
Figure~\ref{fig:2M0434_corner} shows the joint posteriors for 2M0434 across both chemistry frameworks, with and without the veiling component described in Sect.~\ref{sec:forward_model}. In all four retrievals, $\log g$ and metallicity are strongly and positively correlated, with Pearson correlation coefficients (following \citealt{Picos2024}) of $r=0.82$ (equilibrium) and $0.91$ (free chemistry) for the fiducial runs, and $0.78$ and $0.95$ for the runs without veiling. The same correlation is present in all six targets of our sample, with $r = 0.67$--$0.92$ across the twelve fiducial retrievals, bracketing the $r = 0.80$--$0.89$ reported by \citet{Picos2024} for their three young M dwarfs. However, only for 2M0434 does the degeneracy translate into broader posteriors ($\lesssim\pm$0.17 in $\log g$ and $\lesssim$0.2~dex in metallicity vs $\lesssim$0.07 and $\lesssim$0.05 for the other targets), so that its solution can move along the degeneracy between chemistry frameworks. Such broad, correlated posteriors are also seen for the youngest, lowest-gravity object of \citet{Picos2024}, J1200 ($\log g = 3.3\pm{0.2}$, Pearson $r = 0.89$), suggesting they may be characteristic of $K$-band retrievals of very-low-gravity late-M dwarfs rather than for 2M0434 alone. This is a known limitation of $K$-band spectroscopy \citep{Zhang2021_BD}: the limited wavelength coverage and scarcity of gravity-sensitive features make it difficult to disentangle surface gravity from metallicity, as both affect molecular band strengths in similar ways. Breaking this degeneracy would require complementary observations at shorter wavelengths (e.g. $J$ or $H$ band) or independent constraints on surface gravity from dynamical mass measurements \citep{Picos2024}, which are not obtainable for the isolated objects in our sample.

With veiling included the two frameworks agree on their solution for 2M0434: $\log g = 3.64^{+0.15}_{-0.13}$ with $[\mathrm{Fe/H}] = -0.09^{+0.13}_{-0.14}$ (equilibrium) versus $\log g = 3.69^{+0.17}_{-0.14}$ with $[\mathrm{C/H}] = 0.20^{+0.17}_{-0.19}$ (free), consistent within uncertainties. However, without it, the frameworks separate to opposite ends of the degeneracy ridge, with a separation of $\sim1.2$~dex in surface gravity and $\sim$1.6~dex in metallicity. All four retrievals used the adopted sampler configuration (Sect.~\ref{sec:sample_config_convergence}), so this separation is not a consequence of under-sampling. The inset table of Fig.~\ref{fig:2M0434_corner} quantifies the preference for the veiling prescription. The log-likelihood improves by $\Delta\ln\mathcal{L} = +10.76$ (equilibrium) and $+5.83$ (free chemistry), but the Bayesian evidence is only weakly in favour of veiling for equilibrium chemistry ($\ln B = +2.05$) and inconclusive for free chemistry ($-0.98$), likely due to the penalty of additional parameters. We retained the veiling component in our retrievals because it is motivated a priori by the red $J-K$ colour and probable youth of 2M0434 (Sect.~\ref{sec:forward_model}), improves the fit in both frameworks, and is the configuration where the frameworks agree with each other and with the low gravities retrieved for other young late-M dwarfs ($\log g \sim 3.3$--$3.9$ $(\pm0.1)$; \citealt{Picos2024}). Both frameworks retrieve a non-zero veiling continuum ($r_0 = 0.21^{+0.04}_{-0.03}$, $\alpha_\mathrm{veil} = 2.40^{+0.27}_{-0.35}$ in equilibrium; $r_0 = 0.13 \pm 0.02$, $\alpha_\mathrm{veil} = 2.53^{+0.20}_{-0.23}$ in free chemistry), with consistent power-law slopes between frameworks. The physical origin of this excess is uncertain, with 2M0434 classified as a Class III source from the Taurus census of \citet{Esplin_2019}. This classification disfavours an inner dust disk as the source of the veiling, with the red $J-K$ colour partly attributable to extinction ($A_J = 1.61$; \citealt{Esplin_2019}). The retrieved veiling may instead act as an effective continuum component, absorbing potential youth-related effects not captured by our forward model or residual continuum-level systematics, rather than a genuine indication of a circum(sub)stellar disk. Regardless of origin, the non-zero $r_0$ indicates a preference for additional continuum emission, which brings the two frameworks into agreement, although their posteriors remain comparatively broad along the degeneracy. Despite the disagreement in $\log g$ and metallicity without veiling, the $^{12}$CO/$^{13}$CO and C/O ratios are robust, with both frameworks returning values within uncertainties regardless of the presence of the veiling component, and the C/O ratio offsets between chemistry frameworks fully consistent with trends seen across the sample (Sect.~\ref{sec:co_ratio}). 
\end{multicols}

\section{Retrieved pressure-temperature profiles}
\label{apdx:PT}
\begingroup
\setcaptype{figure}
\centering
\includegraphics[width=\textwidth]{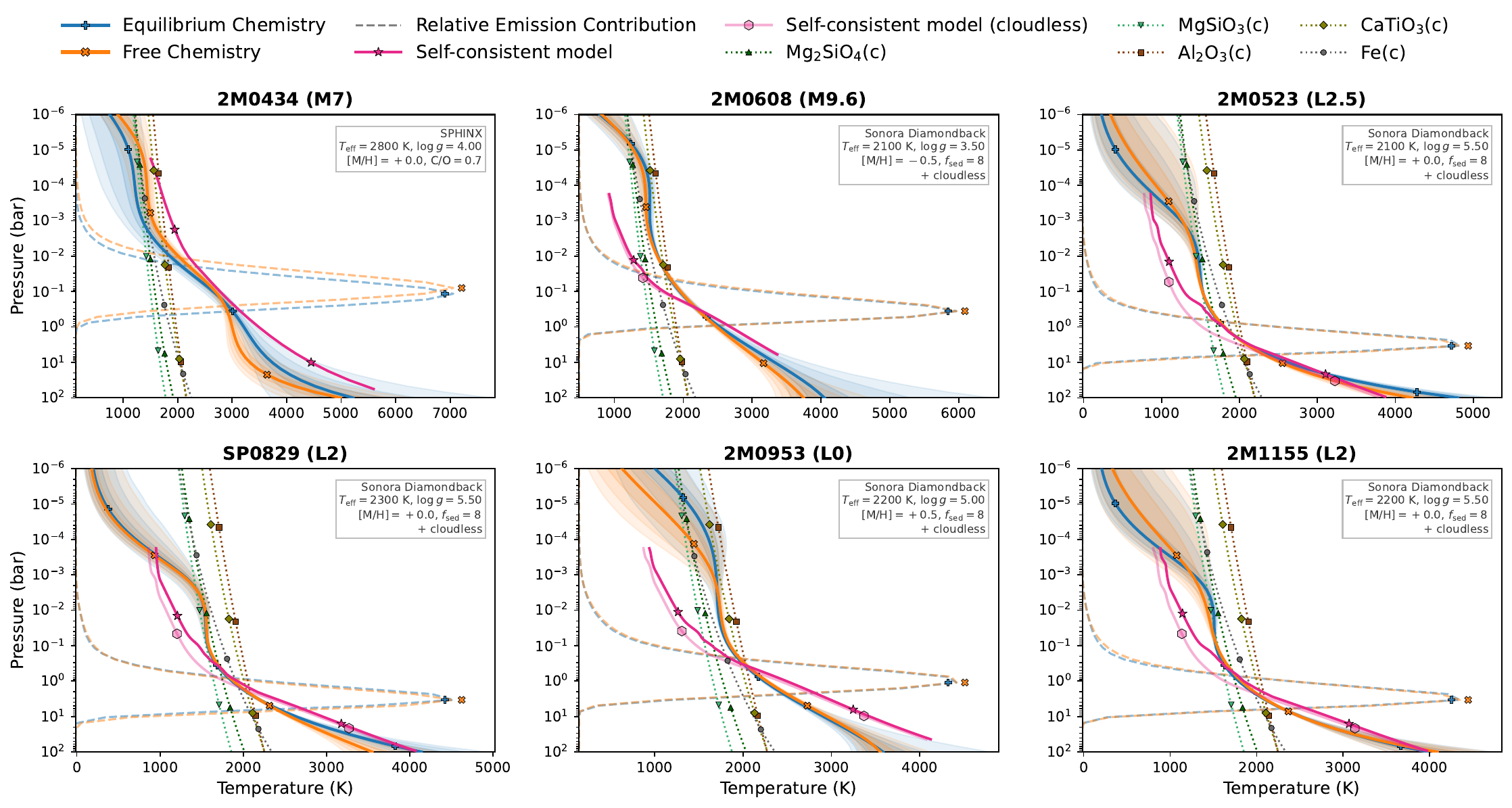}
\caption{Retrieved P-T profiles of the six targets from the equilibrium- (blue) and free-chemistry (orange) retrievals. The solid lines represent the median profiles, and the shaded envelopes the 1, 2, and 3$\sigma$ uncertainties of the profile(s). The dashed curves show the normalised emission contribution, with a symbol near the peak marking the pressure at which $T_\mathrm{phot}$ was computed. The magenta line in each target panel shows a self-consistent model profile from the cloudless SPHINX grid (\citealt{SPHINX_Iyer_2023}) for the warmer M7 2M0434, and a cloudy Sonora Diamondback \citep{SonoraDiamondback_Morley2024} profile for the other targets, with a cloudless Sonora Diamondback profile in light pink. These self-consistent models were selected by proximity to the effective temperature of each target (Table~\ref{tab:target_properties}). For 2M0434 and 2M0953, we used $T_\mathrm{eff} = 2788$~K and 2213~K from the spectral-type relations of \citet{Sanghi_2023} for young (low gravity) and field objects, respectively, and the retrieved equilibrium-chemistry $\log g$ and [Fe/H] (and C/O for SPHINX). We used $f_\mathrm{sed}=8$ for the cloudy Sonora Diamondback models. The selected model parameters are indicated in each panel. For 2M0434 we used the SPHINX grid floor of $\log g = 4.00$, the closest to its retrieved $\log g = 3.64^{+0.15}_{-0.13}$. The Sonora Diamondback grid is computed at solar C/O only ($=0.458$; \citealt{Lodders_2010_C/O, SonoraDiamondback_Morley2024}). Its $f_\mathrm{sed}$ is the cloud parametrisation of \citet{AckermanMarley2001} and does not correspond directly to the $f_\mathrm{sed}$ of our grey cloud model (Eq. \ref{eq:cloud}). The dotted curves represent the condensation curves of Fe, Mg$_2$SiO$_4$, MgSiO$_3$ \citep{Visscher2010}, Al$_2$O$_3$ and CaTiO$_3$ \citep{Wakeford_2016}, evaluated at the retrieved equilibrium-chemistry metallicity of each target. The curve-specific symbols are staggered along all curves to aid in identifying them.}
\label{fig:Pt_profs}
\par\endgroup

\clearpage
\section{Detection significance via cross-correlation} \label{apdx:CCF_overall}
\begingroup
\noindent
\begin{minipage}[t]{0.48\textwidth}
\vspace{0pt}
\setcaptype{table}
\centering
\caption{CCF species detection significance for the equilibrium- and free-chemistry models.}
\label{tab:ccf_snr}
\scriptsize
\setlength{\tabcolsep}{3pt}
\renewcommand{\arraystretch}{1.0}
\begin{tabular}{l c c c c c c}
\hline\hline
Species & 2M0434 & 2M0608 & 2M0523 & SP0829 & 2M0953 & 2M1155 \\
 & (M7) & (M9.6) & (L2.5) & (L2) & (L0) & (L2) \\
\hline
H$_2$O &  &  &  &  &  & \\
\hspace{1em}Free & $\mathbf{76.8}$ & $\mathbf{120.3}$ & $\mathbf{144.4}$ & $\mathbf{108.0}$ & $\mathbf{69.8}$ & $\mathbf{150.4}$ \\
\hspace{1em}Eqbm & $\mathbf{75.1}$ & $\mathbf{119.8}$ & $\mathbf{140.7}$ & $\mathbf{104.5}$ & $\mathbf{70.0}$ & $\mathbf{145.7}$ \\
H$_2^{18}$O &  &  &  &  &  & \\
\hspace{1em}Free & $1.9$ & $2.2$ & $2.2$ & $3.4$ & $2.9$ & $1.7$ \\
\hspace{1em}Eqbm & $2.0$ & $2.2$ & $2.6$ & $3.1$ & $3.1$ & $1.4$ \\
$^{12}$CO &  &  &  &  &  & \\
\hspace{1em}Free & $\mathbf{50.7}$ & $\mathbf{60.4}$ & $\mathbf{92.1}$ & $\mathbf{68.4}$ & $\mathbf{39.8}$ & $\mathbf{93.7}$ \\
\hspace{1em}Eqbm & $\mathbf{49.2}$ & $\mathbf{61.1}$ & $\mathbf{92.8}$ & $\mathbf{68.4}$ & $\mathbf{36.9}$ & $\mathbf{91.3}$ \\
$^{13}$CO &  &  &  &  &  & \\
\hspace{1em}Free & $3.8$ & $\mathbf{5.6}$ & $3.6$ & $2.1$ & $1.2$ & $3.9$ \\
\hspace{1em}Eqbm & $4.1$ & $\mathbf{5.6}$ & $2.7$ & $1.9$ & $1.0$ & $3.5$ \\
C$^{18}$O &  &  &  &  &  & \\
\hspace{1em}Free & $-0.6$ & $0.3$ & $-0.3$ & $0.6$ & $-0.0$ & $-0.2$ \\
\hspace{1em}Eqbm & $-0.3$ & $-0.2$ & $0.7$ & $0.2$ & $-0.1$ & $-0.4$ \\
C$^{17}$O &  &  &  &  &  & \\
\hspace{1em}Free & $0.6$ & $-0.5$ & $-0.0$ & $0.0$ & $-0.1$ & $-0.6$ \\
\hspace{1em}Eqbm & $0.7$ & $-0.5$ & $0.8$ & $0.2$ & $-0.1$ & $-0.5$ \\
HF &  &  &  &  &  & \\
\hspace{1em}Free & $2.3$ & $\mathbf{5.5}$ & $3.5$ & $3.9$ & $2.5$ & $4.9$ \\
\hspace{1em}Eqbm & $2.5$ & $\mathbf{5.4}$ & $3.8$ & $3.8$ & $2.5$ & $4.7$ \\
Ca &  &  &  &  &  & \\
\hspace{1em}Free & $\mathbf{17.6}$ & $\mathbf{33.8}$ & $\mathbf{22.5}$ & $\mathbf{34.8}$ & $14.5^{(a)}$ & $2.9$ \\
\hspace{1em}Eqbm & $\mathbf{17.3}$ & $\mathbf{33.6}$ & $\mathbf{25.3}$ & $\mathbf{21.9}$ & $14.8^{(a)}$ & $4.1$ \\
Sc &  &  &  &  &  & \\
\hspace{1em}Free & $2.4$ & $0.6$ & $-1.1$ & $-0.1$ & $-0.7$ & $-0.3$ \\
\hspace{1em}Eqbm & $2.3$ & $0.3$ & $-1.2$ & $-0.1$ & $-0.5$ & $-0.2$ \\
AlO &  &  &  &  &  & \\
\hspace{1em}Free & $3.1$ & $3.1$ & $1.6$ & $1.0$ & $0.5$ & $1.8$ \\
\hspace{1em}Eqbm & $3.3$ & $3.1$ & $1.8$ & $1.2$ & $0.4$ & $1.9$ \\
\hline
Mg &  &  &  &  &  & \\
\hspace{1em}Free & -- & $0.8$ & $-0.9$ & $0.3$ & $0.3$ & $-0.4$ \\
\hspace{1em}Eqbm & -- & $1.5$ & ${5.6}^{(b)}$ & $4.7$ & $1.9$ & ${7.3}^{(b)}$ \\
Na &  &  &  &  &  & \\
\hspace{1em}Free & $\mathbf{6.2}$ & $4.2$ & -- & -- & $0.7$ & -- \\
\hspace{1em}Eqbm & $\mathbf{6.2}$ & $4.1$ & -- & -- & $0.8$ & -- \\
Ti &  &  &  &  &  & \\
\hspace{1em}Free & $\mathbf{5.1}$ & $1.9$ & -- & -- & $1.8$ & -- \\
\hspace{1em}Eqbm & $\mathbf{5.0}$ & $2.1$ & -- & -- & $1.7$ & -- \\
SH &  &  &  &  &  & \\
\hspace{1em}Free & $2.0$ & $2.7$ & -- & -- & $0.8$ & -- \\
\hspace{1em}Eqbm & $2.2$ & $2.9$ & -- & -- & $0.7$ & -- \\
H$_2$S &  &  &  &  &  & \\
\hspace{1em}Free & -- & $0.6$ & $-1.3$ & $0.5$ & $2.0$ & $-0.7$ \\
\hspace{1em}Eqbm & -- & $0.9$ & $-0.9$ & $0.3$ & $1.0$ & $-1.1$ \\
AlH &  &  &  &  &  & \\
\hspace{1em}Free & -- & $-2.7$ & $-1.4$ & -- & $-1.5$ & -- \\
\hspace{1em}Eqbm & -- & $-2.6$ & $-1.5$ & -- & $-2.3$ & -- \\
NH$_3$ &  &  &  &  &  & \\
\hspace{1em}Free & -- & -- & $-0.9$ & -- & -- & $0.1$ \\
\hspace{1em}Eqbm & -- & -- & $-0.5$ & -- & -- & $0.3$ \\
\hline
\end{tabular}
\tablefoot{The values listed are the S/Ns computed at zero velocity offset, with $\mathbf{S/N \geq 5}$ highlighted in bold as significant detections \citep{Landman2024}. We computed the CCF S/N on a velocity grid $v \in [-500, +500]$~km\,s$^{-1}$ in steps of 1~km\,s$^{-1}$, using the ACF-subtracted CCF (Sect.~\ref{sec:ccf_methods}) at $|v| > 100$~km\,s$^{-1}$ to estimate noise, for the majority of targets. The exceptions are 2M0953, for which we used a velocity grid $v \in [-2000, +2000]$~km\,s$^{-1}$ and $|v| > 500$~km\,s$^{-1}$, and SP0829, for which we used a velocity grid $v \in [-1200, +1200]$~km\,s$^{-1}$ and $|v| > 200$~km\,s$^{-1}$, due to high retrieved rotational velocity and associated spectral broadening for these targets.\\
\tablefoottext{a}{For 2M0953 the Ca CCF peak at the systemic velocity is exceeded by a stronger feature (S/N~$\approx 20$) near $v \approx 800$~km\,s$^{-1}$, so we do not regard it as a detection despite S/N~$> 5$ at RV~$= 0$ (Sect.~\ref{sec:abundances}).}\tablefoottext{b}{For 2M0523 and 2M1155 the equilibrium-chemistry Mg CCF peaks reach S/N~$> 5$, but the free-chemistry models account for the same absorption without Mg. Therefore, we regard Mg as tentative rather than as detected (Sect.~\ref{sec:abundances}).}}
\end{minipage}\hfill
\begin{minipage}[t]{0.48\textwidth}
\vspace{0pt}
\setcaptype{figure}
\centering
\includegraphics[width=\textwidth, height=0.70\textheight, keepaspectratio]{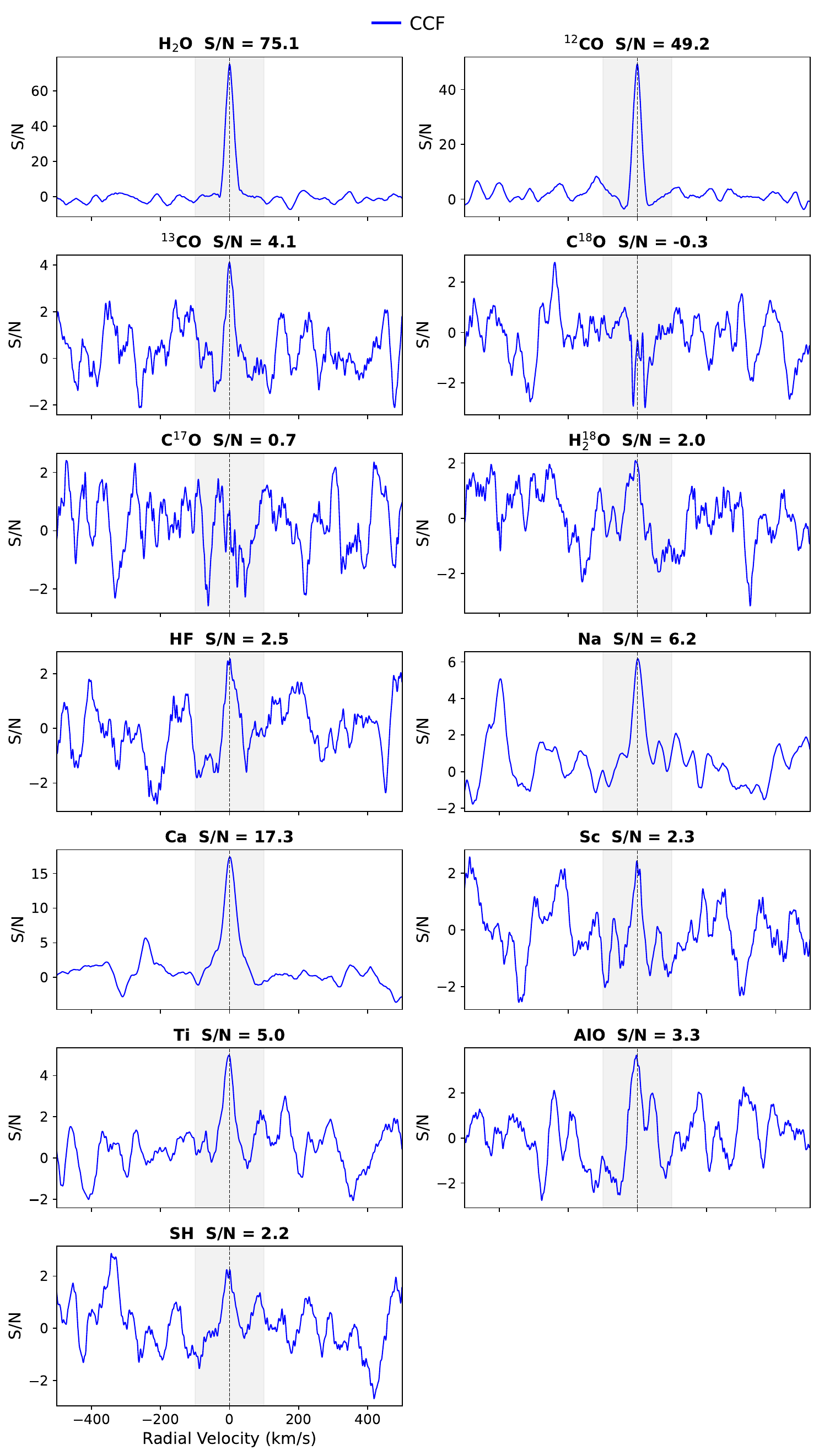}
\caption{CCFs for the 13 species included in the 2M0434 equilibrium-chemistry retrieval. Each panel plots the species total CCF summed over all orders and detectors as a function of radial velocity, computed according to Eq. \ref{eq:ccf}. The shaded band marks the $\pm100\;\mathrm{km\,s^{-1}}$ window excluded when estimating the noise level for 2M0434. The vertical dashed line indicates the systemic velocity of 2M0434. The obtained S/Ns are displayed above each species panel. Significant detections (S/N $\geq 5$; \citealt{Landman2024}) of H$_2^{16}$O, $^{12}$CO, Na, Ca, and Ti are visible with peaks centred at the systemic velocity. A marginal detection is visible for $^{13}$CO (S/N $= 4.1$), and only weak hints of HF (2.5) and AlO (3.3) marked by peaks above the noise centred at the systemic velocity.}
\label{fig:ccf_2m0434}
\end{minipage}
\par\endgroup

\clearpage
\section{Anomalous H$_2^{18}$O signals in 2M0953 and SP0829}
\label{appdx:H218O_appendix}
\begingroup
\setcaptype{figure}
\noindent\sidecaption
\includegraphics[width=11.95cm]{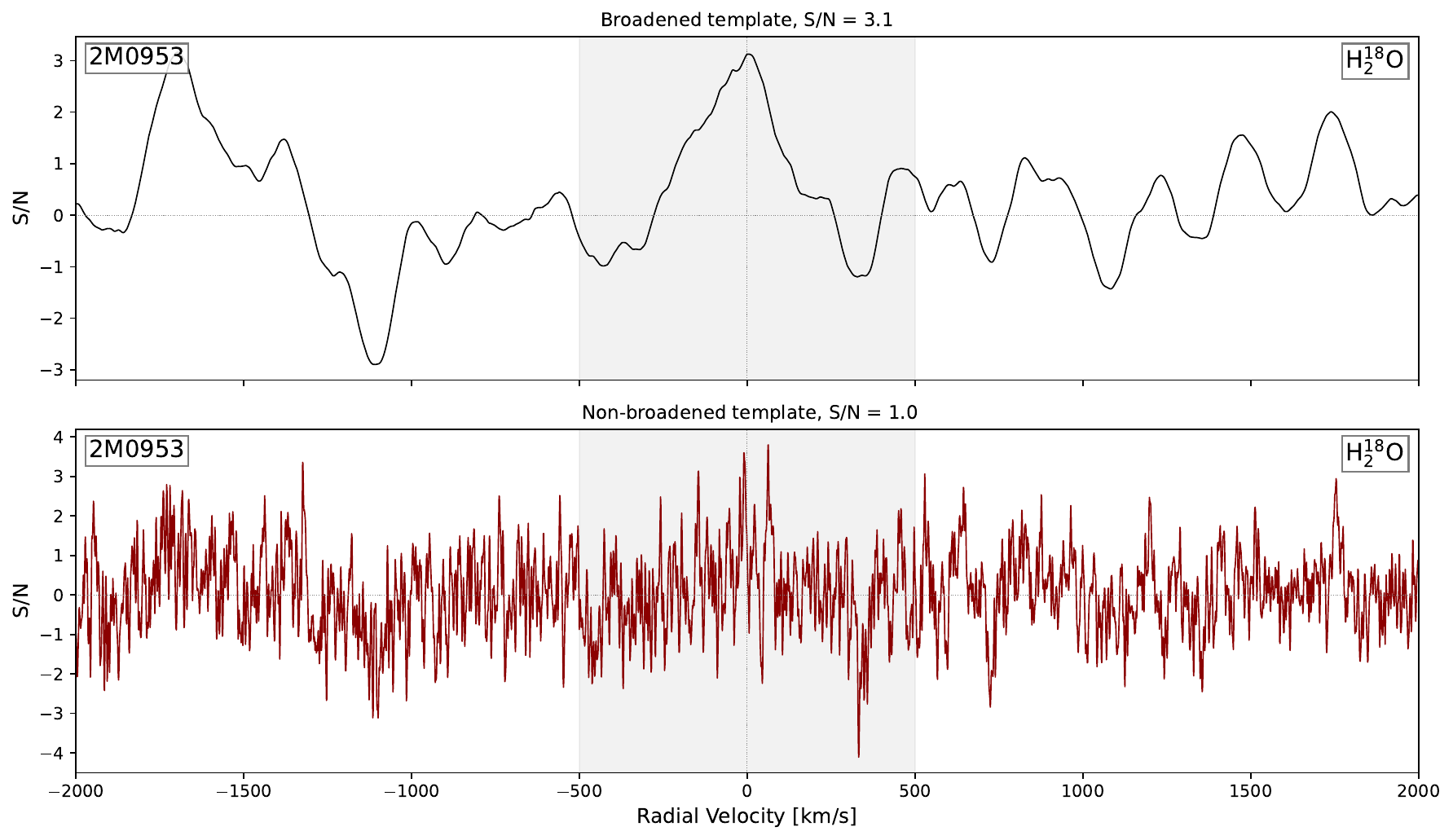}
    \caption{H$_2^{18}$O CCF for 2M0953 computed over $v \in [-2000, +2000]$~km\,s$^{-1}$, with noise estimated from $|v| > 500$~km\,s$^{-1}$ (shaded). \textit{Top}: CCF using the rotationally broadened template with a retrieved $v\sin i = 86.89^{+0.48}_{-0.61}$~km\,s$^{-1}$. \textit{Bottom}: CCF using a non-broadened template ($v\sin i = 0.01$~km\,s$^{-1}$), which would reveal telluric contamination as a narrow amplified peak. The absence of significant peak amplification in the lower panel suggests the signal is not driven by residual telluric contamination. The broadened CCF shows extended structure at amplitudes comparable to the central peak (S/N~$= 3.1$), inconsistent with a genuine molecular detection.}
    \label{fig:h218o_telluric}
\par\endgroup

\begin{multicols}{2}
For the 2M0953 and SP0829 retrievals, we additionally assessed H$_2^{18}$O detection significance first via Bayesian model comparison. We performed an exclusion retrieval with an identical setup to the retrieved model, but with the species effectively removed, and the Bayesian evidence of each exclusion model was compared to that of the complete model, where $\ln B_{\mathrm{H}_2^{18}\mathrm{O}} = \ln \mathcal{Z} - \ln\mathcal{Z}_{\mathrm{w/o}\,\mathrm{H}_2^{18}\mathrm{O}}$. For 2M0953, the Bayesian exclusion retrieval yields $\ln B_{\mathrm{H}_2^{18}\mathrm{O}} = +0.004$, and for SP0829 $-3.52$. For 2M0953, the small positive value is inconclusive, whereas for SP0829 the negative value indicates the exclusion model is moderately preferred. We discuss below the results of the equilibrium-chemistry retrievals and their CCF analysis, with free chemistry showing identical behaviour throughout. 

Both targets have high retrieved rotational velocities: $v\sin i = 86.89^{+0.48}_{-0.61}$~km\,s$^{-1}$ for 2M0953 and $29.48^{+0.40}_{-0.38}$~km\,s$^{-1}$ for SP0829, which broaden all spectral features and can smear CCF power from the central peak into extended wings. For the standard velocity grid, the noise was estimated from $|v| > 100$~km\,s$^{-1}$ (Sect.~\ref{sec:ccf_methods}), a region that may not be signal-free for fast rotators. We therefore expanded the velocity grids to $v \in [-2000, +2000]$~km\,s$^{-1}$ with noise from $|v| > 500$~km\,s$^{-1}$ for 2M0953, and $v \in [-1200, +1200]$~km\,s$^{-1}$ with noise from $|v| > 200$~km\,s$^{-1}$ for SP0829 to obtain H$_2^{18}$O CCF S/N at RV~$= 0$ of 3.1 for 2M0953 and SP0829 (Table~\ref{tab:ccf_snr}). On the expanded grids, both CCFs show extended structure at large velocities far from RV~$= 0$, at amplitudes comparable to the central peak. This pattern is already inconsistent with a genuine detection. We describe below further robustness tests performed to identify the origin and reliability of the signal, ultimately arriving at the conclusion that it is spurious in both cases. We focus on 2M0953 as the more extreme case, with SP0829 showing analogous behaviour throughout.

Telluric absorption features do not share the target's radial velocity, so their CCF signature would appear as a narrow, unbroadened peak at the telluric rest frame in a CCF computed with a non-rotationally broadened template, amplified relative to the rotationally broadened counterpart. We computed the H$_2^{18}$O CCF using both the standard broadened template with retrieved $v\sin i$ and a non-broadened ($v\sin i = 0.01$~km\,s$^{-1}$) version of the same template. The non-broadened CCF shows no significantly amplified peak at the telluric rest frame for 2M0953 (Fig.~\ref{fig:h218o_telluric}), indicating the signal is not driven by telluric residuals. The broadened CCF over the extended $\pm 2000$~km\,s$^{-1}$ range also shows the extended off-centre structure noted above, challenging the validity of a tentative detection.

\end{multicols}

\begingroup
\setcaptype{figure}
\noindent\sidecaption
\includegraphics[width=11.95cm]{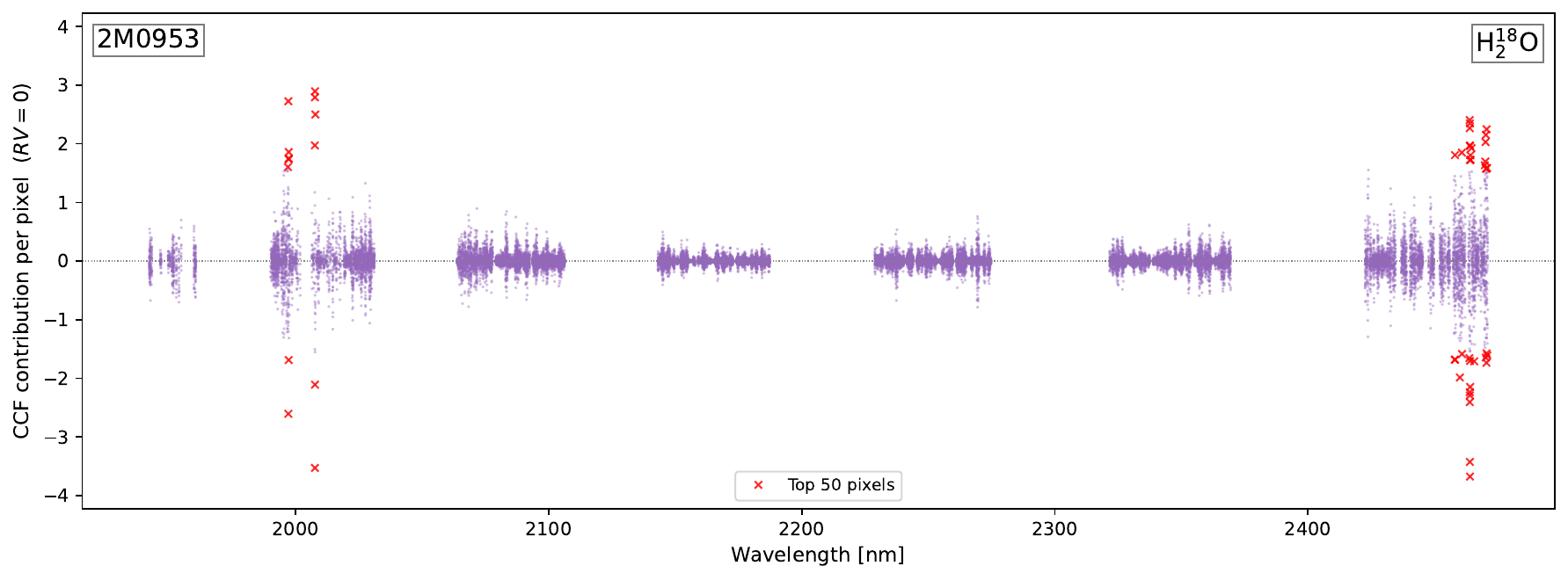}
     \caption{Noise-scaled covariance-weighted H$_2^{18}$O CCF contribution per pixel at RV~$= 0$ for 2M0953, defined as $\vec{T} \times [\Sigma^{-1}\,\vec{R}]$ where $T$ is the species template and $\vec{R}$ is the data-minus-excluded-model residual vector (see Sect.~\ref{sec:ccf_methods}). The red crosses mark the top~50 pixels ranked by absolute contribution. Of these, 37 are concentrated in the last (seventh) spectral order and 13 in the second order.}
     \label{fig:h218o_pixel_contrib}
\par\endgroup

\clearpage
\begingroup
\setcaptype{figure}
\noindent\sidecaption
\vbox{%
  \hbox{\includegraphics[width=11.95cm]{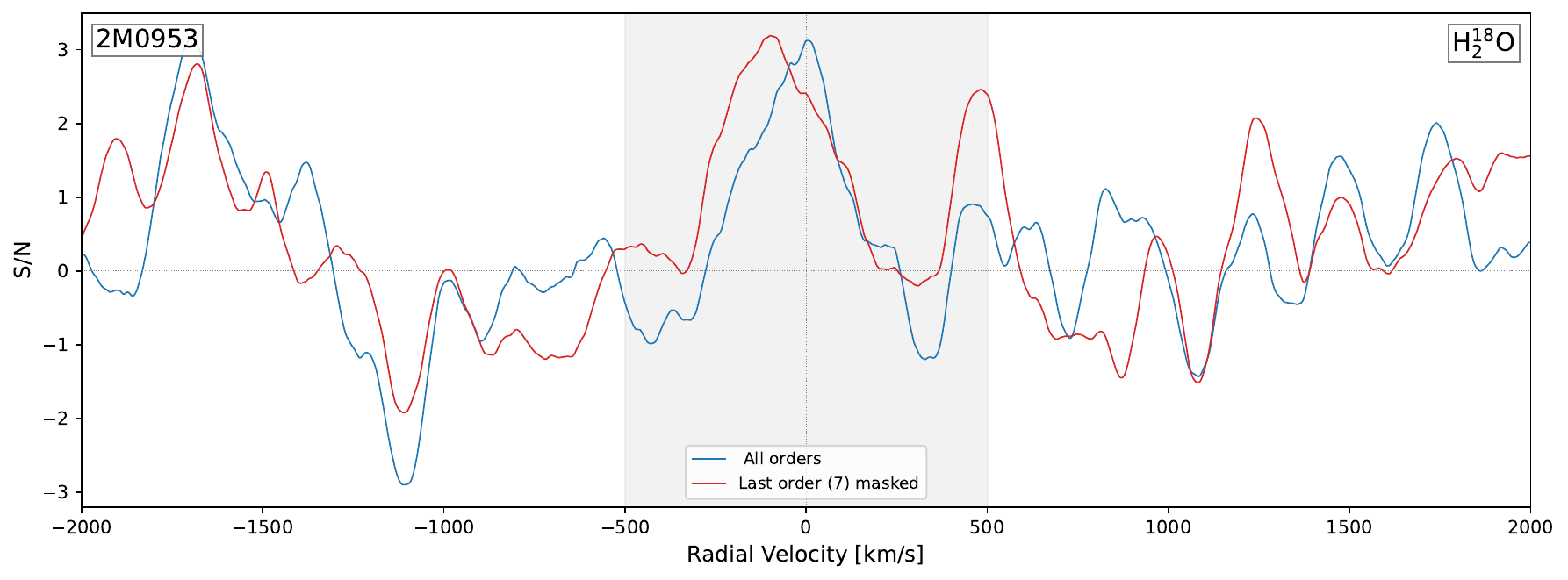}}
  \hbox{\includegraphics[width=11.95cm]{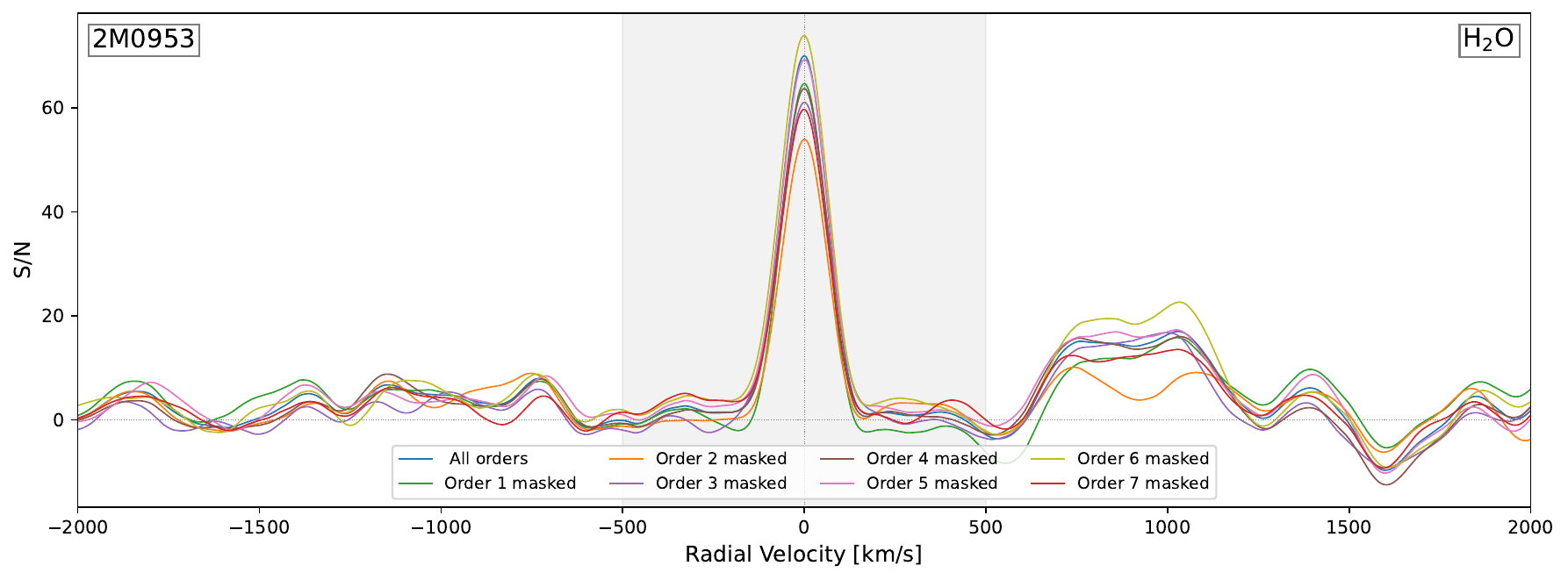}}}
    \caption{\textit{Top:} H$_2^{18}$O CCF with all spectral orders included (blue) and with the last (seventh) order masked (red) for the fiducial 2M0953 retrieval. \textit{Bottom:} H$_2^{16}$O CCF with all spectral orders included (blue) and with each individual spectral order masked for the same retrieval, demonstrating the behaviour of a genuine detection under these tests as opposed to the spurious H$_2^{18}$O detection. Masking the last order reduces H$_2^{18}$O S/N at RV~$= 0$ from $\sim$3.1 to 2.4 and shifts the peak off centre, confirming the disproportionate contribution of the last spectral order as the primary driver of the apparent H$_2^{18}$O signal. This contrasts with the behaviour seen for the H$_2^{16}$O S/N with its peak remaining centred after masking across all individual spectral orders.}
    \label{fig:h218o_last_order}
\par\endgroup

\begin{multicols}{2}
H$_2^{18}$O opacity is relatively uniform across the K2166 band (see Fig.~3 of \citealt{Grasser2025}), so a genuine detection would be expected to draw CCF contribution across spectral orders. To identify which pixels drive the marginal peak at RV~$= 0$, we computed the per-pixel contribution to the CCF S/N, i.e. the individual terms summed in Eq.~\ref{eq:ccf}. For a minor species such as H$_2^{18}$O, the sign of these terms is not diagnostic. The signal in each of the individual pixels lies far below the noise, so the individual terms only reveal a detection in their sum, with the inverse covariance additionally mixing in the residuals of neighbouring pixels. We therefore used this map only to locate where the largest contributions sit for later masking out individual orders from the CCF computation. For 2M0953, the top~50 contributing pixels ranked by absolute contribution are concentrated ($37/50$) in the last spectral order (Fig.~\ref{fig:h218o_pixel_contrib}), with $13/50$ pixels outside this last order.

To investigate the last spectral order as the primary driver of the tentative signal, we recomputed the CCF with all pixels in that order masked. For 2M0953, masking the last order reduces the S/N at RV~$= 0$ from $\sim$3.1 to 2.4 and shifts the peak off centre (Fig.~\ref{fig:h218o_last_order}), a behaviour sharply contrasted by H$_2^{16}$O, whose peak remains centred when each of the seven orders is masked in turn, as expected for its similarly uniform opacity structure (see Fig.~3 of \citealt{Grasser2025}). For SP0829, the behaviour is similar: H$_2^{18}$O S/N drops from 3.1 to 2.4 and a secondary peak near $v \approx 300$~km\,s$^{-1}$ becomes the dominant feature in the masked CCF. In both cases, the disproportionate impact of a single spectral order after masking confirms that the last order is primarily responsible for the apparent H$_2^{18}$O signal.

These tests point to the conclusion that H$_2^{18}$O signals in 2M0953 and SP0829 are spurious, driven by anomalous residuals concentrated in the last spectral order rather than a coherent isotopologue detection across the spectrum or telluric residual contamination. For an ultra-fast rotator such as 2M0953, rotational broadening alone can suppress the CCF amplitude, so the CCF S/N alone is not a decisive diagnostic. The last-order masking test is therefore the more critical evidence: it probes how the signal is distributed across the spectrum independently of its peak amplitude, and indicates the apparent signal originates from anomalous residuals in the last spectral order rather than a genuine atmospheric detection. We report H$_2^{18}$O in 2M0953 and SP0829 as non-detections, with the retrieved isotope ratio values of log\,H$_2^{16}$O/H$_2^{18}$O~$= 1.86^{+0.08}_{-0.07}$ (2M0953) and $2.10^{+0.11}_{-0.09}$ (SP0829) serving as lower limits rather than constrained ratios.
\end{multicols}
\end{appendix}

\end{document}